\PassOptionsToPackage{unicode}{hyperref}
\PassOptionsToPackage{hyphens}{url}
\PassOptionsToPackage{dvipsnames,svgnames,x11names}{xcolor}
\documentclass[
  12pt,
]{article}
\usepackage{amsmath,amssymb}
\usepackage{setspace}
\usepackage{iftex}
\ifPDFTeX
  \usepackage[T1]{fontenc}
  \usepackage[utf8]{inputenc}
  \usepackage{textcomp} % provide euro and other symbols
\else % if luatex or xetex
  \usepackage{unicode-math} % this also loads fontspec
  \defaultfontfeatures{Scale=MatchLowercase}
  \defaultfontfeatures[\rmfamily]{Ligatures=TeX,Scale=1}
\fi
\usepackage{lmodern}
\ifPDFTeX\else
\fi
\IfFileExists{upquote.sty}{\usepackage{upquote}}{}
\IfFileExists{microtype.sty}{% use microtype if available
  \usepackage[]{microtype}
  \UseMicrotypeSet[protrusion]{basicmath} % disable protrusion for tt fonts
}{}
\makeatletter
\@ifundefined{KOMAClassName}{% if non-KOMA class
  \IfFileExists{parskip.sty}{%
    \usepackage{parskip}
  }{% else
    \setlength{\parindent}{0pt}
    \setlength{\parskip}{6pt plus 2pt minus 1pt}}
}{% if KOMA class
  \KOMAoptions{parskip=half}}
\makeatother
\usepackage{xcolor}
\usepackage[margin=1in]{geometry}
\usepackage{setspace}
\usepackage{longtable,booktabs,array}
\usepackage{calc} % for calculating minipage widths
\usepackage{etoolbox}
\makeatletter
\patchcmd\longtable{\par}{\if@noskipsec\mbox{}\fi\par}{}{}
\makeatother
\IfFileExists{footnotehyper.sty}{\usepackage{footnotehyper}}{\usepackage{footnote}}
\makesavenoteenv{longtable}
\usepackage{graphicx}
\makeatletter
\def\maxwidth{\ifdim\Gin@nat@width>\linewidth\linewidth\else\Gin@nat@width\fi}
\def\maxheight{\ifdim\Gin@nat@height>\textheight\textheight\else\Gin@nat@height\fi}
\makeatother
\setkeys{Gin}{width=\maxwidth,height=\maxheight,keepaspectratio}
\makeatletter
\def\fps@figure{htbp}
\makeatother
\providecommand{\tightlist}{%
  \setlength{\itemsep}{0pt}\setlength{\parskip}{0pt}}
\NewDocumentCommand\citeproctext{}{}
\NewDocumentCommand\citeproc{mm}{%
  \begingroup\def\citeproctext{#2}\cite{#1}\endgroup}
\makeatletter
 \let\@cite@ofmt\@firstofone
 \def\@biblabel#1{}
 \def\@cite#1#2{{#1\if@tempswa , #2\fi}}
\makeatother
\newlength{\cslhangindent}
\newlength{\csllabelwidth}
\newenvironment{CSLReferences}[2] % #1 hanging-indent, #2 entry-spacing
 {\begin{list}{}{%
  \setlength{\itemindent}{0pt}
  \setlength{\leftmargin}{0pt}
  \setlength{\parsep}{0pt}
  \ifodd #1
   \setlength{\leftmargin}{\cslhangindent}
   \setlength{\itemindent}{-1\cslhangindent}
  \fi
  \setlength{\itemsep}{#2\baselineskip}}}
 {\end{list}}
\usepackage{calc}

\usepackage{float}
\usepackage{paralist}
\usepackage{fancyhdr}
\usepackage{lastpage}
\usepackage{dcolumn}
\usepackage[font={small}]{caption}
\usepackage{hyperref}
\usepackage{caption}
\usepackage{rotating}
\usepackage{booktabs}
\usepackage[nottoc, numbib]{tocbibind}
\usepackage{graphicx}

\usepackage{algorithm2e, algpseudocode}
\RestyleAlgo{ruled}
\newcounter{algoline}
\AtBeginEnvironment{algorithm}{\setcounter{algoline}{0}}
\usepackage[bottom]{footmisc}
\usepackage{booktabs}

\usepackage{mathtools}
\usepackage{bbm}
\usepackage{booktabs}
\usepackage{longtable}
\usepackage{array}
\usepackage{multirow}
\usepackage{wrapfig}
\usepackage{float}
\usepackage{colortbl}
\usepackage{pdflscape}
\usepackage{threeparttable}
\usepackage{threeparttablex}
\usepackage[normalem]{ulem}
\usepackage{makecell}
\usepackage{xcolor}
\ifLuaTeX
  \usepackage{selnolig}  % disable illegal ligatures
\fi
\usepackage{bookmark}
\IfFileExists{xurl.sty}{\usepackage{xurl}}{} % add URL line breaks if available
\hypersetup{
  colorlinks=true,
  linkcolor={blue},
  filecolor={Maroon},
  citecolor={Blue},
  urlcolor={blue},
  pdfcreator={LaTeX via pandoc}}

\makeatletter
\g@addto@macro\normalsize{%
  \setlength\abovedisplayskip{6pt}
  \setlength\belowdisplayskip{6pt}
  \setlength\abovedisplayshortskip{3pt}
  \setlength\belowdisplayshortskip{3pt}
}
\makeatother

\title{Heterogeneous Peer Effects

with Endogenous Network Formation}
\author{Duong Trinh\footnote{University of Graz, Department of Economics. Universitätsstraße 15 Bauteil, Graz, Austria, 8010. Email: \href{mailto:duong.trinh@uni-graz.at}{\nolinkurl{duong.trinh@uni-graz.at}}} \qquad\qquad Santiago Montoya-Blandón\footnote{University of Glasgow, Adam Smith Business School. Room 620, ASBS Building, 2 Discovery Place, Glasgow, United Kingdom, G11 6EY. Email: \href{mailto:santiago.montoya-blandon@glasgow.ac.uk}{\nolinkurl{santiago.montoya-blandon@glasgow.ac.uk}}}}
\date{}

\begin{document}
\maketitle

\setstretch{1.5}
\begin{abstract}
This paper introduces a new econometric framework for modeling social interactions with heterogeneous peer responses, addressing endogenous link formation. Our Selection-corrected Heterogeneous Spatial Autoregressive (SCHSAR) approach jointly models link formation and outcome determination. We incorporate a finite mixture structure to capture heterogeneity in peer effects and account for unobserved individual-specific factors driving both network formation and outcome equations, addressing network endogeneity for credible estimation of heterogeneous spillover effects. We propose a fully Bayesian data augmentation approach for estimation and inference, overcoming challenges posed to standard likelihood-based methods. A simulation study validates our approach. Our empirical application to an innovation network among U.S. firms reveals significant positive, yet heterogeneous, peer effects on corporate R\&D investments, after accounting for endogenous network formation. The findings highlight varying firm behaviors in response to exogenous R\&D policy shocks and and quantify firm-level direct and spillover effects, offering valuable insights for evidence-based and targeted policy design.

\medskip


\medskip

\end{abstract}

% \newpage
\pagenumbering{arabic}

\section{Introduction}\label{SCHSARsection1}

Modeling social interactions and spillovers is a significant topic in economics, as understanding how individuals' outcomes are shaped by those of their peers within networks is essential for both theoretical insights and policy design, with applications ranging from education (\citeproc{ref-patacchini2017heterogeneous}{Patacchini et al. 2017}; \citeproc{ref-tincani2018heterogeneous}{Tincani 2018}) and labor markets (\citeproc{ref-ioannides2004job}{Ioannides and Loury 2004}) to industrial organization (\citeproc{ref-de2020consumption}{De Giorgi et al. 2020}; \citeproc{ref-fainmesser2016pricing}{Fainmesser and Galeotti 2016}) and beyond. More often than not, networks exhibit substantial heterogeneity in spillover patterns across agents. While weighted or directed networks can capture variations in exposure or interaction intensity, they cannot account for the differences in how agents translate peer behavior into their own decisions, which results in heterogeneous marginal responses to peers and thus heterogeneous peer effects. Capturing such heterogeneity and distinguishing between those who amplify versus those who absorb spillovers is thus crucial for avoiding unintended policy consequences (\citeproc{ref-hong2013heterogeneous}{Hong and Raudenbush 2013}; \citeproc{ref-leung2022causal}{Leung 2022}; \citeproc{ref-bargagli2025heterogeneous}{Bargagli-Stoffi et al. 2025}; \citeproc{ref-auerbach2026local}{Auerbach et al. 2026}). The challenge is compounded by the fact that individuals may select into relationships based on factors that also affect outcomes, creating endogeneity problems. For instance, in education, students form friendship networks based on both observed and unobserved traits, and peer influences on academic outcomes vary across individuals (\citeproc{ref-calvo2009peer}{Calvó-Armengol et al. 2009}). Similarly, in innovation, firms strategically choose R\&D collaboration partners, and the impact of these alliances on innovation activities is not uniform across firms (\citeproc{ref-powell1996interorganizational}{Powell et al. 1996}). Labor market outcomes are shaped by professional networks that workers selectively form, with effects on job search and wages differing by individual characteristics (\citeproc{ref-bayer2008place}{Bayer et al. 2008}), while in agricultural technology adoption, farmers' advice networks influence their adoption decisions heterogeneously (\citeproc{ref-conley2010learning}{Conley and Udry 2010}). Critically, failing to account for endogenous formation of links or substantial heterogeneity can produce severely incorrect inferences about the peer effects. As detailed network data becomes increasingly available, the demand for credible econometric frameworks has intensified, yet existing methods remain inadequate for accommodating the complex realities of economic and social interactions.\footnote{Recent surveys (see, e.g., \citeproc{ref-bramoulle2020peer}{Bramoullé et al. 2020}; \citeproc{ref-kline2020econometric}{Kline and Tamer 2020}) have acknowledged empirical challenges in estimating peer effects in networks.}

This paper addresses these longstanding methodological challenges by proposing a unified framework, the Selection-corrected Heterogeneous Spatial Autoregressive (hereafter referred to as SCHSAR) model, that simultaneously tackles both heterogeneity and endogeneity concerns when modeling network interactions. At its core, the SCHSAR framework jointly models network formation and individual outcomes through two stages, corresponding to two equations. The network formation equation explicitly expresses dyadic links as the realization of decision-making processes influenced by observed characteristics and unobservable individual-specific factors. The outcome equation features a mixture structure in which individuals belong to finite latent types, each governed by distinct set of parameters when they interact with peers in a network. This data-driven grouping captures the rich heterogeneity of peer effects, allowing both the magnitude and direction of the effects to vary across unobserved clusters. Notably, network endogeneity may arise due to unobservable factors affecting both stages, creating the need to correct potential selection bias. The SCHSAR model addresses this issue by accounting for these factors that link the two equations. Overall, SCHSAR offers a flexible and robust approach to draw credible conclusions about peer effects in settings where both individual heterogeneity and network endogeneity are present.

We develop a Bayesian Markov chain Monte Carlo (MCMC) algorithm to estimate the proposed SCHSAR model, offering a straightforward implementation with several practical advantages. First, the algorithm leverages Bayesian data augmentation to achieve computational efficiency while maintaining statistical rigor. By sampling latent variables alongside model parameters from the joint posterior distribution, we overcome the intractable high-dimensional integration problem typically faced in maximum likelihood estimation. This eliminates the numerical optimization challenges while seamlessly incorporating model constraints through prior specifications. The unified Bayesian approach facilitates one-step inference across all unknown quantities, providing principled uncertainty quantification. Second, our implementation is able to deliver rich individual-level insights through posterior prediction. The algorithm generates complete posterior predictive distributions for the individual latent type indicators, enabling a data-driven probabilistic classification of individuals into latent clusters of heterogeneous interaction effects. Moreover, this facilitates an examination of how latent heterogeneity correlates with observable characteristics, enriching the interpretability of the model. Last but not least, the Bayesian framework naturally streamlines inference on policy-relevant quantities that are challenging to analyze from a frequentist perspective. For example, spillin and spillout effects, which are highly nonlinear functions of the model parameters, can be computed directly from posterior samples as estimation byproducts. Comprehensive simulation evidence validates the computational tractability and convergence of our Bayesian implementation. Our Monte Carlo experiments also demonstrate favorable frequentist properties, showing that the SCHSAR model delivers valid inference and improved performance relative to more naive approaches that either ignore endogenous network formation or assume homogeneous interaction effects.

Having developed valid tools for estimation, we apply the proposed SCHSAR framework to examine the formation of an innovation collaboration network among U.S. firms and heterogeneous interaction effects on firm performance, with a focus on their R\&D efforts. Using a sample of \(1,\!150\) firms, the empirical analysis reveals significantly positive, yet heterogeneous, peer effects on corporate R\&D investments, even after controlling for selection bias due to endogenous network formation. Allowing for heterogeneity uncovers two latent firm types: approximately \(34\) percent constitute a peer-driven type with a larger estimated network effect of around \(0.215\) and weaker own-price elasticity near \(-2.2\), while the remaining \(66\) percent of firms belong to a self-driven type with a smaller network effect of around \(0.127\) and stronger own-price elasticity near \(-9.5\). These elasticities capture how sensitive a firm's R\&D investment is to changes in the R\&D tax price. The peer-driven type is more responsive to changes in peers' R\&D investments, while the self-driven type is more sensitive to its own cost structure. Additionally, we disentangle notable transmitters and absorbers in response to exogenous R\&D policy shocks -- firms that excel at absorbing peer innovation are rarely those best at transmitting it -- highlighting distinct roles within the network. Furthermore, this framework facilitates the quantification of firm-level direct, spillin, and spillout effects, offering valuable insights for policy design: targeting firms with high total spillout effects -- often highly central in high-tech sectors -- can accelerate network-wide innovation diffusion under budget constraints.

Our methodology builds on and contribute to three strands of the literature in econometrics. \textit{First}, our strategy to correct selection bias stemming from network endogeneity aligns with subsequent seminal works that integrate structural models of network interaction and formation (e.g., \citeproc{ref-goldsmith2013social}{Goldsmith-Pinkham and Imbens 2013}; \citeproc{ref-hsieh2016social}{Hsieh and Lee 2016}; \citeproc{ref-johnsson2021estimation}{Johnsson and Moon 2021}; \citeproc{ref-qu2015estimating}{Qu and Lee 2015}; and \citeproc{ref-auerbach2022identification}{Auerbach 2022})\footnote{These developments are naturally linked to evolving literature on the econometrics of network formation (see, e.g., \citeproc{ref-chandrasekhar2016econometrics}{Chandrasekhar 2016}; \citeproc{ref-graham2015methods}{Graham 2015}), and researchers are increasingly leveraging recent insights from this literature to expand the Spatial Autoregressive (SAR) framework to account for network endogeneity.}. Particularly, Goldsmith-Pinkham and Imbens (\citeproc{ref-goldsmith2013social}{2013}) and Hsieh and Lee (\citeproc{ref-hsieh2016social}{2016}) introduce Bayesian joint modeling approach that uses latent variables to control for unobserved factors influencing both network formation and outcome. Our SCHSAR framework goes beyond these foundational contributions in several ways. Most notably, we relax the assumption of homogeneous peer effects by incorporating a finite mixture structure that enables heterogeneous responses to peers. Broadly speaking, our model nests Goldsmith-Pinkham and Imbens (\citeproc{ref-goldsmith2013social}{2013}) as a special case when the number of mixture components is reduced to one. Additionally, unlike prior work that focuses solely on unobserved homophily, our framework also allows unobserved degree heterogeneity to drive network formation. This again provides a richer form of individual heterogeneity, facilitating a more robust analysis. Lastly, we adopt a probit instead of logit link function for modeling network formation via introducing normally distributed dyadic shocks. This closely resembles the multivariate probit structure and allows for the use of data augmentation from a Bayesian perspective (\citeproc{ref-ding2023bayesian}{Ding et al. 2023}).

\textit{Second}, our model connects to a broader literature on heterogeneous spatial models (e.g., \citeproc{ref-aquaro2021estimation}{Aquaro et al. 2021}; \citeproc{ref-cornwall2017embracing}{Cornwall and Parent 2017}; \citeproc{ref-lesage2016interpreting}{LeSage and Chih 2016}, \citeproc{ref-lesage2018bayesian}{2018}). We draws inspiration from Cornwall and Parent (\citeproc{ref-cornwall2017embracing}{2017}), who introduce a finite mixture approach to spatial econometric modeling. Their Spatial Autoregressive Mixture (henceforth, SAR-M) model marks an innovative step toward integrating spatial dependence and heterogeneity. However, SAR-M and related models share a common and significant limitation: the assumption that the spatial weights matrix (spatial version of the network adjacency matrix) is exogenously given and fixed. This restricts their empirical applicability, particularly in settings where weights are constructed from data-driven choices or based on strategic decisions by units (for example, GDP, income, or trade flows). In such cases, the endogeneity of the spatial weights matrix becomes a first-order concern, as emphasized in critical works (see \citeproc{ref-han2016bayesian}{Han and Lee 2016}; \citeproc{ref-kelejian2014estimation}{Kelejian and Piras 2014}; \citeproc{ref-qu2015estimating}{Qu and Lee 2015}). Our SCHSAR framework addresses this limitation by endogenizing the network formation process, allowing the adjacency (or spatial weight) matrix to be driven by unobserved factors. In doing so, our model advances Cornwall and Parent's (\citeproc{ref-cornwall2017embracing}{2017}) SAR-M model in a cross-sectional setup, and the insights may be extended to heterogeneous models utilizing panel data, such as in LeSage and Chih (\citeproc{ref-lesage2018bayesian}{2018}).

\textit{Third}, our paper also speaks to the emerging literature on the heterogeneity of peer effects. Several prior studies (e.g., \citeproc{ref-arduini2020identification}{Arduini et al. 2020a}; \citeproc{ref-arduini2020treatment}{b}; \citeproc{ref-beugnot2019gender}{Beugnot et al. 2019}; \citeproc{ref-houndetoungan2026count}{Houndetoungan 2026}) have attempted to incorporate a certain degree of heterogeneity by allowing peer effects to vary across predefined groups based on observable characteristics such as gender or ethnicity. In contrast, our framework offers a more flexible alternative, delivering data-driven unobserved heterogeneity. This is related to the work of Masten (\citeproc{ref-masten2018random}{2018}) and Peng (\citeproc{ref-peng2019heterogeneous}{2019}), who extend the Spatial Autoregressive (SAR) model using random coefficients. Nonetheless, unlike our approach, these studies essentially require the network to be exogenously given -- a considerably skeptical assumption among economists, raising the question of whether relaxing it has major implications for estimates of heterogeneous peer effects.

To the best of our knowledge, econometric models simultaneously accommodating heterogeneous peer effects and accounting for the endogeneity of network or spatial structure remain underdeveloped due to the intertwined challenges. By bridging recent advancements in distinct lines of research, SCHSAR provides a unified and flexible framework that can be readily applied to both social networks and spatial data contexts.

The remainder of this paper is organized as follows. In Section \ref{SCHSARsection2}, we formally present the Selection-corrected Heterogeneous Spatial Autoregressive (SCHSAR) framework, accounting for heterogeneous peer effects and endogenous network formation. In Section \ref{SCHSARsection3}, we develop Bayesian MCMC algorithms to estimate the model and conduct inferences. Next, we evaluate the performance of our method using simulations in Section \ref{SCHSARsection4} and apply the proposed approach to study U.S. firms' innovation network in Section \ref{SCHSARsection5}. Finally, we conclude the paper in Section \ref{SCHSARsection6} with brief remarks on the method and policy recommendations based on our results.

\section{Selection-corrected Heterogeneous Spatial Autoregressive Model}\label{SCHSARsection2}

\subsection{Interaction in Network with Heterogeneous Peer Effects}\label{section2.1}

We consider a network consisting of \(N\) individuals, each indexed by \(i \in \{1,\ldots,N\}\). This network can be described by an \(N \times N\) adjacency matrix \(\mathbf{W}\), whose entries \(w_{ij}\) equal one if individuals \(i\) and \(j\) are connected, and zero otherwise. By definition, the diagonal elements of \(\mathbf{W}\) are always zero. Each individual has an associated outcome of interest, collectively represented by an \(N\)-dimensional vector \(\mathbf{Y} \coloneqq [Y_1, \ldots, Y_N]^\top\). In addition, \(K\) individual characteristics are observed and stacked in an \(N\times K\) matrix \(\mathbf{X}\), with \(k\)-th column denoted by \(X^k\) and \(i\)-th row denoted by \(X_i^\top\).

The general spatial autoregressive (SAR) model for studying interactions in networks is specified as
\begin{equation}
\label{SAR}
Y_i = \lambda \sum_{j=1}^{N} w_{ij}Y_j + X_i^\top \beta + \sum_{j=1}^{N} w_{ij} X_j^\top \delta + u_i,
\end{equation}
where the scalar \(\lambda\) captures the endogenous peer effect (also called network or spillover effect), as an individual's outcome may depend on those of their peers through the connections \(w_{i1}, \ldots, w_{in}\); the collection of \(K\)-dimensional vectors \(\beta \coloneqq [\beta^1,\ldots,\beta^K]^\top\) captures the influence from the individuals' exogenous characteristics on the outcome; \(\delta \coloneqq [\delta^1,\ldots,\delta^K]^\top\) captures the average influence of exogenous characteristics of others mediated through the network (also known as the contextual effects); and \(\mathbf{u} \coloneqq [u_1,\ldots,u_N]^\top\) is a vector of stochastic errors whose elements we assume to be i.i.d. with zero mean and constant variance \(\sigma_u^2\). The interaction model \(\eqref{SAR}\) has been widely studied (\citeproc{ref-bramoulle2009identification}{Bramoullé et al. 2009}; \citeproc{ref-lee2010specification}{Lee et al. 2010}; \citeproc{ref-lin2010identifying}{Lin 2010}, among many others) and is also referred to as the linear-in-means network model. Despite many available variants in the literature, these models typically assume that the peer effect parameter \(\lambda\) is constant and common to all individuals. This means that all individuals are influenced in the same way by an exposure summary of their peers' outcomes (i.e., \(\sum_{j=1}^{N} w_{ij}Y_j\)). However, assuming homogeneous effects is restrictive when some individuals may be highly susceptible to peer effects, while others remain relatively autonomous in their behaviors.

To accommodate such heterogeneity, we assume that our sample is representative of a population composed of a finite number of latent types of individuals, indexed by \(g \in \mathcal{G} \coloneqq \{1,\ldots,G\}\). Each type \(g\) is characterized by its own set of parameters \(\{\lambda_g,\beta_g,\delta_g,\sigma^2_{ug}\}\), which are similar for all individuals of the same type but vary across types. The type is assigned independently among individuals, following a multinomial distribution with probability \(\boldsymbol{\pi} \coloneqq [\pi_1,\ldots,\pi_G]\), where \(0<\pi_g\leq 1\) and \(\sum_{g=1}^G \pi_g = 1\). We denote \(z_{ig}\) as the individual-type indicators such that \(z_{ig}=1\) if \(i\) belongs to type \(g\) and \(0\) otherwise. Combine \(z_i \coloneqq  [z_{i1},z_{i2},\ldots,z_{iG}]\) and stack them in an \(N\times G\) allocation matrix \(\mathbf{z} \coloneqq [z_1^\top, \ldots, z_N^\top]^\top\). These indicators are stochastic and unobserved, and the probability that the latent type of \(i\) is \(g\) is \(\mathbbm{P}(z_{ig} =1) = \pi_{g}\). We denote \(\mathcal{I}_g\) as the set of type-\(g\) individuals and \(N_g\) as the corresponding cardinality. The sets \(\mathcal{I}_1,\ldots,\mathcal{I}_G\) thus form a partition of \(\{1,\ldots,N\}\) such that \(\sum_{g=1}^{G} N_g = N\). In this way, all parameters in the original model \(\eqref{SAR}\) become type-specific as follows\footnote{From a microfounded perspective, we first consider \(G\) independent groups where individuals play a simultaneous-move game within each group, with the linear-quadratic payoff for \(i \in \mathcal{I}_g\) given by
  \[
  U_i(Y_i,Y_{-i}) \coloneqq \lambda_g Y_i\Big(\sum_{j\in\mathcal{I}_g} w_{ij}^{(g)}Y_j\Big) + Y_i\Big(X_i^\top\beta_g+\sum_{j\in\mathcal{I}_g}w_{ij}^{(g)}X_j^\top \delta_g + u_i\Big) -\frac{1}{2}Y_i^2.
  \]
  The first term captures the effect of strategic complements (substitutes) from \(i\)'s peers when \(\lambda_g>0\) (\(\lambda_g<0\)). The unique best response function derived from utility maximization leads to the standard SAR specification for each group \(g\). Next, we allow for linkages both within and between groups in the realized network \(\mathbf{W}\). The peers' exposure summaries thus include all individuals in the network, regardless of their groups, i.e.~\(\sum_{j=1}^N w_{ij}Y_j\) and \(\sum_{j=1}^N w_{ij}X_j\). Under the same linear--quadratic preferences, the resulting best-response function yields exactly the heterogeneous SAR specification in \(\eqref{HSAR1}\).}
\begin{equation}
\label{HSAR1}
Y_i = \lambda_g \sum_{j=1}^{N} w_{ij}Y_j + X_i^\top \beta_g + \sum_{j=1}^{N} w_{ij} X_j^\top \delta_g + u_i, \quad u_i \sim \mathcal{N}(0, \sigma^2_{ug}),
\end{equation}
for each individual \(i \in \mathcal{I}_g\). Collect type-specific parameters into \(G\)-dimensional vectors
\begin{equation*}
\begin{gathered}
\check\lambda=[\lambda_1,\ldots,\lambda_G]^\top,\quad
\check\sigma_u^2=[\sigma^2_{u1},\ldots,\sigma^2_{uG}]^\top,\\
\check\beta^k=[\beta_1^k,\ldots,\beta_G^k]^\top,\quad
\check\delta^k=[\delta_1^k,\ldots,\delta_G^k]^\top, \quad k=1,\ldots,K.
\end{gathered}
\end{equation*}
The corresponding \(N\)-dimensional vectors of individual-specific parameters are obtained through the mapping
\[
\tilde\lambda=\mathbf z\check\lambda,\quad
\tilde\sigma_u^2=\mathbf z\check\sigma_u^2,\quad
\tilde\beta^k=\mathbf z\check\beta^k,\quad
\tilde\delta^k=\mathbf z\check\delta^k.
\]
Accordingly, a heterogeneous version of the spatial autoregressive model (HSAR) can be written as
\begin{equation}
\label{HSAR2}
Y_i = \tilde{\lambda}_i \sum_{j=1}^N w_{ij}Y_j + X_i^\top \tilde{\beta}_i + \sum_{j=1}^N w_{ij} X_j^\top \tilde{\delta}_i + u_i, \quad u_i \sim \mathcal{N}(0, \tilde{\sigma}^2_{u,i}),
\end{equation}
where \(\tilde{\lambda}_i\), \(\tilde{\beta}_i\), \(\tilde{\delta}_i\), and \(\tilde{\sigma}^2_{u,i}\) are individual-specific parameters for agent \(i\), which depend on what type he/she is assigned and parameters of that type.

Let \(\mathcal{L} \coloneqq \text{diag}(\tilde\lambda)\) be an \(N\times N\) diagonal matrix of individual peer-effect parameters. Given that the outcomes for all individuals \(i=1,\ldots,N\) obey equations analogous to \(\eqref{HSAR2}\), the system of equations can be compactly written in matrix notation as
\begin{equation}
\label{HSAR3}
\mathbf{Y} = \mathcal{L}\mathbf{W}\mathbf{Y} + \sum_{k=1}^K\tilde{\beta}^k \circ X^k + \sum_{k=1}^K \tilde{\delta}^k \circ (\mathbf{W}X^k)
+ \mathbf{u},
\end{equation}
where \(\circ\) denotes the Hadamard (element-wise) multiplication. This can be further expressed in the reduced form below
\begin{equation}
\label{HSAR4}
\mathbf{Y} = \left[\mathbf{I}_N - \mathcal{L}\mathbf{W}\right]^{-1}\left(\sum_{k=1}^K\tilde{\beta}^k \circ X^k + \sum_{k=1}^K \tilde{\delta}^k \circ (\mathbf{W}X^k) + \mathbf{u}\right).
\end{equation}
We define \(\mathbf{\tilde{Y}} \coloneqq \left[\mathbf{I}_N - \mathcal{L} \mathbf{W}\right] \mathbf{Y}\), which can be interpreted as the outcome with endogenous peer effects removed. Additionally, \(\bar{Y}_{(i)} \coloneqq \sum_{j=1}^N w_{ij}Y_j\) and \(\bar{X}_{(i)}^\top \coloneqq \sum_{j=1}^N w_{ij} X_j^\top\) are the summary measures of peers' outcomes and peers' characteristics for individual \(i\). These transformations facilitate the derivation of the likelihood function in the later sections\footnote{We note how our setup with latent types directly leads to a finite mixture approach. Knowing the latent individual-type indicator \(z_{ig} = 1\) informs us: \(\tilde{Y}_i\sim \mathcal{N}\bigl(X_i^\top\beta_g + \bar{X}_{(i)}^\top\delta_g,\sigma^2_{ug}\bigr)\). Hence, marginalizing over the type indicator \(z_i\), we obtain the mixture distribution: \(p\left( \tilde{Y}_i  \mid \mathbf{X}, \{\lambda_g,\beta_g,\delta_g,\sigma^2_{ug},\pi_g\}_{g=1}^G \right) \propto \sum_{g=1}^{G} \pi_g \mathcal{N}(X_i^\top\beta_g + \bar{X}_{(i)}^\top\delta_g,\sigma^2_{ug}),\text{ where } \sum_{g=1}^{G}\pi_g = 1.\)}.

\emph{Remark 1.} The HSAR specification offers greater flexibility in modeling peer effects than the standard SAR. This relaxes the assumption that peer effects must be uniformly positive or negative across all individuals. Some may be positively influenced, adopting behaviors or attitudes from their peers, whereas others may react negatively. Furthermore, the HSAR specification accounts for varying intensities of peer influence, as represented by the magnitude of \(\vert\tilde{\lambda}_i\vert\). It is important to note that, when introducing various latent types, the model permits unrestricted interactions both within and between types in the network. Individuals who tend to share the same type, by definition, exhibit similar ``receiving rates,'' but this is not strictly tied to their network positions or any predetermined characteristics (\emph{ex-ante}). We will demonstrate later that, by placing priors for both type indicators (\(\mathbf{z}\)) and assignment probabilities (\(\boldsymbol{\pi}\)), their updated posteriors produce a data-driven probabilistic assignment of individuals to types (groups).

\emph{Remark 2.} For simplicity, we present here a local-aggregate model specification (i.e., the adjacency matrix \(\mathbf{W}\) is not row-normalized), but the HSAR is also applicable to a local-average model (i.e., the matrix \(\mathbf{W}\) is row-normalized such that each row sums to unity).\footnote{Aggregate and average models are different in terms of behavioral foundations, which entails different interpretations for the peer effect parameter \(\lambda\) (\citeproc{ref-liu2014endogenous}{Liu et al. 2014}).} The existence and uniqueness of the Nash equilibrium require that \(\mathbf{I}_N - \mathcal{L}\mathbf{W}\)
to be invertible, which holds when the spectral radius of \(\mathcal{L}\mathbf{W}\)
is less than one. Following Kelejian and Prucha (\citeproc{ref-kelejian2010specification}{2010}), we restrict the value of all \(\lambda_g\) to the interval \((-1/\tau^\ast,1/\tau^\ast)\), where \(\tau^\ast \coloneqq \min\left\{\underset{1\leq i\leq N}{\max}\sum_{j=1}^N \vert w_{ij}\vert,\underset{1\leq j\leq N}{\max}\sum_{i=1}^N \vert w_{ij}\vert\right\}\). When \(\mathbf{W}\) is a row-normalized matrix, the simpler condition \(\vert\lambda_g\vert < 1\) for all \(g=1,\ldots,G\) is sufficient. We impose this restriction throughout the paper and on the estimation procedure for \(\lambda_g\).

While the HSAR model delivers a flexible approach to heterogeneous peer effects, it treats the adjacency matrix \(\mathbf W\) as exogenous -- a common yet strong assumption. In social and economic networks, however, links are chosen, and the same unobserved attributes that shape linking decisions may also affect individual outcomes. When this occurs, \(\mathbf W\) becomes endogenous in the outcome equation, generally inducing bias in the HSAR estimation. Motivated by this concern, we next model \(\mathbf W\) explicitly as the outcome of a strategic network formation process. This provides an explicit selection mechanism that rationalizes \(\mathbf W\) and clarifies how unobservables generate correlation between peer variables and outcome disturbances. In doing so, we can correct for selection bias and obtain credible estimates of heterogeneous peer effects.

\subsection{Strategic Network Formation}\label{section2.2}

Network formation is modeled through a strategic choice framework, where individuals decide to form a link based on perceived utility gain. We consider each dyad \((i,j)\) composed of individuals \(i\) and \(j\), with \(i\neq j\). The respective marginal utilities\footnote{That is, the difference in utility between two options, linking or not.} individuals \(i\) and \(j\) receive from forming a link are
\begin{equation}
\label{utility}
U_{ij}(\epsilon^{\ast}_{ij}) = \psi_{ij} + \epsilon^{\ast}_{ij} \quad \text{and} \quad U_{ji}(\epsilon^{\ast}_{ji}) = \psi_{ji} + \epsilon^{\ast}_{ji},
\end{equation}
where \(\psi_{ij}\) and \(\psi_{ji}\) are score functions that depend on individual attributes of \(i\) and \(j\), and \((\epsilon^{\ast}_{ij},\epsilon^{\ast}_{ji})\) is a pair of idiosyncratic shocks that captures the uncertainty in their connection decision.

We assume the score function to be symmetric and deterministic across dyads. In addition, while each dyad can fully observe their characteristics when making decisions, there exists a component unobservable to the researcher. Specifically,
\begin{equation}
\label{linkscore}
\psi_{ij} = \psi_{ji} = C_{ij}^\top\gamma + f(a_{i}, a_{j}).
\end{equation}
In equation \(\eqref{linkscore}\), \(C_{ij}\) is an \(L\)-dimensional vector of dyad-specific regressors derived from the observed exogenous characteristics of individuals \(i\) and \(j\). For example, \(C_{ij}^l=\vert c_{i}^l- c_{j}^l\vert\) reflects homophily in observables, one of the key features in the network formation literature to acknowledge that individuals prefer linking to similar others. Furthermore, the unobserved component, \(f(a_{i}, a_{j})\), is a function of individual unobserved characteristics. Here, we have introduced \(a_i\) as a scalar summary of individual \(i\)'s unobserved characteristics, which is also regarded as an individual random effect. We assume \(a_i\) are independent across individuals with a commonly used parametric distribution \(p(a_i\vert\theta_a)\). The specification of \(f(a_{i}, a_{j})\) depends on how \(a_i\) drives link preferences, and thus could be in the form of either unobserved degree heterogeneity or unobserved homophily. The choice is determined by which feature is more likely to be predominant.

When unobserved homophily is exhibited, the larger the difference in unobservables between individuals \(i\) and \(j\), the less likely they are to connect. To capture homophily on unobserved dyad attributes, we use the conventional specification in the literature (see, e.g., \citeproc{ref-goldsmith2013social}{Goldsmith-Pinkham and Imbens 2013}; \citeproc{ref-hsieh2016social}{Hsieh and Lee 2016}).
\begin{equation}
f(a_{i}, a_{j}) = - \vert a_{i} - a_{j} \vert .
\end{equation}
Apart from homophily, Graham (\citeproc{ref-graham2017econometric}{2017}) highlighted that individual heterogeneity in the number of links (a.k.a. degree), is another common feature of social networks. This is due to the fact that, the degree varies among individuals in many social networks -- a few individuals serve as prominent ``hubs'' with numerous links, whereas others only own a few links. In addressing unobserved degree heterogeneity, we build on studies that accommodate this feature in network formation models (e.g., \citeproc{ref-ding2023bayesian}{Ding et al. 2023}; \citeproc{ref-dzemski2019empirical}{Dzemski 2019}) and incorporate individual-specific heterogeneity in an additive manner
\begin{equation}
f(a_{i}, a_{j}) = a_{i} + a_{j}.
\end{equation}
The unobserved individual characteristic \(a_i\) can be interpreted as social capital, which enhances the likelihood of forming a link. Depending on the context, these characteristics may include trustworthiness, socioeconomic status, or charisma. These intangible factors also tend to affect individual activity outcomes, as evidenced in previous studies combining network formation and interaction (\citeproc{ref-han2021spatial}{Han et al. 2021}; \citeproc{ref-johnsson2021estimation}{Johnsson and Moon 2021}; \citeproc{ref-weng2023beyond}{Weng and Parent 2023}, among others).

\emph{Remark 3.} To facilitate the discussion of the identification and estimation procedure that will be addressed later, we assume \(a_i \sim \mathcal{N}(0,\sigma_a^2)\) in the context of unobserved degree heterogeneity, and \(a_i \sim \mathcal{C}at(\rho_1,\ldots,\rho_S)\) when considering unobserved homophily.

A link between individuals \(i\) and \(j\) is formed if and only if it improves the average utility of \(i\) and \(j\) given by
\begin{equation}
w^{\ast}_{ij} \coloneqq \frac{U_{ij}(\epsilon^{\ast}_{ij}) + U_{ji}(\epsilon^{\ast}_{ji})}{2} = \psi_{ij} + \frac{\epsilon^{\ast}_{ij}+\epsilon^{\ast}_{ji}}{2}\geq 0.
\end{equation}
Thus, binary link indicators representing the network are realizations of these latent average utilities
\begin{equation}
w_{ij} = w_{ji} = \begin{cases}
1 \quad \text{if } w^{*}_{ij} \geq 0,
\\
0 \quad \text{if } w^{*}_{ij} < 0.
\end{cases}
\quad \text{for } i = 1,\ldots,N; j = 1,\ldots,i-1.
\end{equation}

By defining the average of dyad-level shocks \(\epsilon_{ij} \coloneqq \frac{\epsilon^{\ast}_{ij}+\epsilon^{\ast}_{ji}}{2}\), we can write the network formation model succinctly as follows
\begin{equation}
\label{NF}
\begin{split}
w^{*}_{ij} &= C^\top_{ij}\gamma + f(a_{i}, a_{j}) + \epsilon_{ij},\\
w_{ij} &= w_{ji} = \mathbbm{1}(w^{*}_{ij} \geq 0 ), \quad \text{for } i = 1,\ldots,N; j = 1,\ldots,i-1;\\
\end{split}
\end{equation}
where \(\mathbbm{1}(\cdot)\) is the indicator function. Next, we define \(A_{ij} \coloneqq f(a_i,a_j)\) for \(i=1,\ldots,N\) and \(j=1,\ldots,N\) with \(i\neq j\). Substituting into the representation of latent dyadic utilities \(w^{*}_{ij}\), we obtain
\begin{equation}
w_{ij}^\ast = C_{ij}^\top \gamma + A_{ij} + \epsilon_{ij}.  
\end{equation}
Following the ways the elements of the adjacency matrix \(\mathbf{W} = [w_{ij}]\) are indexed, we can stack across row \(i\) for a given \(j\) as detailed below
\begin{equation}
w_{-j}^\ast = C_{-j} \gamma + A_{-j} + \epsilon_{-j}, \quad j = 1,\ldots, N;
\end{equation}
where \(w_{-j}^\ast\), \(A_{-j}\), and \(\epsilon\) are \((N-1)\)-dimensional vectors, \(C_{-j}\) is a \((N-1)\times L\) matrix
\[
\begin{aligned}
w^*_{-j} &=
\begin{bmatrix}
w^*_{1j}\\
\vdots\\
w^*_{j-1,j}\\
w^*_{j+1,j}\\
\vdots\\
w^*_{Nj}
\end{bmatrix},
C_{-j} =
\begin{bmatrix}
C_{1j}^\top\\
\vdots\\
C_{j-1,j}^\top\\
C_{j+1,j}^\top\\
\vdots\\
C_{Nj}^\top
\end{bmatrix},
A_{-j} =
\begin{bmatrix}
A_{1j}\\
\vdots\\
A_{j-1,j}\\
A_{j+1,j}\\
\vdots\\
A_{Nj}
\end{bmatrix},
\epsilon_{-j} =
\begin{bmatrix}
\epsilon_{1j}\\
\vdots\\
\epsilon_{j-1,j}\\
\epsilon_{j+1,j}\\
\vdots\\
\epsilon_{Nj}
\end{bmatrix}.
\end{aligned}
\]

We can then stack across the index \(j\) to obtain
\begin{equation}
\label{NF2}
\mathbf{w^\ast} = \mathbf{C}\gamma + \mathbf{A} + \boldsymbol{\epsilon},
\end{equation}
where \(\mathbf{w^\ast}\), \(\mathbf{A}\), and \(\boldsymbol{\epsilon}\) are \(N(N-1)\)-dimensional vectors and \(\mathbf{C}\) is a \(N(N-1)\times L\) matrix
\[ 
\begin{aligned}
\mathbf{w}^\ast &=
\begin{bmatrix}
w^*_{-1}\\
\vdots\\
w^*_{-N}
\end{bmatrix},
\mathbf{C} =
\begin{bmatrix}
C_{-1}\\
\vdots\\
C_{-N}
\end{bmatrix},
\mathbf{A} =
\begin{bmatrix}
A_{-1}\\
\vdots\\
A_{-N}
\end{bmatrix},
\boldsymbol{\epsilon} =
\begin{bmatrix}
\epsilon_{-1}\\
\vdots\\
\epsilon_{-N}
\end{bmatrix}.
\end{aligned}
\]

As the uncertainty in the linking decision comes solely through \(\epsilon_{ij}\) in \eqref{NF}, we lastly assume that this idiosyncratic error is i.i.d across dyads with a standard normal distribution such that
\begin{equation}
\label{probitshocks}
p(\boldsymbol{\epsilon}) = \prod_{i=1}^{N} \prod_{j<i}\phi(\epsilon_{ij}),
\end{equation}
where \(\phi(\cdot)\) denotes the standard normal density. This assumption is equivalent to setting the distributions of the original \(\epsilon^{\ast}_{ij}\) as \(\mathcal{N}(0,2)\) independently across dyads. Fixing the variance of shocks, which is unity due to our normalization, can be seen as an identifying restriction. This is standard in models with binary dependent variables (see, e.g., \citeproc{ref-cameron2005microeconometrics}{Cameron and Trivedi 2005}; \citeproc{ref-chan2019bayesian}{Chan et al. 2019}). Indeed, as we only observe whether a specific link was formed or not, all scalings of the idiosyncratic shocks will be observationally equivalent. Furthermore, while the distribution of \(\epsilon_{ij}\) is often specified as logistic in the network formation literature, the advantages of \(\eqref{probitshocks}\) are demonstrated in Ding et al. (\citeproc{ref-ding2023bayesian}{2023}). They advocate that this assumption leads to a normal likelihood for the latent variables, which facilitates the incorporation of other elements into their sampling scheme in a Bayesian context.

Let \(\mathbf a \coloneqq [a_1,\ldots,a_N]^\top\). Conditional on \(\left(\mathbf{C},\mathbf{a}\right)\), the augmented likelihood\footnote{Without introducing \(\mathbf{w}\) as latent data, the likelihood of observing network \(\mathbf{W}\) is given by: \(p(\mathbf{W}\mid \mathbf{C}, \mathbf{a}, \gamma ) = \prod_{i=1}^{N}\prod_{j<i}[1-\Phi(\psi_{ij})]^{1-w_{ij}}[\Phi(\psi_{ij})]^{w_{ij}}.\) When used in a Bayesian setting, it does not yield conjugate full conditionals, making sampling from the posterior distribution of parameters computationally expensive and less efficient.} function for parameters in the network formation model in \eqref{NF} is associated with the joint distribution of the observed network and latent link utilities \((\mathbf{W},\mathbf{w}^\ast)\) and can be expressed as
\begin{equation}
\label{NetworkLik}
\begin{split}
p\bigl(\mathbf{W},\mathbf{w}^\ast &\mid \mathbf{C}, \mathbf{a}, \gamma\bigr) = p\bigl(\mathbf{W}\mid \mathbf{w}^\ast) \cdot p(\mathbf{w}^\ast\mid \mathbf{C}, \mathbf{a}, \gamma\bigr)\\
&= \prod_{i=1}^{N} \prod_{j<i}\left[\mathbbm{1}(w_{ij}^\ast \geq 0)\mathbbm{1}(w_{ij}=1)+\mathbbm{1}(w_{ij}^\ast < 0)\mathbbm{1}(w_{ij}=0)\right] \varphi\bigl(w_{ij}^\ast; \psi_{ij},1\bigr),
\end{split}
\end{equation}
where \(\varphi(\cdot; \mu, \sigma^2)\) denotes the normal density with mean \(\mu\) and variance \(\sigma^2\). Recalling that \(\psi_{ij} = C_{ij}^\top\gamma - f(a_i,a_j)\), this expression immediately implies that the conditional posterior distributions of preference parameters (\(\gamma\)) will only depend on the networks through values of \(\mathbf{w}^\ast\).

\subsection{SCHSAR -- A Joint Modeling Approach}\label{schsar-a-joint-modeling-approach}

The network formation model in \eqref{NF} highlights a central source of network endogeneity: the individual characteristics that are not observable to the researcher \(a_i\) enters the linking incentives through the function \(f(a_i,a_j)\). Because this same latent trait may also affect the outcome \(Y_i\), omitting it from the outcome equation induces correlation between the composite error term and the network-dependent regressors constructed from \(\mathbf W\). To illustrate the mechanism more clearly, we reconsider the HSAR interaction model in \eqref{HSAR1} while incorporating this component of unobserved heterogeneity
\begin{equation}
Y_i = \lambda_g \sum_{j=1}^{N} w_{ij}Y_j 
+ X_i^\top \beta_g 
+ \sum_{j=1}^{N} w_{ij} X_j^\top \delta_g 
+ a_i \kappa_g + u_i,
\qquad 
u_i \sim \mathcal{N}(0, \sigma^2_{ug}),
\end{equation}
for each individual \(i \in \mathcal{I}_g\), where \(g = 1,\ldots,G\). Similar to the coefficients \(\beta_g^k\), the parameter \(\kappa_g\) is type-specific and captures the magnitude and direction of the latent factor \(a_i\) on individual outcomes. We define the \(G\)-dimensional vector \(\kappa \coloneqq [\kappa_1,\ldots,\kappa_G]^\top\) and the corresponding individual-specific vector \(\tilde{\kappa} \coloneqq \mathbf{z}\kappa\). If one ignores the unobserved factor \(\mathbf{a}\), the disturbance term in the outcome equation becomes \(v_i = a_i \kappa_g + u_i\). Because \(a_i\) also enters the network formation process, this misspecification induces correlation between \(v_i\) and the peer variables, including \(\bar{Y}_{(i)} = \sum_{j=1}^N w_{ij}Y_j\) and \(\bar{X}_{(i)} = \sum_{j=1}^N w_{ij}X_j\). Consequently, estimation of the HSAR model without accounting for \(\mathbf a\) suffers from omitted-variable bias (or selection bias), even when individuals are correctly classified into their respective types. In the special case where \(\kappa_g = 0\) for all \(g\), \(a_i\) and \(v_i\) would be uncorrelated, thus eliminating the endogeneity of \(\mathbf{W}\). In that ideal scenario, the HSAR model could be accurately estimated by treating \(\mathbf{W}\) as exogenously given. Otherwise, the endogenous selection issue should be addressed.

Therefore, we propose a joint modeling framework with two stages. In the first stage, individuals strategically form links, leading to the adjacency matrix \(\mathbf W\) through the network formation process described in Section \ref{section2.2}. In the second stage, individuals choose actions (outcomes) conditional on the realized network structure, generating heterogeneous peer effects across individuals.\footnote{Individuals are assumed to be myopic in the sense that when forming links they do not internalize future equilibrium behavior that arises through peer interactions.} We allow the same latent heterogeneity \(\mathbf a\) to appear in both stages and combine the network formation equation and the outcome equation into a unified Selection-corrected Heterogeneous Spatial Autoregressive (SCHSAR) specification. Under this framework, the outcome equation has the following matrix representation
\begin{equation}
\label{SCHSAR}
\mathbf{Y} = \mathcal{L}\mathbf{W}\mathbf{Y} 
+ \sum_{k=1}^K\tilde{\beta}^k \circ X^k 
+ \sum_{k=1}^K \tilde{\delta}^k \circ (\mathbf{W}X^k) 
+ \tilde{\kappa} \circ \mathbf{a} 
+ \mathbf{u}.
\end{equation}

By explicitly introducing \(\mathbf a\), the SCHSAR outcome equation in \eqref{SCHSAR} generalizes the HSAR specification in \eqref{HSAR3} and provides a structural correction for the endogeneity arising from latent individual heterogeneity. Conditional on \((\mathbf W,\mathbf X,\mathbf a)\), the augmented likelihood for the parameters in the outcome equation is associated with the joint distribution of the observed outcomes and latent type indicators \((\mathbf Y,\mathbf z)\) and takes the form\footnote{Recall that $p(\mathbf z\mid\boldsymbol\pi)=\pi_1^{N_1}\cdots\pi_G^{N_G}$, where $N_g$ denotes the number of type-$g$ individuals.}
\begin{equation}
\label{OutcomeLik2}
\begin{split}
&p\left(\mathbf{Y}, \mathbf{z} \mid \mathbf{W}, \mathbf{X}, \mathbf{a}, 
\{\lambda_g,\beta_g,\delta_g,\kappa_g,\sigma^2_{ug},\pi_g\}_{g=1}^G\right)
\\
&\quad = \left|\mathbf I_N-\mathcal L\mathbf W\right|
\prod_{i=1}^N\prod_{g=1}^G
\left[
\varphi\!\left(
\tilde Y_i;\,
X_i^\top\beta_g+\bar X_{(i)}^\top\delta_g+a_i\kappa_g,\,
\sigma_{ug}^2
\right)
\right]^{z_{ig}}
\prod_{g=1}^G \pi_g^{N_g},
\end{split}
\end{equation}
where \(\tilde Y_i = Y_i - \lambda_g \bar Y_{(i)}\).

\medskip

Putting everything together, the complete-data likelihood for all model parameters corresponds to the joint distribution of observed and latent data \(\left(\mathbf{W},\mathbf{w}^\ast,\mathbf{Y}, \mathbf{z}, \mathbf{a}\right)\) and can be written as
\begin{equation}
\label{JointLik}
\begin{split}
&p\left(\mathbf{W},\mathbf{w}^\ast,\mathbf{Y}, \mathbf{z}, \mathbf{a} 
\mid \mathbf{C}, \mathbf{X},\theta_a,\gamma,
\{\lambda_g,\beta_g,\delta_g,\kappa_g,\sigma^2_{ug},\pi_g\}_{g=1}^G\right)
\\
&\quad= p\left(\mathbf{W},\mathbf{w}^\ast\mid \mathbf{C}, \mathbf{a}, \gamma\right)
\cdot p\left(\mathbf{Y}, \mathbf{z}\mid \mathbf{W}, \mathbf{X}, \mathbf{a},
\{\lambda_g,\beta_g,\delta_g,\kappa_g,\sigma^2_{ug},\pi_g\}_{g=1}^G\right)
\cdot p\left(\mathbf{a}\mid \theta_a \right).
\end{split}
\end{equation}

The likelihood factorizes into two parts: one associated with \((\mathbf{W},\mathbf{w}^\ast,\mathbf{Y},\mathbf{z})\) conditional on \(\mathbf a\), and another describing the distribution of \(\mathbf a\) given its parameters. The first part further factorizes into separate contributions from the network and the outcome, as derived in \eqref{NetworkLik} and \eqref{OutcomeLik2}, respectively. To obtain the marginal likelihood function in terms of the observed data \((\mathbf{W},\mathbf{Y})\)\footnote{That is, the observed-data likelihood.}, we need to integrate out all latent variables \(\left(\mathbf{w}^\ast,\mathbf{z},\mathbf{a}\right)\) from \(\eqref{JointLik}\). Maximum likelihood estimation of this likelihood is impeded by the resulting high-dimensional integration. Instead, we employ the Bayesian approach for estimation and inference, which is particularly advantageous in this setting. First, Bayesian data augmentation technique (\citeproc{ref-albert1993bayesian}{Albert and Chib 1993}; \citeproc{ref-tanner1987calculation}{Tanner and Wong 1987}) can be adopted to sample parameters together with latent variables from the joint posterior distribution. This circumvents the need to work with the marginal posterior distribution that lacks a closed-form expression. Second, the full Bayesian framework facilitates inference without relying on asymptotic approximations. By utilizing the posterior sample, we can directly perform inference on various functions of the model parameters, such as direct and indirect effects. It is noteworthy that these effects are computed as own- and cross-partial derivatives for the response of outcome \(Y\) to changes in the explanatory variable \(X^k\), which exhibit a nonlinear relationship with the model parameters.

\emph{Remark 4.} Our Bayesian joint modeling approach to correct for selection on unobservables aligns with the classical control function method based on correlated unobservables (see, i.e, \citeproc{ref-heckman1979sample}{Heckman 1979}; \citeproc{ref-heckman1985alternative}{Heckman and Robb Jr 1985}; \citeproc{ref-navarro2010control}{Navarro 2010}), which has been extended to broader settings of social interaction models (see \citeproc{ref-blume2015linear}{Blume et al. 2015}; \citeproc{ref-johnsson2021estimation}{Johnsson and Moon 2021} for recent treatments). The key mechanism of the control function approach is to identify and recover a proxy or function of the unobserved variable that drives endogeneity, using its statistical relationship with observed data. This allows us to condition on this function -- termed a ``control function'' -- to eliminate bias from omitted variables. In our context, the source of network endogeneity is unobserved individual heterogeneity, \(a_i\), which concurrently influences both the network formation process and the individual outcomes. If \(a_i\) were observed, it would be straightforward to include it directly in the estimation and resolve the endogeneity issue. Since \(a_i\) is unobserved, we instead specify a joint Bayesian model where \(a_i\) enters explicitly as a latent variable in both the network formation equation and the outcome equation. Through this structure, the posterior distribution of \(a_i\) is updated based on the observed links and outcomes, effectively recovering a probabilistic representation of the latent heterogeneity. By incorporating \(a_i\) into the system, we convert a structural endogeneity problem into one of missing data, solvable through the complete-data likelihood. Model parameters can thus be estimated properly within a coherent inferential framework. Notably, this approach does not require external instruments -- especially valuable when valid instruments are weak or unavailable. Moreover, the Bayesian approach is particularly advantageous in our setting, where heterogeneous peer effects are modeled via latent types in a finite mixture structure, further complicating the applicability of traditional instrumental variable methods. Broadly speaking, our SCHSAR approach inherits the Bayesian recipe by Goldsmith-Pinkham and Imbens (\citeproc{ref-goldsmith2013social}{2013}) and Hsieh and Lee (\citeproc{ref-hsieh2016social}{2016}), but nests their models as a special case when the number of mixture components \(G\) is set to one.

The proposed SCHSAR model inherently faces standard identification challenges related to its two-stage structure but these can be addressed by leveraging previous results as special cases. In the network formation equation, identification is complicated by the binary nature of observed links and the presence of unobserved heterogeneity encapsulated within a single-index specification. Parameters in the deterministic linear index are semiparametrically identified up to scale (e.g., \citeproc{ref-ichimura1993semiparametric}{Ichimura 1993}). Full identification for all parameters can thus be achieved with additional parametric assumptions and constraints on the distribution of the stochastic component, which includes individual unobserved heterogeneity \(a_i\) and idiosyncratic errors \(\epsilon_{ij}\), as specified in subsection \hyperref[section2.2]{2.2}. With sufficient variation (i.e., the single index spanning the support of the latent distribution), parameters are point identified and the moments of the underlying unobserved heterogeneity can be recovered. In the outcome equation, identification is challenged by the spatially lagged dependent variable along with peer effects typical of SAR models, by individual unobservables \(a_i\), and by the latent group structure. Specifically, identification of parameters in SAR relies on full-rank conditions for conditional expectations of regressors given the network. If the unobserved heterogeneity \(a_i\) is identified from the network formation equation, treating it as an additional regressor and extending the full-rank condition addresses the network endogeneity. With multiple latent groups (\(G>1\)), further issues such as permutation invariance of the likelihood arise and can be resolved using insight from Frühwirth-Schnatter (\citeproc{ref-fruhwirth2006finite}{2006}) on mixtures of Gaussian regression models. We assume that type groups have pairwise different sizes and restricting the feasible parameter space by ordering the mixing weights: \(1 > \pi_g > \pi_{g'} > 0\) for all \(1 \leq g < g' \leq G\).

\section{Bayesian Estimation}\label{SCHSARsection3}

We implement Bayesian inference to simultaneously handle two key elements of the proposed SCHSAR framework -- heterogeneity and endogeneity -- with latent variables playing a crucial role, while allowing for seamless incorporation of model constraints. The procedure begins with a prior probabilistic belief about unknown parameters, collected into \(\boldsymbol{\theta}\), and systematically updates this belief using observed sample data, denoted by \(\boldsymbol{\mathcal{D}}\). Formally, Bayes' rule gives us the posterior distribution as follows
\[
p(\boldsymbol{\theta}\mid \boldsymbol{\mathcal{D}})\propto p(\boldsymbol{\mathcal{D}}\mid\boldsymbol{\theta}) \times p(\boldsymbol{\theta}),
\]
where \(p(\boldsymbol{\mathcal{D}}\mid\boldsymbol{\theta})\) is the likelihood function, and \(p(\boldsymbol{\theta})\) is the prior distribution. Throughout, we define \(\boldsymbol{\theta}\) to encompass all model parameters -- \(\{\lambda_g,\beta_g,\delta_g,\kappa_g,\sigma^2_{u,g},\pi_g\}_{g=1}^G\), \(\gamma\), \(\theta_a^2\) -- augmented with latent variables comprising individual random effects \(\mathbf{a}\), latent individual type indicators \(\mathbf{z}\), and latent network link utilities \(\mathbf{w}^\ast\). The observed outcomes, network data, and exogenous characteristics are compiled into \(\boldsymbol{\mathcal{D}} = (\mathbf{W},\mathbf{Y},\mathbf{C},\mathbf{X})\). This unified Bayesian approach is computational tractability and enables one-step inference across all unknown quantities, offering principled uncertainty quantification.

Having specified the likelihood function characterizing our econometric model in \eqref{JointLik}, we proceed with providing a prior specification for posterior inference. For ease of exposition, we initially focus on the case involving unobserved degree heterogeneity, where the individual random effects influence network formation according to the linear additive form \(f(a_i,a_j) = a_i+a_j\). Unobserved homophily, characterized by \(f(a_i,a_j) = - \vert a_i-a_j\vert\), can be accommodated similarly within our Bayesian framework, albeit with minor modifications. This extension is elaborated at the end of this section.

\subsection{Prior Specification}\label{prior-specification}

We specify prior distributions for all unknown quantities in our model. For computational convenience, we employ the conjugate priors commonly used in the Bayesian literature. Regarding the individual random effect \(a_i\), we assume a normal distribution with prior mean of zero and the variance \(\sigma_a^2\), where this variance parameter itself follows an Inverse Gamma distribution and will be updated during our estimation procedure:
\begin{flalign}
\qquad a_i \mid \sigma^2_{a} &\overset{iid}{\sim} \mathcal{N}(0, \sigma^2_{a}),
&&\\
\qquad \sigma^2_{a} &\sim \mathcal{I}\mathcal{G}(\underline{s}_{a},\underline{r}_{a}) \Leftrightarrow p(\sigma_{a}^2) = \frac{\underline{r}_{a}^{\underline{s}_{a}}}{\Gamma(\underline{s}_{a})}(\sigma^2_{a})^{-(\underline{s}_{a}+1)} \exp\left(-\frac{\underline{r}_{a}}{\sigma^2_{a}}\right).
&&
\end{flalign}

For the network formation equation, we consider a multivariate normal prior for parameters \(\gamma\):
\begin{flalign}
\qquad\gamma \sim \mathcal{N}_{L}(\underline{\nu}_{\gamma}, \underline{\Sigma}_{\gamma}).
&&
\end{flalign}

For the outcome equation, we assume standard priors independently for \(\beta_g\), \(\delta_g\), \(\kappa_g\), and \(\sigma^2_{ug}\). Specifically, for each \(g = 1,\ldots,G\),
\begin{flalign}
\qquad \beta_g &\sim \mathcal{N}_K(\underline{\nu}_{\beta}, \underline{\Sigma}_{\beta}),
&&\\
\qquad \delta_g &\sim \mathcal{N}_K(\underline{\nu}_{\delta}, \underline{\Sigma}_{\delta}),
&&\\
\qquad \kappa_g &\sim \mathcal{N}(\underline{\nu}_{\kappa}, \underline{\Sigma}_{\kappa}),
&&\\
\qquad \sigma^2_{ug} &\sim \mathcal{I}\mathcal{G}(\underline{s}_{u},\underline{r}_{u}) \Leftrightarrow p(\sigma_{ug}^2) \propto \frac{\underline{r}_{u}^{\underline{s}_{u}}}{\Gamma(\underline{s}_{u})} (\sigma^2_{ug})^{-(\underline{s}_{u}+1)} \exp\left(-\frac{\underline{r}_{u}}{\sigma^2_{ug}}\right).
&&
\end{flalign}

% We also assume \(\lambda_1,\ldots,\lambda_G\) are independent a priori. The prior for each parameter \(\lambda_g\) is assumed to follow a Beta distribution centered on zero, \(\mathcal{B}(d,d)\), which is introduced by LeSage and Parent (\citeproc{ref-lesage2007bayesian}{2007}) to represent an alternative to the uniform prior on the interval \((-1,1)\):
% \begin{flalign}
% \label{betaprior}
% \qquad \lambda_g \sim \mathcal{B}(d,d) \Leftrightarrow p(\lambda_g) = \frac{1}{\text{Beta}(d,d)} \frac{(1+\lambda_g)^{d-1}(1-\lambda_g)^{d-1}}{2^{2d-1}},
% &&
% \end{flalign}
% where \(\text{Beta}(d,d) = \int_0^1 t^{d-1}(1-t)^{d-1}dt\) is the Beta function. Specifically, values for hyperparameter \(d_o\) close to unity induce a relatively uninformative prior that places zero prior weight on end points of the interval for \(\lambda_g\). In our setting, we initialize \(\underline{d} = 1.01\). Recall that in our model setup, the individual-type indicator vector \(z_i\) follows a Multinomial distribution that depends on a vector of assignment probabilities \(\boldsymbol{\pi}\). We complete by assigning a Dirichlet prior on the distribution of \(\boldsymbol{\pi}\):\footnote{We requires an explicit specification regarding the number of types (i.e., G, the number of mixture components). This could be done by employing model selection criteria along with diagnostic plots.}
We also assume \(\lambda_1,\ldots,\lambda_G\) are independent a priori. To ensure that the support of each parameter \(\lambda_g\) is restricted to the admissible stability region, we assign a symmetric transformed Beta prior on \((-1/\tau^\ast,1/\tau^\ast)\), following the approach discussed by LeSage and Parent (\citeproc{ref-lesage2007bayesian}{2007}). Specifically,
\begin{flalign}
\label{betaprior}
\qquad \lambda_g \sim \mathcal{B}(d,d) \Leftrightarrow p(\lambda_g) = \frac{1}{\text{Beta}(d,d)} \frac{(1+\tau^\ast\lambda_g)^{d-1}(1-\tau^\ast\lambda_g)^{d-1}}{2^{2d-1}},
&&
\end{flalign}
where \(\text{Beta}(d,d) = \int_0^1 t^{d-1}(1-t)^{d-1}dt\) is the Beta function. This prior places support only on values satisfying the stability condition \(\vert \lambda_g \vert < 1/\tau^\ast\), which guarantees that the matrix \(\mathbf{I}_N - \mathcal{L}\mathbf{W}\) is nonsingular. Values for hyperparameter \(d\) close to unity induce a nearly uniform prior over the admissible parameter space while avoiding concentration near the boundary points. In our implementation, we set \(d = 1.01\), producing a weakly informative prior. Recall that in our model setup, the individual-type indicator vector \(z_i\) follows a Multinomial distribution that depends on a vector of assignment probabilities \(\boldsymbol{\pi}\). We complete by assigning a Dirichlet prior on the distribution of \(\boldsymbol{\pi}\):\footnote{We requires an explicit specification regarding the number of types (i.e., G, the number of mixture components). This could be done by employing model selection criteria along with diagnostic plots.}
\begin{flalign}
\qquad z_i\mid \boldsymbol{\pi} &\overset{iid}{\sim} \mathcal{M}ult(1, \boldsymbol{\pi}) \Leftrightarrow
p(\mathbf{z} \mid \boldsymbol{\pi}) = \prod_{i=1}^n p(z_i \mid \boldsymbol{\pi}) = \prod_{i=1}^n \prod_{g=1}^G \pi_g^{z_{gi}}, \text{ where } \boldsymbol{\pi} = [\pi_1,\ldots,\pi_G]^\top,
&&\\
\qquad \boldsymbol{\pi} &\sim \mathcal{D}ir(\underline{\alpha}_{1},\underline{\alpha}_{2},\ldots,\underline{\alpha}_{G}) \Leftrightarrow p(\boldsymbol{\pi}) \propto \pi_1^{\underline{\alpha}_{1}-1}\pi_2^{\underline{\alpha}_{2}-1}\ldots\pi_G^{\underline{\alpha}_{G}-1}.
&&
\end{flalign}
In our implementation for the simulation study and the empirical application in subsequent sections, we set the prior hyperparameters to standard non-informative values as follows: \(\underline{\nu}_{\gamma} = \mathbf{0}_{L}, \ \underline{\Sigma}_{\gamma} = 10^4\mathbf{I}_L, \ \underline{\nu}_{\beta} = \mathbf{0}_{K}, \ \underline{\Sigma}_{\beta} = 10^4\mathbf{I}_K, \ \underline{\nu}_{\kappa} = 0, \ \underline{\Sigma}_{\kappa} = 10^4\), \(\underline{r}_{a} = \underline{s}_{a} \approx 0\), \(\underline{r}_{u} = \underline{s}_{u} \approx 0\), and \(\underline{\alpha}_1 = \ldots = \underline{\alpha}_G = 1/G\).

\subsection{Posterior Analysis}\label{posterior-analysis}

From Bayes' theorem, the augmented joint posterior density of interest can be expressed as
\begin{equation}
\label{jointposterior}
\begin{split}
p(\boldsymbol{\theta}\mid \boldsymbol{\mathcal{D}}) 
&\propto p\bigl(\mathbf{W},\mathbf{w}^\ast,\mathbf{Y}, \mathbf{z}, \mathbf{a} \mid \mathbf{C}, \mathbf{X},\sigma_a,\gamma,\{\lambda_g,\beta_g,\delta_g,\kappa_g,\sigma^2_{u,g},\pi_g\}_{g=1}^G\bigr)\\
&\quad \times p(\lambda,\beta,\delta,\kappa,\sigma^2_u,\boldsymbol{\pi},\gamma,\sigma_a^2).
\end{split}
\end{equation}
The first term on the right-hand side is the likelihood function defined previously in \eqref{JointLik}, which factorizes further into
\begin{equation}
\begin{split}
&p\bigl(\mathbf{W},\mathbf{w}^\ast\mid \mathbf{C}, \mathbf{a}, \gamma\bigr) \times p\bigl(\mathbf{Y}, \mathbf{z}\mid \mathbf{W}, \mathbf{X}, \mathbf{a}, \{\lambda_g,\beta_g,\delta_g,\kappa_g,\sigma^2_{u,g},\pi_g\}_{g=1}^G\bigr) \times p\bigl(\mathbf{a}\mid \sigma_a^2\bigr)\\
&= \bigl[p\bigl(\mathbf{W}\mid \mathbf{w}^\ast) \times p(\mathbf{w}^\ast\mid \mathbf{C}, \mathbf{a}, \gamma\bigr)\bigr]\\
&\qquad \times \bigl[p\bigl(\mathbf{Y} \mid \mathbf{z},\mathbf{W}, \mathbf{X}, \sigma_a,\gamma,\{\lambda_g,\beta_g,\delta_g,\kappa_g,\sigma^2_{u,g},\pi_g \bigl) \times p\,\bigl(\mathbf{z} \mid \boldsymbol{\pi}\bigr)\bigr] \times p\bigl(\mathbf{a}\mid \sigma_a^2\bigr),\\
\end{split}
\end{equation}
and the second term in \eqref{jointposterior} gives the joint prior of the model parameters and factorizes into independent priors as follows
\begin{equation}
p(\lambda,\beta,\delta,\kappa,\sigma^2_u,\boldsymbol{\pi},\gamma,\sigma_a^2) = p(\lambda) \times p(\beta) \times p(\delta) \times p(\kappa) \times p(\sigma^2_u) \times p(\boldsymbol{\pi}) \times p(\gamma) \times p(\sigma_a^2).
\end{equation}

With this representation, we estimate the model using the data-augmented Markov chain Monte Carlo (MCMC) approach. Specifically, our MCMC scheme treats latent variables \((\mathbf{w}^\ast,\mathbf{z},\mathbf{a})\) as parameters to be estimated, and sample unknowns from the joint posterior distribution by cycling through three blocks of conditional distributions. The first block of conditionals is used for updating the individual random effects \(\{a_i\}_{i=1}^N\) and their variance \(\sigma_a^2\). The second block of conditionals is for other parameters and latent network utilities associated with the network formation equation. The third block of conditionals is for those associated with the outcome equation, including latent mixture indicators. These blocks respectively are
\begin{equation}
\begin{split}
\mathbf{a}, \sigma_a^2 &\mid \underline{s}_{a}, \underline{r}_{a}, \mathbf{w}^\ast,\gamma, \mathbf{z},\{\lambda_g,\beta_g,\delta_g,\kappa_g,\sigma^2_{u g},\pi_g\}_{g=1}^G, \boldsymbol{\mathcal{D}};\\
\mathbf{w}^\ast,\gamma &\mid \underline{\nu}_{\gamma}, \underline{\Sigma}_{\gamma},\mathbf{a}, \boldsymbol{\mathcal{D}};\\
\mathbf{z},\{\lambda_g,\beta_g,\delta_g,\kappa_g,\sigma^2_{u g},\pi_g\}_{g=1}^G &\mid \underline{d}, \underline{\nu}_{\beta}, \underline{\Sigma}_{\beta}, \underline{\nu}_{\delta}, \underline{\Sigma}_{\delta}, \underline{\nu}_{\kappa}, \underline{\Sigma}_{\kappa}, \underline{s}_{u}, \underline{r}_{u},  \mathbf{a}, \boldsymbol{\mathcal{D}}.
\end{split}
\end{equation}
For notational simplicity, let \(\boldsymbol{\theta}_{-\theta_1}\) denote the set of all parameters (including latent variables) in \(\boldsymbol{\theta}\) excluding the component \(\theta_1\).

To facilitate efficient sampling in random‑effect and network blocks, we introduce the following quantities relevant to the network formation equation
\[
\begin{aligned}
\mathbf{w} &= [w_{-1}^\top,\ldots,w_{-N}^\top]^\top, \quad w_{-j} = [w_{1j},\ldots, w_{j-1,j}, w_{j+1,j},\ldots,w_{Nj}]^\top, \quad j = 1,\ldots,J;
\\
\mathbf{F} &= [F_{-1},\ldots,F_{-N}]^\top, \quad F_{-j} = [e_{1}^\top,\ldots,e_{j-1}^\top, e_{j+1}^\top,\ldots,e_{N}^\top]^\top, \quad j = 1,\ldots,J;
\\
\mathbf{E} &= \mathbf{I}_N \otimes \iota_{N-1};
\\
\mathbf{H} &= \mathbf{F} + \mathbf{E};
\end{aligned}
\]
where \(\otimes\) denotes the Kronecker product. Then, we can rewrite \(\mathbf{A} = \mathbf{H}\mathbf{a}\) and replace into equation \eqref{NF}
\begin{equation}
\mathbf{w^\ast} = \mathbf{C}\gamma + \mathbf{H}\mathbf{a} + \boldsymbol{\epsilon}.
\end{equation}
Consequently, the full conditional posterior distributions for parameters and latent variables in the random-effects and network formation blocks can be written explicitly as follows
\begin{flalign}
\label{posta}
\qquad \mathbf{a} \mid \boldsymbol{\theta}_{-a_i}, \boldsymbol{\mathcal{D}} &\sim \mathcal{N}\bigl(\overline{\nu}_{a},\overline{\Sigma}_{a}\bigr),
&&\\
\label{postsigmaa}
\qquad \sigma_a^2 \mid \boldsymbol{\theta}_{-\sigma_a^2},\boldsymbol{\mathcal{D}} &\sim \mathcal{I}\mathcal{G}(\overline{s}_{a}, \overline{r}_{a}),
&&\\
\label{postw}
\qquad \mathbf{w}^\ast \mid \boldsymbol{\theta}_{-\mathbf{w}^\ast},\boldsymbol{\mathcal{D}} &\sim TM\mathcal{N}_{\mathcal{S}}\left(\mathbf{C}\gamma + \mathbf{H}\mathbf{a},\mathbf{I}_{N(N-1)}\right),
&&\\
\label{postgamma}
\qquad \gamma \mid \boldsymbol{\theta}_{-\gamma},\boldsymbol{\mathcal{D}} &\sim \mathcal{N}_{L}(\overline{\nu}_{\gamma}, \overline{\Sigma}_{\gamma}),
\end{flalign}
Here, \(TM\mathcal{N}_{\mathcal{S}}\) denotes a multivariate normal distribution truncated to the region \(\mathcal{S}\) implied by the binary adjacency vector \(\mathbf{w}\) above.\footnote{Specifically, the truncated region is \([0,+\infty)\) for elements corresponding to observed network links (\(w_{ij}=1\)), and \((-\infty,0)\) otherwise (\(w_{ij}=0\)). Sampling from this truncated multivariate normal distribution can be implemented efficiently using the algorithm proposed by Botev (\citeproc{ref-botev2017normal}{2017}).}. The posterior hyperparameters governing other conditional distributions are defined as
\begin{flalign}
\nonumber
\qquad\overline{\Sigma}_{a} &\coloneqq \left[ \sigma_a^{-2}\mathbf{I}_N + \mathbf{H}^\top\mathbf{H} + \tilde\kappa\tilde\kappa^\top \circ \text{diag}(\tilde\sigma^{-2}_u) \right]^{-1},
&&\\\nonumber
\qquad\overline{\nu}_{a} &\coloneqq \overline{\Sigma}_{a} \left[ \mathbf{H}^\top(\mathbf{w}^\ast - \mathbf{C}\gamma) + \tilde\kappa\circ\tilde\sigma^{-2}_u\circ \left( \mathbf{\tilde{Y}} - \sum_{k=1}^K X^k \circ \tilde{\beta}^k - \sum_{k=1}^K (\mathbf{W} X^k) \circ \tilde{\delta}^k \right) \right],
&&\\\nonumber
\qquad \overline{s}_{a} &\coloneqq \underline{s}_{a} + \frac{N}{2},
&&\\\nonumber
\qquad \overline{r}_{a} &\coloneqq \underline{r}_{a} + \frac{\mathbf{a}^\top\mathbf{a}}{2},
&&\\\nonumber
\qquad\overline{\Sigma}_{\gamma} &\coloneqq \left[\underline{\Sigma}_{\gamma}^{-1} + \mathbf{C}^\top\mathbf{C} \right]^{-1},
&&\\\nonumber
\qquad\overline{\nu}_{\gamma} &\coloneqq \overline{\Sigma}_{\gamma} \left[\underline{\Sigma}_{\gamma}^{-1}\underline{\nu}_{\gamma} + \mathbf{C}^\top(\mathbf{w}^\ast - \mathbf{H}\mathbf{a})\right].
&&
\end{flalign}

Regarding the outcome block, at each iteration, given the latent group assignments \(\mathbf{z}\), the data are partitioned into \(G\) mixture components. For each group \(g=1\ldots,G\), let \(\mathcal{I}_g\) index observations assigned to group \(g\) and \(N_g \coloneqq \sum_{i=1}^N z_{ig}\) denote the group size. We define \(\tilde{Y}_g \coloneqq \{\tilde{Y}_i\}_{i \in \mathcal{I}_g}\), \(X_g \coloneqq \{X_i\}_{i \in \mathcal{I}_g}\), \(\tilde{X}_g \coloneqq \{\bar{X}_{(i)}\}_{i \in \mathcal{I}_g}\), and \(a_g \coloneqq \{a_i\}_{i \in \mathcal{I}_g}\). The conditional posteriors of group-specific parameters including \(\lambda_g\), \(\beta_g\), \(\delta_g\), \(\kappa_g\), and \(\sigma_{ug}^2\), can be characterized accordingly. Conversely, conditional on the parameters of the mixture, allocation \(z_i\) follows an independent Multinomial distribution with the classification weights (i.e., the probability that each observation \(i\) belongs to a given group \(g\)) can be calculated from the predictive densities. Finally, the conditional posterior for the \(G\)-dimensional vector of component probability \(\pi\) follows a Dirichlet distribution. The full conditional posteriors for the parameters are given as follows
\begin{flalign}
\label{postlambda}
& p(\lambda_g \mid \boldsymbol{\theta}_{-\lambda_g},\boldsymbol{\mathcal{D}}) \propto p\left(\lambda_g\right) \cdot \vert \mathbf{I}_N - \mathcal{L}\mathbf{W}\vert \cdot (2\pi\sigma_{ug}^2)^{-N_{g}/2}
&&\\\nonumber
&\quad \times \prod_{g=1}^{G}\exp \left\{ -\frac{1}{2\sigma^2_{ug}}(\tilde{Y}_g - X_g\beta_g - \tilde{X}_g\delta_g - a_g \kappa_g)^\top(\tilde{Y}_g - X_g\beta_g - \tilde{X}_g\delta_g - a_g \kappa_g) \right\}, g = 1,\ldots,G,
&&
\end{flalign}
which is the only one that does not belong to a known class of distributions, whereas the conditional posteriors for other parameters follow standard conjugate results
\begin{flalign}
\label{postbeta}
\qquad \beta_g &\mid \boldsymbol{\theta}_{-\beta_g},\boldsymbol{\mathcal{D}} \sim \mathcal{N}_K\bigl(\overline{\nu}_{\beta g},\overline{\Sigma}_{\beta g}\bigr), \quad g = 1,\ldots,G,
&&\\
\label{postdelta}
\qquad \delta_g &\mid \boldsymbol{\theta}_{-\delta_g},\boldsymbol{\mathcal{D}} \sim \mathcal{N}_K\bigl(\overline{\nu}_{\delta g},\overline{\Sigma}_{\delta g}\bigr), \quad g = 1,\ldots,G,
&&\\
\label{postkappa}
\qquad  \kappa_g &\mid \boldsymbol{\theta}_{-\kappa_g}, \boldsymbol{\mathcal{D}} \sim \mathcal{N}\bigl(\overline{\nu}_{\kappa g},\overline{\Sigma}_{\kappa g}\bigr), \quad g = 1,\ldots,G,
&&\\
\label{postsigmau}
\qquad \sigma^2_{ug} &\mid \boldsymbol{\theta}_{-\sigma^2_{ug}}, \boldsymbol{\mathcal{D}} \sim \mathcal{I}\mathcal{G}\bigl(\overline{s}_{ug},\overline{r}_{ug}\bigr), \quad g = 1,\ldots,G,
&&\\
\label{postz}
z_i &\mid \boldsymbol{\theta}_{-z_i}, \boldsymbol{\mathcal{D}}
\sim \mathcal{M}ult \left(1, [\omega_{i1}, \omega_{i2}, \ldots, \omega_{iG} ]\right), \quad i = 1,\ldots,N,
&&\\
\label{postpi}
\mathbf{\pi} &\mid \boldsymbol{\theta}_{-\mathbf{\pi}}, \boldsymbol{\mathcal{D}} \sim \mathcal{D}ir(\overline{\alpha}_1, \overline{\alpha}_2,\ldots,\overline{\alpha}_G),
&&
\end{flalign}
with posterior hyperparameters (for \(i = 1,\ldots,N\) and \(g = 1,\ldots,G\)) are
\begin{flalign}
\nonumber
\qquad\overline{\Sigma}_{\beta g} &\coloneqq \left(\underline{\Sigma}_{\beta}^{-1} + \sigma^{-2}_{ug}X_g^{\top} X_g\right)^{-1},
&&\\\nonumber
\overline{\nu}_{\beta g} &\coloneqq \overline{\Sigma}_{\beta g} \left[\underline{\Sigma}_{\beta}^{-1}\underline{\nu}_{\beta} + \sigma^{-2}_{ug}X_g^{\top}(\tilde{Y}_g - \tilde{X}_g\delta_g -  a_g \kappa_g)\right],
&&\\\nonumber
\qquad\overline{\Sigma}_{\delta g} &\coloneqq \left(\underline{\Sigma}_{\delta}^{-1} + \sigma^{-2}_{ug}\tilde{X}_g^{\top}\tilde{X}_g \right)^{-1},
&&\\\nonumber
\overline{\nu}_{\delta g} &\coloneqq \overline{\Sigma}_{\delta g} \left[\underline{\Sigma}_{\delta}^{-1}\underline{\nu}_{\delta} + \sigma^{-2}_{ug}\tilde{X}_g^{\top}(\tilde{Y}_g - X_g\beta_g - a_g \kappa_g)\right],
&&\\\nonumber
\qquad\overline{\Sigma}_{\kappa g} &\coloneqq \left(\underline{\Sigma}_{\kappa}^{-1} + \sigma^{-2}_{ug}a_g^{\top} a_g\right)^{-1},
&&\\\nonumber
\qquad\overline{\nu}_{\kappa g} &\coloneqq \overline{\Sigma}_{\kappa g} \left[\underline{\Sigma}_{\kappa}^{-1}\underline{\nu}_{\kappa} + a_g^{\top}(\tilde{Y}_g - \tilde{X}_g\delta_g - X_g\beta_g)\right],
&&\\\nonumber
\qquad\overline{s}_{ug} &\coloneqq \underline{s}_{u} + \frac{N_g}{2},
&&\\\nonumber
\qquad\overline{r}_{ug} &\coloneqq \underline{r}_{u} + \frac{1}{2}\left( \tilde{Y}_g - X_g\beta_g - \tilde{X}_g\delta_g - a_g \kappa_g \right)^{\top}\left( \tilde{Y}_g - X_g\beta_g - \tilde{X}_g\delta_g - a_g \kappa_g\right),
&&\\\nonumber
\qquad\omega_{ig} &\coloneqq \text{Pr}\bigl(z_{ig} = 1 \mid \boldsymbol{\theta}_{-z_{ig}},\boldsymbol{\mathcal{D}}\bigr) = \frac{\pi_gq_{ig}}{\sum_{g=1}^G \pi_gq_{ig}},
&&\\\nonumber
\qquad q_{ig} &\coloneqq (2\pi\sigma^2_{ug})^{-1/2} \exp\left\{  -\frac{1}{2\sigma^2_{ug}}(Y_i - \lambda_g \overline{Y}_{(i)} - X_i^\top\beta_g - \bar{X}_{(i)}^\top\delta_g - a_i \kappa_g)^2\right\},
&&\\\nonumber
\qquad\overline{\alpha}_g &\coloneqq \underline{\alpha}_g + N_g.
&&
\end{flalign}

\subsection{Markov Chain Monte Carlo (MCMC) Algorithm}\label{markov-chain-monte-carlo-mcmc-algorithm}

As in the preceding posterior analysis, with conjugate priors, \(\{a_i\}\), \(\{w^*_{ij}\}\), \(\gamma\), \(\{z_{ig}\}\), \(\{\beta_{g}\}\), \(\{\delta_{g}\}\), \(\{\kappa_{g}\}\), \(\{\sigma^2_{ug}\}\), \(\{\pi_{g}\}\) can be sampled straightforwardly via Gibbs sampling. By contrast, the conditional posterior distribution of \(\{\lambda_{g}\}\) shown in \eqref{postlambda} does not conform to a standard form. To address this issue, a Metropolis-Hastings (M-H) step with an important twist is incorporated into the procedure, resulting in Metropolis-within-Gibbs sampling. Specifically, at the \(t\)-th iteration, for each group \(g = 1,\ldots,G\), sampling \(\lambda_g\) from \(p(\lambda_g \mid \boldsymbol{\theta}_{-\lambda_g},\boldsymbol{\mathcal{D}})\) involves two main steps:

\begin{enumerate}
\def\labelenumi{\arabic{enumi}.}
\tightlist
\item
  We generate a candidate value for \(\lambda_g\) by perturbing the current value using a proposal distribution:\\
  Propose \(\lambda^{\star}_g =  \lambda^{(t-1)}_g + \mathcal{N}(0,\tau_g)\), where the scaling parameter (i.e., proposal increment shape) \(\tau_g\) is tuned according to Adaptive Scaling Metropolis (ASM) algorithm (\citeproc{ref-andrieu2008tutorial}{Andrieu and Thoms 2008}; \citeproc{ref-vihola2022}{Vihola 2022}), with details can be found in Appendix \ref{SCHSAR-APDX_AdaptiveMCMC}. Compared with the standard random-walk Metropolis sampler, the key idea is to implement an automatic adjustment of \(\tau_g\) based on monitoring the acceptance rates and stepsizes during the MCMC sampling procedure. As a result, this adaptation can learn from the historical MCMC draws (accepted draws) to make the proposal distribution better suited to the target distribution, thereby improving the efficiency and convergence of the algorithm.
\end{enumerate}

\begin{enumerate}
\def\labelenumi{\arabic{enumi}.}
\setcounter{enumi}{1}
\tightlist
\item
  We compute the acceptance rate using the ratio of the posterior densities at the proposed and current values:\\
  Let \(\mathcal{L}^{\star}_{\cdot\mid g} = \text{diag}\left(\mathbf{z}[\lambda_1,\ldots,\lambda_{g-1},\lambda^{\star}_g,\lambda_{g+1},\ldots,\lambda_{G}]^\top\right)\), \(\mathbf{\tilde{Y}}_g^{\star}=\left[\mathbf{I}_N - \tilde{\mathcal{L}}^{\star}_{\cdot\mid g}\mathbf{W}\right] \mathbf{Y} \circ z_g\)\\
  and \(\mathcal{L}^{(t-1)}_{\cdot\mid g} = \text{diag}\left(\mathbf{z}[\lambda_1,\ldots,\lambda_{g-1},\lambda^{(t-1)}_g,\lambda_{g+1},\ldots,\lambda_{G}]^\top\right)\),\(\mathbf{\tilde{Y}}_g^{(t-1)}=\left[\mathbf{I}_N - \tilde{\mathcal{L}}^{(t-1)}_{\cdot\mid g}\mathbf{W}\right] \mathbf{Y} \circ z_g\).\\
  Also,
  \[
  \begin{aligned}
  u^{\star}_{\cdot\mid g} &= \mathbf{\tilde{Y}}_g^{\star} - X_g\beta_g - \tilde{X}_g\delta_g - a_g \kappa_g\\
  u^{(t-1)}_{\cdot\mid g} &= \mathbf{\tilde{Y}}_g^{(t-1)} - X_g\beta_g - \tilde{X}_g\delta_g - a_g \kappa_g\\
  \end{aligned}
  \]
  Then, recall that \(p(\lambda_g)\) is the density function of the Beta prior distribution defined in \eqref{betaprior}, with the acceptance rate
  \begin{equation}
  \alpha(\lambda^{\star}_g, \lambda^{(s-1)}_g)
  \coloneqq \text{min} \left\{ \frac{\vert \mathbf{I}_N - \tilde{\mathcal{L}}^{\star}_{\cdot\mid g}\mathbf{W} \vert \exp\left[-u^{\star\top}_{\cdot\mid g}u^{\star}_{\cdot\mid g}/(2\sigma_{ug}^2) \right]}{\mid \mathbf{I}_N - \tilde{\mathcal{L}}^{(t-1)}_{\cdot\mid g}\mathbf{W} \mid \exp\left[-u^{(t-1)\top}_{\cdot\mid g}u^{(t-1)}_{\cdot\mid g}/(2\sigma_{ug}^2) \right]} \times \frac{p(\lambda^{\star}_g)}{p(\lambda^{(t-1)}_g)},1 \right\},
  \end{equation}
  update \(\lambda^{(t)}_g = \lambda^{\star}_g\), else set \(\lambda^{(t)}_g = \lambda^{(t-1)}_g\).
\end{enumerate}

The full implementation of MCMC sampling scheme for the SCHSAR model with unobserved degree heterogeneity is outlined in Algorithm \ref{alg:SCHSAR1}.

\begin{center}
\scalebox{0.95}{
\small
\begin{minipage}{1\linewidth}
\begin{algorithm}[H]
\caption{MCMC Sampler for SCHSAR Model (\emph{Unobserved Degree Heterogeneity})}
\label{alg:SCHSAR1}

\vspace{3mm}

\textbf{Procedure}

\refstepcounter{algoline}\nlset{\thealgoline}\quad Set initial values for all model parameters (augmented with latent variables) in $\boldsymbol{\theta}$.

\refstepcounter{algoline}\nlset{\thealgoline}\quad\textbf{Step 1:} Update individual random effects $\mathbf{a}$ by sampling

\refstepcounter{algoline}\nlset{\thealgoline}\qquad (a1) $\mathbf{a}$ from $\mathcal{N}\bigl(\overline{\nu}_{a},\overline{\Sigma}_{a}\bigr)$, given $\sigma_a^2, \mathbf{w}^\ast,\gamma, \mathbf{z},\{\lambda_g,\beta_g,\delta_g,\kappa_g,\sigma^2_{ug}\}_{g=1}^G, \boldsymbol{\mathcal{D}}$; \Comment{\eqref{posta}}

\refstepcounter{algoline}\nlset{\thealgoline}\qquad (a2) $\sigma_a^2$ from $\mathcal{I}\mathcal{G}(\overline{s}_{a}, \overline{r}_{a})$, given $\mathbf{a}$.\Comment{\eqref{postsigmaa}}

\refstepcounter{algoline}\nlset{\thealgoline}\quad\textbf{Step 2:} Conditional on $\mathbf{a}$, update parameters in the network formation equation by sampling

\refstepcounter{algoline}\nlset{\thealgoline}\qquad  (b1) $\mathbf{w}^\ast$ from $TM\mathcal{N}_{\mathcal{S}}\left(\mathbf{C}\gamma + \mathbf{H}\mathbf{a},\mathbf{I}_{N(N-1)}\right)$, given $\gamma,\mathbf{a},\boldsymbol{\mathcal{D}}$;\Comment{\eqref{postw}}

\refstepcounter{algoline}\nlset{\thealgoline}\qquad  (b2) $\gamma$ from $\mathcal{N}_{L}(\overline{\nu}_{\gamma}, \overline{\Sigma}_{\gamma})$, given $\mathbf{w}^\ast,\mathbf{a},\boldsymbol{\mathcal{D}}$.\Comment{\eqref{postgamma}}

\refstepcounter{algoline}\nlset{\thealgoline}\quad\textbf{Step 3:} Conditional on $\mathbf{a}$ and the allocations $\mathbf{z}$, update parameters in the outcome equation by sampling

\refstepcounter{algoline}\nlset{\thealgoline}\qquad (c1) $\boldsymbol{\pi}$ from $\mathcal{D}ir(\overline{\alpha}_1, \overline{\alpha}_2,\ldots,\overline{\alpha}_G)$, given $\mathbf{z},\{\lambda_g,\beta_g,\delta_g,\kappa_g,\sigma^2_{ug}\},\mathbf{a},\boldsymbol{\mathcal{D}}$;\Comment{\eqref{postpi}}

\refstepcounter{algoline}\nlset{\thealgoline}\qquad (c2) each $\beta_g$, for $g = 1,\ldots, G$, from $\mathcal{N}_K\bigl(\overline{\nu}_{\beta g},\overline{\Sigma}_{\beta g}\bigr)$, given $\mathbf{z},\lambda_g,\delta_g,\kappa_g,\sigma^2_{ug},\mathbf{a},\boldsymbol{\mathcal{D}}$;\Comment{\eqref{postbeta}}

\refstepcounter{algoline}\nlset{\thealgoline}\qquad (c3) each $\delta_g$, for $g = 1,\ldots, G$, from $\mathcal{N}_K\bigl(\overline{\nu}_{\delta g},\overline{\Sigma}_{\delta g}\bigr)$, given $\mathbf{z},\lambda_g,\beta_g,\kappa_g,\sigma^2_{ug},\mathbf{a},\boldsymbol{\mathcal{D}}$;\Comment{\eqref{postdelta}}

\refstepcounter{algoline}\nlset{\thealgoline}\qquad (c4) each $\kappa_g$, for $g = 1,\ldots, G$, from $\mathcal{N}\bigl(\overline{\nu}_{\kappa g},\overline{\Sigma}_{\kappa g}\bigr)$, given $\mathbf{z},\lambda_g,\beta_g,\delta_g,\sigma^2_{ug},\mathbf{a},\boldsymbol{\mathcal{D}}$;\Comment{\eqref{postkappa}}

\refstepcounter{algoline}\nlset{\thealgoline}\qquad (c5) each $\sigma^2_{ug}$, for $g = 1,\ldots, G$, from $\mathcal{I}\mathcal{G}\bigl(\overline{s}_{ug},\overline{r}_{ug}\bigr)$, given $\mathbf{z},\lambda_g,\beta_g,\delta_g,\kappa_g,\mathbf{a},\boldsymbol{\mathcal{D}}$;\Comment{\eqref{postsigmau}}

\refstepcounter{algoline}\nlset{\thealgoline}\qquad (c6) each $\lambda_{g}$, for $g = 1,\ldots, G$, via ASM algorithm, given $\mathbf{z},\beta_g,\delta_g,\kappa_g,\sigma^2_{ug},\mathbf{a},\boldsymbol{\mathcal{D}}$.\Comment{\eqref{postlambda}}

\refstepcounter{algoline}\nlset{\thealgoline}\quad\textbf{Step 4:} For each $i = 1,\ldots,N$, sample the allocation $z_i$ from $\mathcal{M}ult \left(1, [\omega_{i1}, \omega_{i2}, \ldots, \omega_{iG}]\right)$, given $\{\lambda_g,\beta_g,\delta_g,\kappa_g,\sigma^2_{ug},\pi_g\}_{g=1}^G,\mathbf{a},\boldsymbol{\mathcal{D}}$.\Comment{\eqref{postz}}

\refstepcounter{algoline}\nlset{\thealgoline}\quad Repeat Step 1-4 using the most recently updated values until convergence.

\textbf{end procedure}

\end{algorithm}
\end{minipage}%
}
\end{center}

% \subsection{Extension}\label{SCHSARhomophily}
\subsection{Extension to unobserved homophily}\label{SCHSARhomophily}

We develop Bayesian estimation for the SCHSAR framework in the case of unobserved homophily, where the individual random effects influence network formation via the function \(f(a_i,a_j) = -\vert a_i - a_j\vert\). In this setting, we assume that the unobserved types \(a_i\) are assumed to take values in \(S\) ordered categories, \(\{1,\ldots,S\}\). Specifically,
\begin{equation}
a_{i} \overset{iid}{\sim} \mathcal{C}at\left(\boldsymbol{\rho}\right) \forall i = 1,\ldots,N,
\end{equation}
where \(\boldsymbol{\rho} = (\rho_1,\ldots,\rho_S)\) is a \(S\)-dimensional vector of probabilities satisfying \(\sum_{s=1}^S \rho_s = 1\). A Dirichlet prior is placed on \(\boldsymbol{\rho}\),
\[
\boldsymbol{\rho} \sim \mathcal{D}ir(\underline{\xi}_1,\ldots,\underline{\xi}_S).
\]
Due to the absolute difference structure \(\vert a_i - a_j\vert\) in the network formation, the likelihood is invariant to reversing the ordering of the latent categories and simultaneously changing the signs of \(\kappa_g\). In particular, the transformation
\[
(a_i, \kappa_g) \rightarrow (S+1 - a_i, -\kappa_g)
\] for all \(i\) and \(g\) leaves the likelihood unchanged (up to an intercept shift in the outcome equation). To resolve this issue, we restrict \(\kappa_g\) to be nonnegative for all \(g\), which can be achieved seamlessly within the MCMC sampling procedure. A careful initialization for \(\mathbf{a} = (a_1,\ldots,a_N)\) using spectral clustering algorithms applied to the observed adjacency matrix \(W\), as described in Appendix \ref{SCHSAR-APDX_CommunityDetection}, further improves computational efficiency.

The conditional posteriors for \(a_i\) and \(\boldsymbol{\rho}\) are
\begin{flalign}
\label{postanew}
\qquad a_i \mid \boldsymbol{\theta}_{-a_i}, \boldsymbol{\mathcal{D}} &\sim  \mathcal{C}at\left(\bar{\rho}_{i1},\ldots,\bar{\rho}_{iS}\right), \quad  i = 1,\ldots,N, 
&&\\
\label{postrho}
\boldsymbol{\rho} \mid \boldsymbol{\theta}_{-\boldsymbol{\rho}}, \boldsymbol{\mathcal{D}} &\sim \mathcal{D}ir\left(\bar{\xi}_1,\ldots,\bar{\xi}_S\right),
\end{flalign}
where for each \(i = 1,\ldots,N\) and \(s = 1,\ldots,S\), the posterior hyperparameters are defined as
\begin{flalign}
\nonumber
\bar{\rho}_{is} &\coloneqq \frac{P_{is}}{\sum_{s=1}^S P_{is}}, \text{ with }
&&\\\nonumber
P_{is} &\coloneqq \rho_s
\times \text{exp}\left\{ -\sum_{g=1}^G \frac{z_{gi}(Y_i - \lambda_g \overline{Y}_{(i)} - X_i^\top\beta_g - \bar{X}_{(i)}^\top\delta_g - s\kappa_g)^2}{2\sigma^2_{g}} - \sum_{j\neq i}\frac{(w^*_{ij}-C_{ij}\gamma+\vert s-a_j \vert)^2}{2}\right\},
&&\\\nonumber
\bar{\xi}_s &\coloneqq \underline{\xi}_s + \sum_{i=1}^N \mathbbm{1}\{a_i = s\}.
\end{flalign}

The conditional posterior distributions for the other parameters remain the same as derived previously, with adjustments required specifically for \(\mathbf{w}^\ast,\gamma,\kappa_g\). Explicitly, they become
\begin{flalign}
\label{postwnew}
\qquad \mathbf{w}^\ast \mid \boldsymbol{\theta}_{-\mathbf{w}^\ast},\boldsymbol{\mathcal{D}} &\sim TM\mathcal{N}_{\mathcal{S}}\left(\mathbf{C}\gamma + \mathbf{A},\mathbf{I}_{N(N-1)}\right),
&&\\
\label{postgammanew}
\qquad\gamma \mid \boldsymbol{\theta}_{-\gamma},\boldsymbol{\mathcal{D}} &\sim \mathcal{N}\bigl(\overline{\nu}_{\gamma}, \overline{\Sigma}_{\gamma}\bigr),
&&\\
\label{postkappanew}
\qquad \kappa_g \mid \boldsymbol{\theta}_{-\kappa_g},\boldsymbol{\mathcal{D}} &\sim \mathcal{TN}_{[0,+\infty)}\bigl(\overline{\nu}_{\kappa g},\overline{\Sigma}_{\kappa g}\bigr),
&&
\end{flalign}
where \(\mathcal{TN}_{[0,+\infty)}\) denotes a truncated normal distribution constrained to \(\kappa_g \geq 0\), and
\begin{flalign}
\nonumber
\qquad\overline{\Sigma}_{\gamma} &\coloneqq \left[ \underline{\Sigma}_{\gamma}^{-1} + \mathbf{C}^{\top}\mathbf{C} \right]^{-1},
&&\\\nonumber
\qquad\overline{\nu}_{\gamma} &\coloneqq \overline{\Sigma}_{\gamma} \left[ \underline{\Sigma}_{\gamma}^{-1}\underline{\nu}_{\gamma} + \mathbf{C}^{\top}(\mathbf{w}^\ast - \mathbf{A})\right],
&&\\\nonumber
\overline{\Sigma}_{\kappa g} &\coloneqq \left[ \underline{\Sigma}_{\kappa g}^{-1} +  \sigma^{-2}_{ug}a_g^{\top}a_g \right]^{-1},
&&\\\nonumber
\overline{\nu}_{\kappa g} &\coloneqq \overline{\Sigma}_{\kappa g} \left[\underline{\Sigma}_{\kappa}^{-1}\underline{\nu}_{\kappa} + a_g^{\top}(\tilde{Y}_g - X_g\beta_g - \tilde{X}_g\delta_g)\right], \quad g=1,\ldots,G.
&&
\end{flalign}

The full implementation of MCMC sampling scheme for the SCHSAR model with unobserved homophily is outlined in Algorithm \ref{alg:SCHSAR2}.

\begin{center}
\scalebox{0.95}{
\small
\begin{minipage}{1\linewidth}
\begin{algorithm}[H]
\caption{MCMC Sampler for SCHSAR Model (\emph{Unobserved Homophily})}
\label{alg:SCHSAR2}

\vspace{3mm}

\textbf{Procedure}

\refstepcounter{algoline}\nlset{\thealgoline}\quad Set initial values for all model parameters (augmented with latent variables) in $\boldsymbol{\theta}$.

\refstepcounter{algoline}\nlset{\thealgoline}\quad\textbf{Step 1:} Sample unobserved component $\mathbf{a}$ by sampling

\refstepcounter{algoline}\nlset{\thealgoline}\qquad (a1) each $a_i$, for $i = 1,\ldots,N$, from $\mathcal{C}at\left(\bar{\rho}\right)$, given $\boldsymbol{\theta}_{-a_i}, \boldsymbol{\mathcal{D}}$. \Comment{\eqref{postanew}}

\refstepcounter{algoline}\nlset{\thealgoline}\qquad (a2) $\boldsymbol{\rho}$ from $\mathcal{D}ir\left(\bar{\xi}_1,\ldots,\bar{\xi}_S\right)$, given $\mathbf{a}$. \Comment{\eqref{postrho}}

\refstepcounter{algoline}\nlset{\thealgoline}\quad\textbf{Step 2:} Conditional on $\mathbf{a}$, sample parameters in the network formation equation by sampling

\refstepcounter{algoline}\nlset{\thealgoline}\qquad  (b1) $\mathbf{w}^\ast$ from $TM\mathcal{N}_{\mathcal{S}}\left(\mathbf{C}\gamma - \mathbf{A},\mathbf{I}_{N(N-1)}\right)$, given $\gamma,\mathbf{a},\boldsymbol{\mathcal{D}}$; \Comment{\eqref{postwnew}}

\refstepcounter{algoline}\nlset{\thealgoline}\qquad  (b2) $\gamma$ from $\mathcal{N}_{L}(\overline{\nu}_{\gamma}, \overline{\Sigma}_{\gamma})$, given $\mathbf{w}^\ast,\mathbf{a},\boldsymbol{\mathcal{D}}$. \Comment{\eqref{postgammanew}}

\refstepcounter{algoline}\nlset{\thealgoline}\quad\textbf{Step 3:} Conditional on $\mathbf{a}$ and the allocations $\mathbf{z}$, sample parameters in the outcome equation by sampling

\refstepcounter{algoline}\nlset{\thealgoline}\qquad (c1) $\boldsymbol{\pi}$ from $\mathcal{D}ir(\overline{\alpha}_1, \overline{\alpha}_2,\ldots,\overline{\alpha}_G)$, given $\mathbf{z},\{\lambda_g,\beta_g,\delta_g,\kappa_g,\sigma^2_{ug}\},\mathbf{a},\boldsymbol{\mathcal{D}}$;\Comment{\eqref{postpi}}

\refstepcounter{algoline}\nlset{\thealgoline}\qquad (c2) each $\beta_g$, for $g = 1,\ldots, G$, from $\mathcal{N}_K\bigl(\overline{\nu}_{\beta g},\overline{\Sigma}_{\beta g}\bigr)$, given $\mathbf{z},\lambda_g, \delta_g, \kappa_g,\sigma^2_{ug},\mathbf{a},\boldsymbol{\mathcal{D}}$;\Comment{\eqref{postbeta}}

\refstepcounter{algoline}\nlset{\thealgoline}\qquad (c3) each $\delta_g$, for $g = 1,\ldots, G$, from $\mathcal{N}_K\bigl(\overline{\nu}_{\delta g},\overline{\Sigma}_{\delta g}\bigr)$, given $\mathbf{z},\lambda_g,\beta_g,\kappa_g,\sigma^2_{ug},\mathbf{a},\boldsymbol{\mathcal{D}}$;\Comment{\eqref{postdelta}}

\refstepcounter{algoline}\nlset{\thealgoline}\qquad (c4) each $\kappa_g$, for $g = 1,\ldots, G$, from $\mathcal{TN}_{[0,+\infty)}\bigl(\overline{\nu}_{\kappa g},\overline{\Sigma}_{\kappa g}\bigr)$, given $\mathbf{z},\lambda_g,\beta_g,\delta_g,\sigma^2_{ug},\mathbf{a},\boldsymbol{\mathcal{D}}$; \Comment{\eqref{postkappanew}}

\refstepcounter{algoline}\nlset{\thealgoline}\qquad (c5) each $\sigma^2_{ug}$, for $g = 1,\ldots, G$, from $\mathcal{I}\mathcal{G}\bigl(\overline{s}_{ug},\overline{r}_{ug}\bigr)$, given $\mathbf{z},\lambda_g,\beta_g,\delta_g,\kappa_g,\mathbf{a},\boldsymbol{\mathcal{D}}$;\Comment{\eqref{postsigmau}}

\refstepcounter{algoline}\nlset{\thealgoline}\qquad (c6) each $\lambda_{g}$, for $g = 1,\ldots, G$, via ASM algorithm, given $\mathbf{z},\beta_g,\delta_g,\kappa_g,\sigma^2_{ug},\mathbf{a},\boldsymbol{\mathcal{D}}$.\Comment{\eqref{postlambda}}

\refstepcounter{algoline}\nlset{\thealgoline}\quad\textbf{Step 4:} For each $i = 1,\ldots,N$, sample the allocation $z_i$ from $\mathcal{M}ult \left(1, [\omega_{i1}, \omega_{i2}, \ldots, \omega_{iG}]\right)$, given $\{\lambda_g,\beta_g,\delta_g,\kappa_g,\sigma^2_{ug},\pi_g\}_{g=1}^G,\mathbf{a},\boldsymbol{\mathcal{D}}$.\Comment{\eqref{postz}}

\refstepcounter{algoline}\nlset{\thealgoline}\quad Repeat Step 1-4 using the most recent values until convergence.

\textbf{end procedure}

\end{algorithm}
\end{minipage}%
}
\end{center}

\subsection{Extension to network intensity}\label{SCHSARintensity}

In the preceding sections, we have focused on the case of binary networks, where a link indicates only the presence or absence of a relationship between two agents. The SCHSAR framework can be extended to accommodate weighted networks, where the observed link \(w_{ij}\) represents the intensity of interaction between nodes \(i\) and \(j\). In this context, we model the network formation process using a censored-normal (Tobit) specification. Recall that \(w_{ij}^{\ast}\) denotes a latent interaction index determined by observable dyadic characteristics, latent agent types, and normally distributed idiosyncratic shocks
\[
w_{ij}^{\ast} = C_{ij}^\top\gamma + f(a_i,a_j) + \epsilon_{ij} = \psi_{ij} + \epsilon_{ij}, \quad \epsilon_{ij} \overset{iid}{\sim} \mathcal{N}(0,\sigma^2_w),
\]
The observed interaction intensity is given by
\[
w_{ij} \coloneqq  w_{ij}^{\ast} \cdot \mathbbm{1}(w_{ij}^{\ast} \geq 0) = \max\{0, w_{ij}^{\ast}\},
\]
where zero corresponds to the absence of a relationship and positive values measure the strength of interaction. Unlike the binary-link specification, the positive interaction intensities are directly observed whenever \(w_{ij} > 0\) and therefore provide information about the dispersion of the latent interaction index. Consequently, the scale parameter \(\sigma_w^2\) is identified from the data and treated as an unknown parameter. The unit-variance normalization imposed in the probit specification is thus no longer required. We assign the conjugate prior
\[
\sigma^2_w \sim \mathcal{I}\mathcal{G}(\underline{s}_w,\underline{r}_w).
\]
The augmented likelihood for the parameters in the network formation model becomes
\[
p(\mathbf{W}, \mathbf{w}^\ast \mid \mathbf{C}, \mathbf{a}, \gamma, \sigma^2_w) = \prod_{i=1}^N \prod_{j < i} \left[ \mathbbm{1}(w_{ij}^\ast \geq 0)\mathbbm{1}(w_{ij} = w_{ij}^\ast) + \mathbbm{1}(w_{ij}^\ast < 0)\mathbbm{1}(w_{ij} = 0) \right] \varphi(w_{ij}^\ast ; \psi_{ij}, \sigma^2_w).
\]

The conditional posterior distributions for the latent interaction index \(\mathbf{w}^\ast\) become
\begin{flalign}
\label{postwnew2}
w_{ij}^\ast \mid \boldsymbol{\theta}_{-w_{ij}^\ast},\boldsymbol{\mathcal{D}} &\sim \begin{cases} \delta_{w_ij}, & \text{if } w_{ij} > 0,\\
T\mathcal{N}_{(-\infty,0]}(\psi_{ij}, \sigma^2_w), & \text{if } w_{ij} = 0. \end{cases},
\end{flalign}
where \(\delta_{w_{ij}}\) denotes a point mass at the observed value \(w_{ij}\), and \(T\mathcal{N}_{(-\infty,0]}\) denotes a normal distribution with mean \(\psi_{ij}\) and variance \(\sigma^2_w\) truncated above at zero. The conditional posteriors for parameters in the network formation equation are also modified accordingly, while the conditional posteriors for those in the outcome equation retain the same functional form as in the binary-network case. Specifically, they become
\begin{flalign}
\label{postgammanew2}
\qquad\gamma \mid \boldsymbol{\theta}_{-\gamma},\boldsymbol{\mathcal{D}} &\sim \mathcal{N}\bigl(\overline{\nu}_{\gamma}, \overline{\Sigma}_{\gamma}\bigr),
&&\\
\label{postsigmaw}
\qquad\sigma^2_w \mid \boldsymbol{\theta}_{-\sigma^2_w},\boldsymbol{\mathcal{D}} &\sim \mathcal{I}\mathcal{G}(\overline{s}_w, \overline{r}_w),
&&
\end{flalign}
where
\begin{flalign}
\nonumber
\qquad\overline{\Sigma}_{\gamma} &\coloneqq \left[ \underline{\Sigma}_{\gamma}^{-1} + \sigma^{-2}_w \mathbf{C}^{\top}\mathbf{C} \right]^{-1},
&&\\\nonumber
\qquad\overline{\nu}_{\gamma} &\coloneqq \overline{\Sigma}_{\gamma} \left[ \underline{\Sigma}_{\gamma}^{-1}\underline{\nu}_{\gamma} + \sigma^{-2}_w\mathbf{C}^{\top}(\mathbf{w}^\ast - \mathbf{A})\right],
&&\\\nonumber
\qquad\overline{s}_w &\coloneqq \underline{s}_w + \frac{N(N-1)}{4},
&&\\\nonumber
\qquad\overline{r}_w &\coloneqq \underline{r}_w + \frac{1}{2}(\mathbf{w}^\ast - \mathbf{C}\gamma - \mathbf{A})^{\top}(\mathbf{w}^\ast - \mathbf{C}\gamma - \mathbf{A}).
&&
\end{flalign}

Regarding the individual random effects \(a_i\) (\(i = 1,\ldots,N\)), the conditional posteriors are still Gaussian and categorical for unobserved degree heterogeneity and unobserved homophily, respectively, but with \(\sigma^2_w\) entering formulas \eqref{posta} and \eqref{postanew}
\begin{flalign}
\nonumber
\qquad\overline{\Sigma}_{a} &\coloneqq \left[ \sigma_a^{-2}\mathbf{I}_N + \sigma^{-2}_w\mathbf{H}^\top\mathbf{H} + \tilde\kappa\tilde\kappa^\top \circ \text{diag}(\tilde\sigma^{-2}_u) \right]^{-1},
&&\\\nonumber
\qquad\overline{\nu}_{a} &\coloneqq \overline{\Sigma}_{a} \left[ \sigma^{-2}_w\mathbf{H}^\top(\mathbf{w}^\ast - \mathbf{C}\gamma) + \tilde\kappa\circ\tilde\sigma^{-2}_u\circ \left( \mathbf{\tilde{Y}} - \sum_{k=1}^K X^k \circ \tilde{\beta}^k - \sum_{k=1}^K (\mathbf{W} X^k) \circ \tilde{\delta}^k \right) \right],
&&\\\nonumber
\bar{\rho}_{is} &\coloneqq \frac{P_{is}}{\sum_{s=1}^S P_{is}}, \text{ with }
&&\\\nonumber
P_{is} &\coloneqq \rho_s
\times \text{exp}\left\{ -\sum_{g=1}^G \frac{z_{gi}(Y_i - \lambda_g \overline{Y}_{(i)} - X_i^\top\beta_g - \bar{X}_{(i)}^\top\delta_g - s\kappa_g)^2}{2\sigma^2_{g}} - \sum_{j\neq i}\frac{(w^*_{ij}-C_{ij}\gamma+\vert s-a_j \vert)^2}{2\sigma^{2}_w}\right\}.
&&
\end{flalign}

The MCMC sampling procedures described in Algorithms \ref{alg:SCHSAR1} and \ref{alg:SCHSAR2} remain largely unchanged. The primary modification concerns the data-augmentation step for the latent network variables. Under the Tobit network formation model, latent interaction indices associated with positive observed links are observed directly and set equal to their observed values, whereas those corresponding to zero-valued links are sampled from the truncated normal distribution in \eqref{postwnew2}. In addition, an extra Gibbs step is introduced to update the variance parameter \(\sigma_w^2\) from its Inverse Gamma conditional posterior distribution in \eqref{postsigmaw}.

% \newpage

\section{Simulation Study}\label{SCHSARsection4}

We conduct a Monte Carlo simulation study to analyze the finite-sample performance of the proposed Bayesian MCMC algorithm for estimating the joint SCHSAR model consisting of the network formation equation in \(\eqref{NF}\) and the outcome equation in \(\eqref{SCHSAR}\).
For the main simulation exercises presented below and the subsequent empirical application, we focus on the baseline SCHSAR specification without contextual effects (i.e., excluding all \(\mathbf{W}X^k\) terms from the outcome equation). This isolates the role of endogenous peer effects and latent heterogeneity while maintaining a parsimonious model structure and reducing the computational demand. It also enables us to examine the extent of estimation bias in the peer effect parameter \(\lambda\) that arises when network endogeneity is ignored. Nevertheless, the proposed estimation procedure in section \ref{SCHSARsection3} readily accommodates contextual effects. To demonstrate this extension, we report additional simulation results for the full SCHSAR model that includes contextual effects in Appendix \ref{SCHSAR-APDX_extra_simulations}. These supplementary experiments confirm that the algorithm remains applicable in richer specifications while highlighting the additional estimation challenges introduced by the inclusion of contextual term.

\subsection{Simulation Design}\label{simulation-design}

The general data-generating process (DGP) is based on the SCHSAR framework in Section \ref{SCHSARsection2}, with the following key elements.

\textit{Generation of the individual random effects}: Each individual \(i=1,\ldots,N\) is assigned an unobserved individual-specific factor \(a_i\), which plays a central role in both the network formation and outcome equations. The form and distribution of \(a_i\) vary across simulation settings (detailed below).

\textit{Generation of the network data $\{w_{ij}\}_{i,j=1}^N$}: We generate observed dyad-specific exogenous variable \(\{C_{ij}\}_{i,j=1}^N\) by first drawing two random variables, \(v_1\) and \(v_2\), from the uniform distribution \(U(0,1)\). We then set \(C_{ij}=1\) if both \(v_1\) and \(v_2\) are below \(0.3\) or above \(0.7\), and set \(C_{ij}=0\) otherwise. The corresponding coefficient of \(C_{ij}\) is set to \(\gamma=1.5\). We simulate each entry of the adjacency matrix \(\mathbf{W}\) based on latent utilities following the network formation equation
\begin{equation}
w_{ij} = w_{ji} = \mathbbm{1}\{C_{ij}\gamma + f(a_{i}, a_{j}) + \epsilon_{ij} \geq 0 \}, \quad \text{for } i = 1,\ldots,N; j = 1,\ldots,i-1,
\end{equation}
with idiosyncratic shock \(\epsilon_{ij}\) drawn either from a standard normal or logistic distribution, depending on the specification.\footnote{In other words, the conditional probability of each \(w_{ij}\) is \(\mathbbm{P}(w_{ij} = 1 \mid \cdot) = \mathcal{H}\left[C^\top_{ij}\gamma + f(a_i,a_j)\right]\), where \(\mathcal{H}\) is the link function, either probit or logit, to be specified later.}

\textit{Generation of the outcome data $\{Y_{i}\}_{i=1}^N$}: Individuals are randomly assigned to one of three latent types (i.e., \(G=3\)) in each simulation with fixed probabilities \(\boldsymbol{\pi} = [0.45, 0.35, 0.2]\). Each individual is endowed with two observed covariates, \(\{X^1_i\}_{i=1}^{N}\) and \(\{X^2_i\}_{i=1}^{N}\), both from \(\mathcal{N}(0, 4)\), whereas their corresponding coefficients are type-specific and set to \(\beta^1 = [-0.5, 0.5, -1.0]^\top\) and \(\beta^2 = [-0.75, 0.8, 1.2]^\top\). The effect of unobserved individual-specific factor \(a_i\) is associated with \(\kappa = [0.8, 0.6, 0.25]^\top\). The type-specific peer effect parameter of interest is \(\lambda = [-0.15, 0.15, 0.3]^\top\). The error term \(u_i\) is normally distributed with variance scaled across types by \(\sigma_u^2 = c_\sigma \times [1,0.75,0.5]\), where \(c_\sigma\) controls the desired signal-to-noise ratio. Combining all components, the outcome variable is generated from the reduced form of the SCHSAR model\footnote{We employ the row-normalized version of \(\mathbf{W}\) in this outcome equation to ensure compatibility with interpretations in the empirical application. This can be implemented without loss of generality within our SCHSAR framework.}

\begin{equation}
\mathbf{Y} = \left[\mathbf{I}_N -\mathcal{L}\mathbf{W}\right]^{-1}\left(\sum_{k=1}^K\tilde{\beta}^k \circ X^k + \tilde{\kappa} \circ \mathbf{a} + \mathbf{u}\right),
\end{equation}
where \(\tilde{\beta}^k = \mathbf{z}\beta^k (k \in \{1,2\})\), and \(\tilde{\kappa} = \mathbf{z}\kappa\), with \(\mathbf{z}\) being the matrix of latent type indicators.

The sample size is kept constant across the simulations and is reflective of a large sample, where \(N=1000\). Meanwhile, the specification of \(a_i\), \(\epsilon_{ij}\), and \(c_\sigma\) are varied in the simulation study to account for different scenarios:

\begin{enumerate}
\def\labelenumi{\arabic{enumi}.}
\item
  The form of the unobserved component in the network formation equation (source of network endogeneity): In the large literature on economic and social network analysis, the latent part associated with the endogeneity between network formation and individual outcomes could present as unobserved degree heterogeneity or unobserved homophily, depending on specific application contexts. Thus, we examine both cases separately.
  \begin{align}
  f(a_i,a_j) &= a_i+a_j, \text{ where } a_i \sim \mathcal{N}(0,\sigma_a^2) \text{ and } \sigma_a^2 = 2;   \\
  \text{or}\quad f(a_i,a_j) &=-\vert a_i-a_j\vert, \text{ where } a_i \sim \mathcal{C}at\left(\frac{1}{2},\frac{1}{2}\right).
  \end{align}
\item
  Possibility of misspecification of the link function: We switch the dyad-shock distribution from standard normal (probit link) to logistic (logit link) to explore the robustness of the estimator under misspecification.
\item
  The signal-to-noise ratio (SNR) in the outcome equation: We set \(c_{\sigma} \in \{0.01,0.1,1\}\) which represents \emph{high}, \emph{medium}, and \emph{low} levels of SNR, respectively.
\end{enumerate}

For each DGP, we generate \(N_{sim} = 100\) independent replications. For each generated dataset, we estimate the following two models:

\begin{enumerate}
\def\labelenumi{\arabic{enumi}.}
\item
  Heterogeneous Spatial Autoregressive (HSAR) Model: This benchmark model ignores the endogeneity of network structure by treating \(\mathbf{W}\) as exogenous. It is conceptually aligned with the SAR-M model introduced by (\citeproc{ref-cornwall2017embracing}{Cornwall and Parent 2017}), which allows for heterogeneous peer effects but assumes a fixed spatial weight matrix.
\item
  Selection-Corrected Heterogeneous Spatial Autoregressive (SCHSAR) Model: This is the fully flexible model we propose, which jointly models network formation and outcomes while accounting for heterogeneity in peer effects and endogeneity in network links through the inclusion of latent variables.
\end{enumerate}

Each MCMC estimation is run for \(5,500\) iterations, with the first \(500\) iterations discarded as burn-in.

\subsection{Simulation Results}\label{simulation-results}

For every simulated dataset, given the posterior distribution of each model parameter resulting from MCMC draws, we derive the posterior mean for a point estimate and compute the corresponding equal-tailed \(95\%\) credible interval. We aggregate the results over \(100\) Monte Carlo replications and evaluate performance through the average bias and the root mean squared error (RMSE) of the point estimates, followed by the coverage rate of the \(95\%\) credible intervals. The simulation results are presented in Table \ref{tab:tab-MCresults-DGPI}--\ref{tab:tab-MCresults-DGPIV}, where the true values of the DGP parameters of interest are also listed in each table for ease of comparison.

Overall, across all data-generating processes (DGP I--IV) the proposed SCHSAR estimators produce near-unbiasedness and nominal coverage. When the signal-to-noise ratio (SNR) is high or medium, the true parameters are recovered very well as posterior means cluster tightly around the truth. Precision falls in low-signal setting as expected, yet coverage remains close to \(0.95\), indicating that intervals widen appropriately. By contrast, the benchmark HSAR estimator, which ignores endogenous network formation, displays severe upward or downward bias and virtually zero coverage for all peer-effect coefficients \(\lambda\). Failing to model the link formation process when the peer effect parameter is of interest therefore renders the statistical inference unreliable.

When the latent component is unobserved degree heterogeneity (DGP I and II), SCHSAR estimators achieve negligible bias for the peer effects (\(\lambda\)) even at low SNR. The variance of the individual random effects, \(\sigma_a^2\), is recovered quite accurately, indicating that the continuous latent heterogeneity is effectively captured leads to the desirable performance of SCHSAR. In contrast, under unobserved homophily (DGP III and IV), the latent mechanism penalises dissimilar types. SCHSAR estimators perform well but both bias and variability of the peer-effect estimates increase, particularly under low signal. A similar pattern arises for the loadings \(\kappa\), suggesting additional challenges added when capturing latent variables in this scenario.

DGP II and IV purposely replace the probit link assumed in estimation with a logit link in the true model. The performance of SCHSAR degrades gracefully: RMSE of the peer-effect estimates (\(\lambda\)) roughly increases relative to correctly specified cases but remains modest. SCHSAR estimators also maintain decent coverage, with a slight drop in the case of unobserved homophily yet stay above \(0.85\). Notably, while estimation for the network-covariate coefficient \(\gamma\) becomes fragile under misspecification, this does not propagate to peer effects, our primary quantities of interest. In short, link-function misspecification is tolerable for peer-effect estimation, indicating SCHSAR is reasonably robust.

In terms of other parameters, SCHSAR estimators for exogenous slope parameters \(\beta\) and group shares \(w\) exhibit virtually zero bias and coverage between \(0.93\) and \(0.97\) across all designs. These results bolster confidence that the mixture structure and covariate effects are recovered faithfully alongside endogenous peer effect parameters. In summary, the Monte Carlo findings collectively demonstrate that the SCHSAR framework delivers reliable estimation and inference.

\newpage

\begin{table}[H]
\centering
\caption{\label{tab:tab-MCresults-DGPI}DGP I: $N = 1000$, Unobserved Degree Heterogeneity}
\centering
\resizebox{\textwidth}{!}{%
{\renewcommand{\arraystretch}{0.85} % Reduce row height
\fontsize{8}{10}\selectfont
\begin{threeparttable}
\begin{tabular}[t]{ccccccccccccc}
\toprule
\multicolumn{3}{c}{\textbf{ }} & \multicolumn{5}{c}{\textbf{SCHSAR}} & \multicolumn{5}{c}{\textbf{HSAR}} \\
\cmidrule(l{3pt}r{3pt}){4-8} \cmidrule(l{3pt}r{3pt}){9-13}
SNR & Parameter & True Value & Mean & Std & Bias & RMSE & Coverage & Mean & Std & Bias & RMSE & Coverage\\
\midrule
 & $\lambda_1$ & -0.15 & -0.151 & 0.010 & -0.001 & 0.011 & 0.92 & -0.881 & 0.101 & -0.731 & 0.738 & 0.00\\

 & $\lambda_2$ & 0.15 & 0.149 & 0.009 & -0.001 & 0.010 & 0.94 & -0.502 & 0.157 & -0.652 & 0.670 & 0.00\\

 & $\lambda_3$ & 0.30 & 0.298 & 0.019 & -0.002 & 0.019 & 0.96 & 0.026 & 0.070 & -0.274 & 0.283 & 0.00\\

 & $\omega_1$ & 0.45 & 0.452 & 0.015 & 0.002 & 0.015 & 0.98 & 0.451 & 0.020 & 0.001 & 0.020 & 0.98\\

 & $\omega_2$ & 0.35 & 0.349 & 0.016 & -0.001 & 0.016 & 0.99 & 0.350 & 0.020 & 0.000 & 0.020 & 0.99\\

 & $\omega_3$ & 0.20 & 0.199 & 0.015 & -0.001 & 0.015 & 0.93 & 0.199 & 0.017 & -0.001 & 0.017 & 0.91\\

 & $\beta_{11}$ & -0.50 & -0.500 & 0.003 & 0.000 & 0.003 & 0.93 & -0.502 & 0.029 & -0.002 & 0.029 & 0.92\\

 & $\beta_{12}$ & 0.50 & 0.498 & 0.017 & -0.002 & 0.017 & 0.86 & 0.496 & 0.026 & -0.004 & 0.026 & 0.92\\

 & $\beta_{13}$ & -1.00 & -0.966 & 0.332 & 0.034 & 0.333 & 0.98 & -1.000 & 0.014 & 0.000 & 0.014 & 0.97\\

 & $\beta_{21}$ & -0.75 & -0.750 & 0.003 & 0.000 & 0.003 & 0.93 & -0.749 & 0.027 & 0.001 & 0.027 & 0.95\\

 & $\beta_{22}$ & 0.80 & 0.800 & 0.004 & 0.000 & 0.004 & 0.95 & 0.801 & 0.021 & 0.002 & 0.021 & 0.96\\

 & $\beta_{23}$ & 1.20 & 1.158 & 0.404 & -0.042 & 0.406 & 0.95 & 1.201 & 0.013 & 0.001 & 0.013 & 0.97\\

 & $\kappa_1$ & 0.80 & 0.798 & 0.005 & -0.002 & 0.005 & 0.90 & -- & -- & -- & -- & --\\

 & $\kappa_2$ & 0.60 & 0.598 & 0.007 & -0.002 & 0.007 & 0.99 & -- & -- & -- & -- & --\\

 & $\kappa_3$ & 0.25 & 0.227 & 0.227 & -0.023 & 0.228 & 0.94 & -- & -- & -- & -- & --\\

 & $\gamma$ & 1.50 & 1.501 & 0.010 & 0.001 & 0.010 & 0.80 & -- & -- & -- & -- & --\\

\multirow{-17}{*}{\centering\arraybackslash High} & $\sigma^2_a$ & 2.00 & 2.001 & 0.083 & 0.001 & 0.083 & 0.99 & -- & -- & -- & -- & --\\
\cmidrule{1-13}
 & $\lambda_1$ & -0.15 & -0.153 & 0.033 & -0.003 & 0.033 & 0.93 & -0.876 & 0.105 & -0.726 & 0.734 & 0.00\\

 & $\lambda_2$ & 0.15 & 0.148 & 0.029 & -0.002 & 0.029 & 0.95 & -0.502 & 0.162 & -0.652 & 0.672 & 0.00\\

 & $\lambda_3$ & 0.30 & 0.300 & 0.031 & 0.000 & 0.031 & 0.96 & 0.028 & 0.081 & -0.272 & 0.284 & 0.01\\

 & $\omega_1$ & 0.45 & 0.452 & 0.015 & 0.002 & 0.016 & 0.97 & 0.451 & 0.020 & 0.001 & 0.020 & 0.97\\

 & $\omega_2$ & 0.35 & 0.348 & 0.016 & -0.002 & 0.016 & 0.98 & 0.349 & 0.021 & -0.001 & 0.021 & 0.98\\

 & $\omega_3$ & 0.20 & 0.200 & 0.014 & 0.000 & 0.014 & 0.94 & 0.200 & 0.017 & 0.000 & 0.017 & 0.92\\

 & $\beta_{11}$ & -0.50 & -0.501 & 0.009 & -0.001 & 0.009 & 0.91 & -0.503 & 0.030 & -0.003 & 0.030 & 0.94\\

 & $\beta_{12}$ & 0.50 & 0.500 & 0.010 & 0.000 & 0.010 & 0.90 & 0.496 & 0.028 & -0.004 & 0.028 & 0.93\\

 & $\beta_{13}$ & -1.00 & -0.999 & 0.008 & 0.002 & 0.008 & 0.97 & -0.998 & 0.017 & 0.002 & 0.017 & 0.96\\

 & $\beta_{21}$ & -0.75 & -0.750 & 0.008 & 0.000 & 0.008 & 0.94 & -0.749 & 0.029 & 0.001 & 0.029 & 0.94\\

 & $\beta_{22}$ & 0.80 & 0.800 & 0.008 & 0.000 & 0.008 & 0.95 & 0.801 & 0.023 & 0.001 & 0.023 & 0.96\\

 & $\beta_{23}$ & 1.20 & 1.200 & 0.011 & 0.000 & 0.011 & 0.95 & 1.202 & 0.016 & 0.002 & 0.016 & 0.96\\

 & $\kappa_1$ & 0.80 & 0.798 & 0.013 & -0.002 & 0.013 & 0.95 & -- & -- & -- & -- & --\\

 & $\kappa_2$ & 0.60 & 0.598 & 0.013 & -0.002 & 0.013 & 0.95 & -- & -- & -- & -- & --\\

 & $\kappa_3$ & 0.25 & 0.249 & 0.014 & -0.001 & 0.014 & 0.96 & -- & -- & -- & -- & --\\

 & $\gamma$ & 1.50 & 1.501 & 0.010 & 0.001 & 0.010 & 0.80 & -- & -- & -- & -- & --\\

\multirow{-17}{*}{\centering\arraybackslash Medium} & $\sigma^2_a$ & 2.00 & 2.002 & 0.083 & 0.002 & 0.083 & 0.99 & -- & -- & -- & -- & --\\
\cmidrule{1-13}
 & $\lambda_1$ & -0.15 & -0.158 & 0.116 & -0.008 & 0.117 & 0.95 & -0.828 & 0.133 & -0.678 & 0.691 & 0.00\\

 & $\lambda_2$ & 0.15 & 0.135 & 0.117 & -0.015 & 0.117 & 0.93 & -0.492 & 0.203 & -0.642 & 0.673 & 0.02\\

 & $\lambda_3$ & 0.30 & 0.293 & 0.132 & -0.007 & 0.132 & 0.95 & 0.011 & 0.175 & -0.289 & 0.338 & 0.50\\

 & $\omega_1$ & 0.45 & 0.450 & 0.018 & 0.000 & 0.018 & 0.97 & 0.449 & 0.022 & -0.001 & 0.022 & 0.99\\

 & $\omega_2$ & 0.35 & 0.348 & 0.022 & -0.002 & 0.022 & 0.96 & 0.346 & 0.026 & -0.004 & 0.027 & 0.98\\

 & $\omega_3$ & 0.20 & 0.203 & 0.018 & 0.003 & 0.018 & 0.95 & 0.205 & 0.022 & 0.005 & 0.023 & 0.93\\

 & $\beta_{11}$ & -0.50 & -0.496 & 0.101 & 0.004 & 0.101 & 0.91 & -0.487 & 0.141 & 0.013 & 0.141 & 0.95\\

 & $\beta_{12}$ & 0.50 & 0.491 & 0.109 & -0.009 & 0.110 & 0.91 & 0.468 & 0.194 & -0.032 & 0.197 & 0.94\\

 & $\beta_{13}$ & -1.00 & -0.990 & 0.033 & 0.010 & 0.035 & 0.96 & -0.977 & 0.137 & 0.024 & 0.139 & 0.95\\

 & $\beta_{21}$ & -0.75 & -0.735 & 0.154 & 0.015 & 0.155 & 0.94 & -0.720 & 0.216 & 0.030 & 0.218 & 0.93\\

 & $\beta_{22}$ & 0.80 & 0.784 & 0.153 & -0.016 & 0.154 & 0.95 & 0.771 & 0.230 & -0.029 & 0.232 & 0.92\\

 & $\beta_{23}$ & 1.20 & 1.202 & 0.037 & 0.002 & 0.037 & 0.92 & 1.194 & 0.076 & -0.006 & 0.076 & 0.94\\

 & $\kappa_1$ & 0.80 & 0.798 & 0.048 & -0.002 & 0.048 & 0.95 & -- & -- & -- & -- & --\\

 & $\kappa_2$ & 0.60 & 0.597 & 0.051 & -0.003 & 0.051 & 0.94 & -- & -- & -- & -- & --\\

 & $\kappa_3$ & 0.25 & 0.246 & 0.062 & -0.004 & 0.062 & 0.97 & -- & -- & -- & -- & --\\

 & $\gamma$ & 1.50 & 1.501 & 0.010 & 0.001 & 0.010 & 0.80 & -- & -- & -- & -- & --\\

\multirow{-17}{*}{\centering\arraybackslash Low} & $\sigma^2_a$ & 2.00 & 2.002 & 0.083 & 0.002 & 0.083 & 0.98 & -- & -- & -- & -- & --\\
\bottomrule
\end{tabular}
% \begin{tablenotes}[para]
% \item \textit{Notes:} This table displays results based on $R=100$ replicates. The values include the average and standard deviation of the point estimates; the average bias (Bias), the Root Mean Squared Error (RMSE), and the coverage rate (Coverage) across replicates; where $\text{Bias}=R^{-1}\sum_{r=1}^R (\hat{\alpha}_r-\alpha),\text{ RMSE}=\sqrt{R^{-1}\sum_{r=1}^R (\hat{\alpha}_r-\alpha)^2},$ and $\text{ Coverage}=R^{-1}\sum_{r=1}^R \mathbbm{1}\{\alpha\in \widehat{CI}_{0.95,r}\}$.
% \end{tablenotes}
\end{threeparttable}}
}
\end{table}

\begin{table}[H]
\centering
\caption{\label{tab:tab-MCresults-DGPII}DGP II: $N = 1000$, Unobserved Degree Heterogeneity, Link Misspecification}
\centering
\resizebox{\textwidth}{!}{%
{\renewcommand{\arraystretch}{0.85} % Reduce row height
\fontsize{8}{10}\selectfont
\begin{threeparttable}
\begin{tabular}[t]{ccccccccccccc}
\toprule
\multicolumn{3}{c}{\textbf{ }} & \multicolumn{5}{c}{\textbf{SCHSAR}} & \multicolumn{5}{c}{\textbf{HSAR}} \\
\cmidrule(l{3pt}r{3pt}){4-8} \cmidrule(l{3pt}r{3pt}){9-13}
SNR & Parameter & True Value & Mean & Std & Bias & RMSE & Coverage & Mean & Std & Bias & RMSE & Coverage\\
\midrule
 & $\lambda_1$ & -0.15 & -0.150 & 0.014 & 0.000 & 0.014 & 0.94 & -0.878 & 0.110 & -0.728 & 0.736 & 0.00\\

 & $\lambda_2$ & 0.15 & 0.151 & 0.013 & 0.001 & 0.013 & 0.96 & -0.590 & 0.217 & -0.740 & 0.771 & 0.00\\

 & $\lambda_3$ & 0.30 & 0.300 & 0.012 & 0.000 & 0.012 & 0.98 & -0.021 & 0.114 & -0.321 & 0.341 & 0.01\\

 & $\omega_1$ & 0.45 & 0.452 & 0.015 & 0.002 & 0.015 & 0.98 & 0.450 & 0.020 & 0.000 & 0.020 & 0.98\\

 & $\omega_2$ & 0.35 & 0.349 & 0.015 & -0.001 & 0.015 & 0.98 & 0.350 & 0.021 & 0.000 & 0.021 & 0.99\\

 & $\omega_3$ & 0.20 & 0.200 & 0.014 & 0.000 & 0.014 & 0.91 & 0.200 & 0.017 & 0.000 & 0.017 & 0.93\\

 & $\beta_{11}$ & -0.50 & -0.500 & 0.003 & 0.000 & 0.003 & 0.97 & -0.502 & 0.031 & -0.002 & 0.031 & 0.94\\

 & $\beta_{12}$ & 0.50 & 0.500 & 0.003 & 0.000 & 0.004 & 0.81 & 0.495 & 0.028 & -0.005 & 0.028 & 0.92\\

 & $\beta_{13}$ & -1.00 & -1.000 & 0.002 & 0.000 & 0.002 & 0.98 & -0.999 & 0.017 & 0.001 & 0.017 & 0.98\\

 & $\beta_{21}$ & -0.75 & -0.750 & 0.003 & 0.000 & 0.003 & 0.96 & -0.749 & 0.028 & 0.001 & 0.028 & 0.96\\

 & $\beta_{22}$ & 0.80 & 0.800 & 0.003 & 0.000 & 0.003 & 0.96 & 0.802 & 0.023 & 0.002 & 0.023 & 0.97\\

 & $\beta_{23}$ & 1.20 & 1.200 & 0.003 & 0.000 & 0.003 & 0.91 & 1.198 & 0.035 & -0.002 & 0.035 & 0.98\\

 & $\kappa_1$ & 0.80 & 1.387 & 0.000 & 0.587 & 0.587 & 0.00 & -- & -- & -- & -- & --\\

 & $\kappa_2$ & 0.60 & 1.041 & 0.000 & 0.441 & 0.441 & 0.00 & -- & -- & -- & -- & --\\

 & $\kappa_3$ & 0.25 & 0.434 & 0.006 & 0.184 & 0.184 & 0.00 & -- & -- & -- & -- & --\\

 & $\gamma$ & 1.50 & 0.864 & 0.000 & -0.636 & 0.636 & 0.00 & -- & -- & -- & -- & --\\

\multirow{-17}{*}{\centering\arraybackslash High} & $\sigma^2_a$ & 2.00 & 0.661 & 0.023 & -1.339 & 1.339 & 0.00 & -- & -- & -- & -- & --\\
\cmidrule{1-13}
 & $\lambda_1$ & -0.15 & -0.153 & 0.042 & -0.003 & 0.042 & 0.94 & -0.872 & 0.115 & -0.722 & 0.731 & 0.00\\

 & $\lambda_2$ & 0.15 & 0.150 & 0.037 & 0.000 & 0.037 & 0.96 & -0.588 & 0.217 & -0.738 & 0.769 & 0.00\\

 & $\lambda_3$ & 0.30 & 0.300 & 0.039 & 0.000 & 0.039 & 0.97 & -0.019 & 0.128 & -0.319 & 0.344 & 0.03\\

 & $\omega_1$ & 0.45 & 0.452 & 0.016 & 0.002 & 0.016 & 0.98 & 0.451 & 0.020 & 0.001 & 0.020 & 0.99\\

 & $\omega_2$ & 0.35 & 0.348 & 0.016 & -0.002 & 0.016 & 0.97 & 0.349 & 0.021 & -0.001 & 0.021 & 0.98\\

 & $\omega_3$ & 0.20 & 0.200 & 0.014 & 0.000 & 0.014 & 0.96 & 0.200 & 0.018 & 0.000 & 0.018 & 0.94\\

 & $\beta_{11}$ & -0.50 & -0.501 & 0.009 & -0.001 & 0.009 & 0.90 & -0.503 & 0.031 & -0.003 & 0.032 & 0.94\\

 & $\beta_{12}$ & 0.50 & 0.500 & 0.010 & 0.000 & 0.010 & 0.86 & 0.496 & 0.029 & -0.004 & 0.029 & 0.94\\

 & $\beta_{13}$ & -1.00 & -0.999 & 0.008 & 0.002 & 0.008 & 0.96 & -0.998 & 0.017 & 0.002 & 0.017 & 0.97\\

 & $\beta_{21}$ & -0.75 & -0.750 & 0.008 & 0.000 & 0.008 & 0.95 & -0.749 & 0.030 & 0.001 & 0.030 & 0.94\\

 & $\beta_{22}$ & 0.80 & 0.799 & 0.008 & -0.001 & 0.008 & 0.95 & 0.801 & 0.025 & 0.001 & 0.025 & 0.96\\

 & $\beta_{23}$ & 1.20 & 1.200 & 0.009 & 0.000 & 0.010 & 0.94 & 1.202 & 0.017 & 0.002 & 0.017 & 0.96\\

 & $\kappa_1$ & 0.80 & 1.390 & 0.024 & 0.590 & 0.591 & 0.00 & -- & -- & -- & -- & --\\

 & $\kappa_2$ & 0.60 & 1.042 & 0.019 & 0.442 & 0.442 & 0.00 & -- & -- & -- & -- & --\\

 & $\kappa_3$ & 0.25 & 0.433 & 0.024 & 0.183 & 0.185 & 0.00 & -- & -- & -- & -- & --\\

 & $\gamma$ & 1.50 & 0.864 & 0.000 & -0.636 & 0.636 & 0.00 & -- & -- & -- & -- & --\\

\multirow{-17}{*}{\centering\arraybackslash Medium} & $\sigma^2_a$ & 2.00 & 0.660 & 0.023 & -1.340 & 1.340 & 0.00 & -- & -- & -- & -- & --\\
\cmidrule{1-13}
 & $\lambda_1$ & -0.15 & -0.160 & 0.148 & -0.011 & 0.148 & 0.95 & -0.816 & 0.144 & -0.666 & 0.681 & 0.00\\

 & $\lambda_2$ & 0.15 & 0.136 & 0.147 & -0.014 & 0.147 & 0.92 & -0.554 & 0.237 & -0.704 & 0.743 & 0.05\\

 & $\lambda_3$ & 0.30 & 0.281 & 0.171 & -0.019 & 0.172 & 0.94 & -0.040 & 0.226 & -0.340 & 0.408 & 0.58\\

 & $\omega_1$ & 0.45 & 0.450 & 0.018 & 0.000 & 0.018 & 0.99 & 0.449 & 0.022 & -0.001 & 0.022 & 0.99\\

 & $\omega_2$ & 0.35 & 0.348 & 0.022 & -0.002 & 0.022 & 0.97 & 0.346 & 0.027 & -0.004 & 0.027 & 0.98\\

 & $\omega_3$ & 0.20 & 0.202 & 0.018 & 0.002 & 0.018 & 0.95 & 0.205 & 0.022 & 0.005 & 0.023 & 0.92\\

 & $\beta_{11}$ & -0.50 & -0.496 & 0.101 & 0.004 & 0.101 & 0.91 & -0.486 & 0.141 & 0.014 & 0.142 & 0.95\\

 & $\beta_{12}$ & 0.50 & 0.491 & 0.109 & -0.009 & 0.110 & 0.90 & 0.481 & 0.155 & -0.019 & 0.156 & 0.95\\

 & $\beta_{13}$ & -1.00 & -0.989 & 0.033 & 0.011 & 0.035 & 0.96 & -0.988 & 0.042 & 0.012 & 0.044 & 0.94\\

 & $\beta_{21}$ & -0.75 & -0.735 & 0.154 & 0.015 & 0.155 & 0.94 & -0.720 & 0.216 & 0.030 & 0.218 & 0.93\\

 & $\beta_{22}$ & 0.80 & 0.784 & 0.153 & -0.016 & 0.154 & 0.95 & 0.768 & 0.227 & -0.032 & 0.229 & 0.93\\

 & $\beta_{23}$ & 1.20 & 1.200 & 0.039 & 0.000 & 0.039 & 0.93 & 1.201 & 0.039 & 0.001 & 0.039 & 0.96\\

 & $\kappa_1$ & 0.80 & 1.391 & 0.082 & 0.591 & 0.596 & 0.00 & -- & -- & -- & -- & --\\

 & $\kappa_2$ & 0.60 & 1.041 & 0.084 & 0.441 & 0.449 & 0.00 & -- & -- & -- & -- & --\\

 & $\kappa_3$ & 0.25 & 0.427 & 0.105 & 0.177 & 0.206 & 0.59 & -- & -- & -- & -- & --\\

 & $\gamma$ & 1.50 & 0.864 & 0.000 & -0.636 & 0.636 & 0.00 & -- & -- & -- & -- & --\\

\multirow{-17}{*}{\centering\arraybackslash Low} & $\sigma^2_a$ & 2.00 & 0.660 & 0.023 & -1.340 & 1.341 & 0.00 & -- & -- & -- & -- & --\\
\bottomrule
\end{tabular}
% \begin{tablenotes}[para]
% \item \textit{Notes:} This table displays results based on $R=100$ replicates. The values include the average and standard deviation of the point estimates; the average bias (Bias), the Root Mean Squared Error (RMSE), and the coverage rate (Coverage) across replicates; where $\text{Bias}=R^{-1}\sum_{r=1}^R (\hat{\alpha}_r-\alpha),\text{ RMSE}=\sqrt{R^{-1}\sum_{r=1}^R (\hat{\alpha}_r-\alpha)^2},$ and $\text{ Coverage}=R^{-1}\sum_{r=1}^R \mathbbm{1}\{\alpha\in \widehat{CI}_{0.95,r}\}$.
% \end{tablenotes}
\end{threeparttable}}
}
\end{table}

\newpage

\begin{table}[H]
\centering
\caption{\label{tab:tab-MCresults-DGPIII}DGP III: $N = 1000$, Unobserved Homophily}
\centering
\resizebox{\textwidth}{!}{%
{\renewcommand{\arraystretch}{0.9} % Reduce row height
\fontsize{8}{10}\selectfont
\begin{threeparttable}
\begin{tabular}[t]{ccccccccccccc}
\toprule
\multicolumn{3}{c}{\textbf{ }} & \multicolumn{5}{c}{\textbf{SCHSAR}} & \multicolumn{5}{c}{\textbf{HSAR}} \\
\cmidrule(l{3pt}r{3pt}){4-8} \cmidrule(l{3pt}r{3pt}){9-13}
SNR & Parameter & True Value & Mean & Std & Bias & RMSE & Coverage & Mean & Std & Bias & RMSE & Coverage\\
\midrule
 & $\lambda_1$ & -0.15 & -0.102 & 0.228 & 0.048 & 0.232 & 0.91 & 0.973 & 0.055 & 1.123 & 1.124 & 0.00\\

 & $\lambda_2$ & 0.15 & 0.184 & 0.170 & 0.034 & 0.172 & 0.92 & 0.983 & 0.030 & 0.833 & 0.834 & 0.00\\

 & $\lambda_3$ & 0.30 & 0.318 & 0.108 & 0.018 & 0.109 & 0.90 & 0.744 & 0.166 & 0.444 & 0.474 & 0.00\\

 & $\omega_1$ & 0.45 & 0.452 & 0.015 & 0.002 & 0.015 & 0.97 & 0.454 & 0.020 & 0.004 & 0.020 & 0.94\\

 & $\omega_2$ & 0.35 & 0.348 & 0.015 & -0.002 & 0.015 & 0.99 & 0.347 & 0.018 & -0.003 & 0.018 & 0.97\\

 & $\omega_3$ & 0.20 & 0.199 & 0.013 & -0.001 & 0.013 & 0.94 & 0.199 & 0.014 & -0.001 & 0.014 & 0.94\\

 & $\beta_{11}$ & -0.50 & -0.500 & 0.003 & 0.000 & 0.003 & 0.95 & -0.490 & 0.071 & 0.010 & 0.071 & 0.92\\

 & $\beta_{12}$ & 0.50 & 0.500 & 0.002 & 0.000 & 0.002 & 0.97 & 0.480 & 0.142 & -0.020 & 0.142 & 0.94\\

 & $\beta_{13}$ & -1.00 & -1.000 & 0.002 & 0.000 & 0.002 & 0.96 & -0.991 & 0.071 & 0.009 & 0.071 & 0.92\\

 & $\beta_{21}$ & -0.75 & -0.750 & 0.003 & 0.000 & 0.003 & 0.94 & -0.716 & 0.236 & 0.034 & 0.238 & 0.94\\

 & $\beta_{22}$ & 0.80 & 0.800 & 0.002 & 0.000 & 0.002 & 0.96 & 0.769 & 0.218 & -0.031 & 0.219 & 0.97\\

 & $\beta_{23}$ & 1.20 & 1.201 & 0.003 & 0.001 & 0.003 & 0.93 & 1.162 & 0.275 & -0.038 & 0.276 & 0.91\\

 & $\kappa_1$ & 0.80 & 0.753 & 0.224 & -0.047 & 0.228 & 0.89 & -- & -- & -- & -- & --\\

 & $\kappa_2$ & 0.60 & 0.567 & 0.166 & -0.034 & 0.168 & 0.91 & -- & -- & -- & -- & --\\

 & $\kappa_3$ & 0.25 & 0.235 & 0.079 & -0.015 & 0.080 & 0.88 & -- & -- & -- & -- & --\\

\multirow{-16}{*}{\centering\arraybackslash High} & $\gamma$ & 1.50 & 1.501 & 0.005 & 0.001 & 0.005 & 0.81 & -- & -- & -- & -- & --\\
\cmidrule{1-13}
 & $\lambda_1$ & -0.15 & -0.105 & 0.251 & 0.045 & 0.254 & 0.91 & 0.966 & 0.052 & 1.116 & 1.117 & 0.00\\

 & $\lambda_2$ & 0.15 & 0.182 & 0.195 & 0.032 & 0.197 & 0.94 & 0.970 & 0.033 & 0.820 & 0.821 & 0.00\\

 & $\lambda_3$ & 0.30 & 0.311 & 0.156 & 0.011 & 0.155 & 0.92 & 0.736 & 0.088 & 0.436 & 0.444 & 0.00\\

 & $\omega_1$ & 0.45 & 0.452 & 0.016 & 0.002 & 0.016 & 0.97 & 0.453 & 0.018 & 0.003 & 0.018 & 0.97\\

 & $\omega_2$ & 0.35 & 0.349 & 0.017 & -0.001 & 0.017 & 1.00 & 0.348 & 0.017 & -0.002 & 0.017 & 0.98\\

 & $\omega_3$ & 0.20 & 0.199 & 0.015 & -0.001 & 0.015 & 0.92 & 0.198 & 0.015 & -0.002 & 0.015 & 0.94\\

 & $\beta_{11}$ & -0.50 & -0.500 & 0.008 & 0.000 & 0.008 & 0.96 & -0.500 & 0.012 & 0.000 & 0.012 & 0.93\\

 & $\beta_{12}$ & 0.50 & 0.498 & 0.014 & -0.002 & 0.014 & 0.96 & 0.499 & 0.013 & -0.001 & 0.013 & 0.96\\

 & $\beta_{13}$ & -1.00 & -0.999 & 0.013 & 0.001 & 0.013 & 0.95 & -0.999 & 0.018 & 0.001 & 0.018 & 0.97\\

 & $\beta_{21}$ & -0.75 & -0.750 & 0.008 & 0.000 & 0.008 & 0.95 & -0.749 & 0.011 & 0.001 & 0.011 & 0.95\\

 & $\beta_{22}$ & 0.80 & 0.800 & 0.009 & 0.000 & 0.009 & 0.96 & 0.800 & 0.010 & 0.000 & 0.010 & 0.97\\

 & $\beta_{23}$ & 1.20 & 1.197 & 0.043 & -0.003 & 0.043 & 0.92 & 1.199 & 0.031 & -0.001 & 0.031 & 0.92\\

 & $\kappa_1$ & 0.80 & 0.754 & 0.232 & -0.046 & 0.235 & 0.89 & -- & -- & -- & -- & --\\

 & $\kappa_2$ & 0.60 & 0.568 & 0.171 & -0.032 & 0.173 & 0.94 & -- & -- & -- & -- & --\\

 & $\kappa_3$ & 0.25 & 0.237 & 0.098 & -0.013 & 0.098 & 0.92 & -- & -- & -- & -- & --\\

\multirow{-16}{*}{\centering\arraybackslash Medium} & $\gamma$ & 1.50 & 1.501 & 0.005 & 0.001 & 0.005 & 0.82 & -- & -- & -- & -- & --\\
\cmidrule{1-13}
 & $\lambda_1$ & -0.15 & -0.111 & 0.308 & 0.039 & 0.309 & 0.95 & 0.879 & 0.059 & 1.029 & 1.031 & 0.00\\

 & $\lambda_2$ & 0.15 & 0.096 & 0.326 & -0.054 & 0.329 & 0.94 & 0.864 & 0.067 & 0.714 & 0.717 & 0.00\\

 & $\lambda_3$ & 0.30 & 0.142 & 0.289 & -0.158 & 0.328 & 0.94 & 0.612 & 0.128 & 0.312 & 0.337 & 0.59\\

 & $\omega_1$ & 0.45 & 0.451 & 0.020 & 0.001 & 0.020 & 0.99 & 0.452 & 0.021 & 0.002 & 0.021 & 0.97\\

 & $\omega_2$ & 0.35 & 0.347 & 0.022 & -0.003 & 0.022 & 0.98 & 0.347 & 0.023 & -0.003 & 0.023 & 0.99\\

 & $\omega_3$ & 0.20 & 0.202 & 0.019 & 0.002 & 0.019 & 0.96 & 0.201 & 0.019 & 0.001 & 0.019 & 0.98\\

 & $\beta_{11}$ & -0.50 & -0.488 & 0.099 & 0.012 & 0.100 & 0.96 & -0.498 & 0.030 & 0.002 & 0.030 & 0.96\\

 & $\beta_{12}$ & 0.50 & 0.489 & 0.104 & -0.011 & 0.104 & 0.94 & 0.500 & 0.030 & 0.000 & 0.030 & 0.95\\

 & $\beta_{13}$ & -1.00 & -1.000 & 0.032 & 0.000 & 0.032 & 0.97 & -1.000 & 0.033 & 0.000 & 0.033 & 0.98\\

 & $\beta_{21}$ & -0.75 & -0.735 & 0.159 & 0.015 & 0.159 & 0.92 & -0.749 & 0.029 & 0.001 & 0.028 & 0.93\\

 & $\beta_{22}$ & 0.80 & 0.782 & 0.157 & -0.018 & 0.157 & 0.92 & 0.797 & 0.031 & -0.003 & 0.031 & 0.92\\

 & $\beta_{23}$ & 1.20 & 1.206 & 0.033 & 0.006 & 0.034 & 0.95 & 1.206 & 0.033 & 0.006 & 0.034 & 0.94\\

 & $\kappa_1$ & 0.80 & 0.754 & 0.230 & -0.046 & 0.233 & 0.95 & -- & -- & -- & -- & --\\

 & $\kappa_2$ & 0.60 & 0.610 & 0.210 & 0.010 & 0.209 & 0.95 & -- & -- & -- & -- & --\\

 & $\kappa_3$ & 0.25 & 0.308 & 0.173 & 0.058 & 0.182 & 0.92 & -- & -- & -- & -- & --\\

\multirow{-16}{*}{\centering\arraybackslash Low} & $\gamma$ & 1.50 & 1.501 & 0.005 & 0.001 & 0.005 & 0.84 & -- & -- & -- & -- & --\\
\bottomrule
\end{tabular}
% \begin{tablenotes}[para]
% \item \textit{Notes:} This table displays results based on $R=100$ replicates. The values include the average and standard deviation of the point estimates; the average bias (Bias), the Root Mean Squared Error (RMSE), and the coverage rate (Coverage) across replicates; where $\text{Bias}=R^{-1}\sum_{r=1}^R (\hat{\alpha}_r-\alpha),\text{ RMSE}=\sqrt{R^{-1}\sum_{r=1}^R (\hat{\alpha}_r-\alpha)^2},$ and $\text{ Coverage}=R^{-1}\sum_{r=1}^R \mathbbm{1}\{\alpha\in \widehat{CI}_{0.95,r}\}$.
% \end{tablenotes}
\end{threeparttable}}
}
\end{table}

\newpage

\begin{table}[H]
\centering
\caption{\label{tab:tab-MCresults-DGPIV}DGP IV: $N = 1000$, Unobserved Homophily, Link Misspecification}
\centering
\resizebox{\textwidth}{!}{%
{\renewcommand{\arraystretch}{0.9} % Reduce row height
\fontsize{8}{10}\selectfont
\begin{threeparttable}
\begin{tabular}[t]{ccccccccccccc}
\toprule
\multicolumn{3}{c}{\textbf{ }} & \multicolumn{5}{c}{\textbf{SCHSAR}} & \multicolumn{5}{c}{\textbf{HSAR}} \\
\cmidrule(l{3pt}r{3pt}){4-8} \cmidrule(l{3pt}r{3pt}){9-13}
SNR & Parameter & True Value & Mean & Std & Bias & RMSE & Coverage & Mean & Std & Bias & RMSE & Coverage\\
\midrule
 & $\lambda_1$ & -0.15 & -0.066 & 0.294 & 0.084 & 0.305 & 0.89 & 0.961 & 0.067 & 1.111 & 1.113 & 0.00\\

 & $\lambda_2$ & 0.15 & 0.208 & 0.220 & 0.059 & 0.226 & 0.89 & 0.974 & 0.044 & 0.824 & 0.825 & 0.00\\

 & $\lambda_3$ & 0.30 & 0.335 & 0.141 & 0.035 & 0.145 & 0.89 & 0.723 & 0.216 & 0.423 & 0.474 & 0.00\\

 & $\omega_1$ & 0.45 & 0.452 & 0.015 & 0.002 & 0.015 & 0.98 & 0.454 & 0.021 & 0.004 & 0.022 & 0.95\\

 & $\omega_2$ & 0.35 & 0.349 & 0.016 & -0.001 & 0.016 & 0.98 & 0.348 & 0.017 & -0.002 & 0.017 & 0.99\\

 & $\omega_3$ & 0.20 & 0.198 & 0.015 & -0.002 & 0.015 & 0.93 & 0.198 & 0.017 & -0.002 & 0.017 & 0.93\\

 & $\beta_{11}$ & -0.50 & -0.500 & 0.003 & 0.000 & 0.003 & 0.93 & -0.475 & 0.112 & 0.025 & 0.114 & 0.86\\

 & $\beta_{12}$ & 0.50 & 0.498 & 0.021 & -0.002 & 0.021 & 0.97 & 0.451 & 0.219 & -0.049 & 0.224 & 0.91\\

 & $\beta_{13}$ & -1.00 & -1.000 & 0.002 & 0.000 & 0.002 & 0.97 & -0.978 & 0.102 & 0.022 & 0.103 & 0.89\\

 & $\beta_{21}$ & -0.75 & -0.750 & 0.003 & 0.000 & 0.003 & 0.95 & -0.663 & 0.375 & 0.087 & 0.383 & 0.91\\

 & $\beta_{22}$ & 0.80 & 0.800 & 0.007 & 0.000 & 0.007 & 0.94 & 0.722 & 0.340 & -0.078 & 0.347 & 0.94\\

 & $\beta_{23}$ & 1.20 & 1.201 & 0.003 & 0.001 & 0.003 & 0.93 & 1.112 & 0.397 & -0.088 & 0.405 & 0.90\\

 & $\kappa_1$ & 0.80 & 0.717 & 0.302 & -0.083 & 0.312 & 0.85 & -- & -- & -- & -- & --\\

 & $\kappa_2$ & 0.60 & 0.539 & 0.224 & -0.061 & 0.231 & 0.89 & -- & -- & -- & -- & --\\

 & $\kappa_3$ & 0.25 & 0.236 & 0.184 & -0.014 & 0.184 & 0.87 & -- & -- & -- & -- & --\\

\multirow{-16}{*}{\centering\arraybackslash High} & $\gamma$ & 1.50 & 1.144 & 0.005 & -0.356 & 0.356 & 0.00 & -- & -- & -- & -- & --\\
\cmidrule{1-13}
 & $\lambda_1$ & -0.15 & -0.080 & 0.287 & 0.070 & 0.294 & 0.88 & 0.956 & 0.060 & 1.106 & 1.107 & 0.00\\

 & $\lambda_2$ & 0.15 & 0.194 & 0.224 & 0.044 & 0.227 & 0.91 & 0.962 & 0.039 & 0.812 & 0.813 & 0.00\\

 & $\lambda_3$ & 0.30 & 0.324 & 0.158 & 0.024 & 0.160 & 0.91 & 0.731 & 0.095 & 0.431 & 0.441 & 0.00\\

 & $\omega_1$ & 0.45 & 0.452 & 0.016 & 0.002 & 0.016 & 0.97 & 0.453 & 0.018 & 0.003 & 0.018 & 0.97\\

 & $\omega_2$ & 0.35 & 0.349 & 0.017 & -0.001 & 0.017 & 1.00 & 0.348 & 0.017 & -0.002 & 0.017 & 0.97\\

 & $\omega_3$ & 0.20 & 0.199 & 0.015 & -0.001 & 0.015 & 0.92 & 0.199 & 0.015 & -0.001 & 0.015 & 0.93\\

 & $\beta_{11}$ & -0.50 & -0.500 & 0.008 & 0.000 & 0.008 & 0.95 & -0.500 & 0.013 & 0.000 & 0.013 & 0.94\\

 & $\beta_{12}$ & 0.50 & 0.498 & 0.013 & -0.002 & 0.013 & 0.94 & 0.500 & 0.011 & 0.000 & 0.011 & 0.96\\

 & $\beta_{13}$ & -1.00 & -0.998 & 0.013 & 0.002 & 0.013 & 0.96 & -1.001 & 0.009 & -0.001 & 0.009 & 0.96\\

 & $\beta_{21}$ & -0.75 & -0.750 & 0.008 & 0.000 & 0.008 & 0.95 & -0.749 & 0.011 & 0.001 & 0.011 & 0.97\\

 & $\beta_{22}$ & 0.80 & 0.800 & 0.009 & 0.000 & 0.008 & 0.94 & 0.800 & 0.010 & 0.000 & 0.010 & 0.97\\

 & $\beta_{23}$ & 1.20 & 1.195 & 0.041 & -0.005 & 0.041 & 0.92 & 1.202 & 0.010 & 0.002 & 0.010 & 0.93\\

 & $\kappa_1$ & 0.80 & 0.726 & 0.292 & -0.074 & 0.299 & 0.88 & -- & -- & -- & -- & --\\

 & $\kappa_2$ & 0.60 & 0.550 & 0.213 & -0.050 & 0.218 & 0.92 & -- & -- & -- & -- & --\\

 & $\kappa_3$ & 0.25 & 0.225 & 0.112 & -0.025 & 0.114 & 0.90 & -- & -- & -- & -- & --\\

\multirow{-16}{*}{\centering\arraybackslash Medium} & $\gamma$ & 1.50 & 1.144 & 0.005 & -0.356 & 0.356 & 0.00 & -- & -- & -- & -- & --\\
\cmidrule{1-13}
 & $\lambda_1$ & -0.15 & -0.053 & 0.375 & 0.097 & 0.385 & 0.89 & 0.865 & 0.064 & 1.015 & 1.017 & 0.00\\

 & $\lambda_2$ & 0.15 & 0.136 & 0.365 & -0.014 & 0.364 & 0.88 & 0.846 & 0.076 & 0.696 & 0.700 & 0.00\\

 & $\lambda_3$ & 0.30 & 0.192 & 0.300 & -0.108 & 0.318 & 0.89 & 0.598 & 0.135 & 0.298 & 0.327 & 0.66\\

 & $\omega_1$ & 0.45 & 0.451 & 0.020 & 0.001 & 0.020 & 0.98 & 0.452 & 0.022 & 0.002 & 0.022 & 0.97\\

 & $\omega_2$ & 0.35 & 0.347 & 0.022 & -0.003 & 0.022 & 0.98 & 0.347 & 0.023 & -0.003 & 0.023 & 1.00\\

 & $\omega_3$ & 0.20 & 0.202 & 0.019 & 0.002 & 0.019 & 0.97 & 0.201 & 0.019 & 0.001 & 0.019 & 0.97\\

 & $\beta_{11}$ & -0.50 & -0.488 & 0.100 & 0.012 & 0.100 & 0.95 & -0.498 & 0.031 & 0.002 & 0.031 & 0.96\\

 & $\beta_{12}$ & 0.50 & 0.489 & 0.104 & -0.011 & 0.104 & 0.93 & 0.500 & 0.030 & 0.000 & 0.030 & 0.96\\

 & $\beta_{13}$ & -1.00 & -1.000 & 0.032 & 0.000 & 0.032 & 0.97 & -0.999 & 0.034 & 0.001 & 0.034 & 0.97\\

 & $\beta_{21}$ & -0.75 & -0.735 & 0.158 & 0.015 & 0.158 & 0.92 & -0.749 & 0.029 & 0.001 & 0.029 & 0.94\\

 & $\beta_{22}$ & 0.80 & 0.782 & 0.157 & -0.018 & 0.158 & 0.92 & 0.797 & 0.032 & -0.002 & 0.032 & 0.93\\

 & $\beta_{23}$ & 1.20 & 1.206 & 0.034 & 0.006 & 0.034 & 0.95 & 1.204 & 0.037 & 0.004 & 0.037 & 0.95\\

 & $\kappa_1$ & 0.80 & 0.691 & 0.340 & -0.109 & 0.356 & 0.88 & -- & -- & -- & -- & --\\

 & $\kappa_2$ & 0.60 & 0.565 & 0.261 & -0.035 & 0.262 & 0.88 & -- & -- & -- & -- & --\\

 & $\kappa_3$ & 0.25 & 0.269 & 0.187 & 0.019 & 0.187 & 0.89 & -- & -- & -- & -- & --\\

\multirow{-16}{*}{\centering\arraybackslash Low} & $\gamma$ & 1.50 & 1.144 & 0.005 & -0.356 & 0.356 & 0.00 & -- & -- & -- & -- & --\\
\bottomrule
\end{tabular}
% \begin{tablenotes}[para]
% \item \textit{Notes:} This table displays results based on $R=100$ replicates. The values include the average and standard deviation of the point estimates; the average bias (Bias), the Root Mean Squared Error (RMSE), and the coverage rate (Coverage) across replicates; where $\text{Bias}=R^{-1}\sum_{r=1}^R (\hat{\alpha}_r-\alpha),\text{ RMSE}=\sqrt{R^{-1}\sum_{r=1}^R (\hat{\alpha}_r-\alpha)^2},$ and $\text{ Coverage}=R^{-1}\sum_{r=1}^R \mathbbm{1}\{\alpha\in \widehat{CI}_{0.95,r}\}$.
% \end{tablenotes}
\end{threeparttable}}
}
\end{table}

\section{Empirical Application}\label{SCHSARsection5}

In this section, we apply the proposed methodology to study market-for-technology network formation and corporate research and development (R\&D) investments in the United States. Using a large-scale firm-level technology collaboration network constructed from patent transaction data, we jointly model link formation and R\&D efforts among firms, quantifying network (peer) effects and assessing how policy instruments shape innovation outcomes.

\subsection{Data}\label{data}

Our dataset spans from 1980 to 2014 and merges multiple sources, which includes firm accounting data from US Compustat, patent trades from the USPTO Patent Assignment Dataset (PAD), and R\&D tax credit information. Specifically, we utilize the Link Compustat -- USPTO Patent Assignment Dataset (PAD)\footnote{Data source: \url{https://zenodo.org/record/6352358}.}, shared publicly by Arqué-Castells and Spulber (\citeproc{ref-arque2022measuring}{2022}). These authors effectively matched assignor/assignee names in the PAD to Compustat GVKEYs, almost directly producing a match between Compustat and patent transactions that took place from \(1980\) to \(2014\). Information on companies corresponding to these GVKEYs and their annual balance sheets from the S\&P North America Annual Compustat is available through Wharton Research Data Services (WRDS). Federal tax information required to build the tax price of R\&D is also acquired from the same study and linked to firms by year\footnote{To calculate the federal component of the firm-specific tax price of R\&D, they make use of the dataset produced by Wilson (\citeproc{ref-wilson2009beggar}{2009}), who calculates the user cost of R\&D faced by a representative firm conducting R\&D within a given state.}. The resulting sample covers \(3,896\) Compustat firms that interact with at least one other firm in the sample and for which the deal has a known execution date. To create a firm-level technology-collaboration network of interest, we restrict attention to innovating firms that have both adopter and provider roles, invest in R\&D, and have complete information on relevant variables such as sales, capital, and employment. We create undirected links among firms, represented by the binary variable \(w_{ij}\) which indicates the presence or absence of at least one transaction in the technology market between firms \(i\) and \(j\). To maintain a manageable sample size, we exploit only the cross-sectional variation between firm dyads and treat the network as time-invariant network. This approach allows for a more stable analysis of relationships between firms during the study period. Firms that are not connected to the network are excluded to ensure that all entities in the final sample have meaningful technological collaborations. This yields a network of \(1,\!150\) firms with a total of \(5,\!576\) links.

\begin{figure}[H]

{\centering \includegraphics[width=0.5\linewidth]{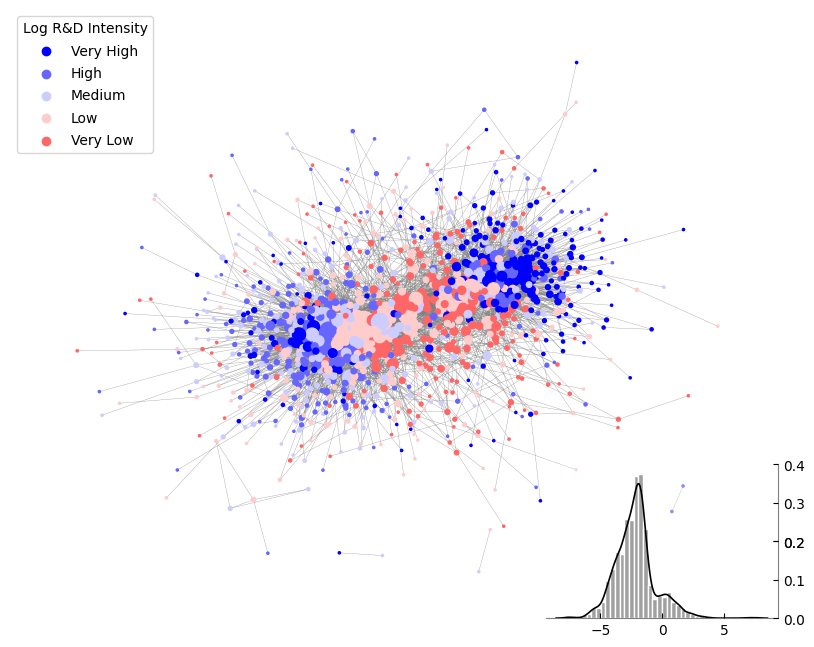} 

}

\caption[Distribution of R\&D intensity among firms]{Distribution of R\&D intensity among firms exhibits visible multimodality.}\label{fig:fig-loggedRDintensity}
\end{figure}

\begin{table}[H]
\centering\centering
\caption{\label{tab:tab-RDNet-summarystats}Data Statistics and Definitions ($N=1150$)}
\centering
\fontsize{9}{11}\selectfont
\begin{threeparttable}
\begin{tabular}[t]{lrrrrrl}
\toprule
Variable & Mean & Med & Std & Min & Max & Description\\
\midrule
RDintensity & 3.075 & 0.113 & 63.804 & 0.000 & 1939.069 & R\&D intensity\\
RDexpense & 252.891 & 31.579 & 764.907 & 0.095 & 7684.677 & Annual R\&D investment\\
Sales & 5381.282 & 304.322 & 18627.362 & 0.005 & 230887.253 & Sales\\
Capital & 321.358 & 13.515 & 1333.668 & 0.007 & 15503.994 & Capital expenditure\\
Employment & 14.606 & 1.244 & 41.260 & 0.004 & 526.483 & Employment\\
EBIT & 468.475 & 16.656 & 1663.241 & -380.792 & 21915.117 & Earnings before interest and taxes\\
Revenue & 4505.899 & 256.781 & 15820.807 & 0.004 & 197726.303 & Revenue\\
Size & 5616.906 & 359.009 & 21191.756 & 0.266 & 380628.074 & Total assets\\
TaxPrice & 0.910 & 0.910 & 0.040 & 0.820 & 1.019 & Federal component of R\&D tax price\\
\bottomrule
\end{tabular}
\begin{tablenotes}[para]
\item \textit{Notes:} This table presents summary statistics for the firm-level dataset, which includes $1150$ firms in total with no missing data. The values of RDexpense, Sales, Capital, Revenue, and Size are in millions of 2010 dollars.
\end{tablenotes}
\end{threeparttable}
\end{table}

The primary firm outcome of interest in our application is the corporate R\&D efforts, measured by the R\&D expenditure-to-sales ratio, often referred to as R\&D intensity. The natural logarithm of R\&D intensity among firms, as depicted in Figure \ref{fig:fig-loggedRDintensity}, reveals a notable multimodal distribution. Table \ref{tab:tab-RDNet-summarystats} presents the definitions and descriptive statistics for essential variables in our final sample, including various firm-level characteristics and the federal component of the firm-specific tax price of R\&D. The latter represents a supply-side shock to corporate R\&D induced by government policy.

\subsection{Model Specification}\label{model-specification}

We adopt the general SCHSAR framework to jointly model the formation of a firm collaboration network and their R\&D efforts in two sequential stages as follows. In the first stage, the network among \(N\) firms is formed according to
\begin{equation}
\begin{split}
w^{*}_{ij} &= \gamma_{0} + \gamma_{1}\cdot \text{sameSIC}_{ij} + \gamma_{2}\cdot \text{sameAAclass}_{ij} + a_i + a_j + \epsilon_{ij}, \quad \epsilon_{ij} \overset{iid}{\sim} \mathcal{N}(0,1);\\
w_{ij} = w_{ji} &= \mathbbm{1}\{w^{*}_{ij} \geq 0 \}; \quad \text{for } i = 1,\ldots,N; \quad j = 1,\ldots,i-1;
\end{split}
\end{equation}
where \(w_{ij}\) indicates whether firms \(i\) and \(j\) have at least one technology-market transaction, and \(w^{*}_{ij}\) represent the latent utility that each of the firms \(i\) and \(j\) gains from forming such a link. In particular, this utility depends on predetermined dyad-specific regressors that include homophily measures such as whether the two firms are in the same industry (\(\text{sameSIC}_{ij}\)), whether they are in the same technology class (\(\text{sameAAclass}_{ij}\)), etc. The intercept \(\gamma_{0}\) captures the fixed cost of maintaining links, while \(a_i\) and \(a_j\) represent firm-specific unobserved degree heterogeneity, concurrent with their ability to create linkages.

Once the network is formed in the first stage, in the second stage, firms choose their actions (e.g., R\&D efforts or other firm outcomes) while accounting for the network structure and the behavior of connected peers. Let \(Y_i\) denote firm \(i\)'s logged R\&D intensity, measured by the natural logarithm of R\&D expenditure-to-Sales ratio, and let \(X_i^k\) be the \(k\)-th observed firm-specific characteristic. The degree heterogeneity \(a_i\) may also affect \(Y_i\) through the unobserved part, making the \(N \times N\) network adjacency matrix \(\mathbf{W}=[w_{ij}]\) potentially endogenous. In accordance with the SCHSAR framework, the outcome equation for all firms can be expressed as in \eqref{SCHSAR}, with the corresponding reduced form given by
\begin{equation}
\mathbf{Y} = \left[\mathbf{I}_N - \mathcal{L}\mathbf{W}\right]^{-1}\left(\sum_{k=1}^K\tilde{\beta}^k \circ X^k + \tilde{\kappa} \circ \mathbf{a} + \mathbf{u}\right).
\end{equation}
This specification not only accommodates the heterogeneous network effects, but also leads to a much richer interpretation of how explanatory variables affect the outcome. Specifically, the marginal effects of a change in the \(k\)-th variable vector, \(X^{k}=\left[X_1^k,\ldots,X_N^k\right]^\top\), are given by the following matrix of partial derivatives
\begin{equation}
\begin{split}
\frac{\partial Y}{\partial X^{k\top}} &= \begin{bmatrix}
\frac{\partial Y_{1}}{\partial X_{1}^k} &\frac{\partial Y_{1}}{\partial X_{2}^k} &\cdots &\frac{\partial Y_{1}}{\partial X_{N}^k}
\\
\frac{\partial Y_{2}}{\partial X_{1}^k} &\frac{\partial Y_{2}}{\partial X_{2}^k} &\cdots &\frac{\partial Y_{2}}{\partial X_{N}^k}
\\
\vdots &\vdots &\ddots&\vdots
\\
\frac{\partial Y_{N}}{\partial X_{1}^k} &\frac{\partial Y_{N}}{\partial X_{2}^k} &\cdots &\frac{\partial Y_{N}}{\partial X_{N}^k}
\end{bmatrix} 
= \left(\mathbf{I}_N - \mathcal{L}\mathbf{W} \right)^{-1}\times \text{diag}\bigl(\tilde{\beta}^k\bigr)\\
&=\begin{pmatrix}
\mathbf{I}_N  -
\begin{bmatrix}
\tilde{\lambda}_1 &0 &\cdots &0
\\
0 & \tilde{\lambda}_2  &\cdots &0
\\
\vdots &\vdots &\ddots&\vdots
\\
0 &0 &\cdots & \tilde{\lambda}_N
\end{bmatrix} \mathbf{W}
\end{pmatrix}^{-1} \times
\begin{bmatrix}
\tilde{\beta}^k_1 &0 &\cdots &0
\\
0 &\tilde{\beta}^k_2 &\cdots &0
\\
\vdots &\vdots &\ddots&\vdots
\\
0 &0 &\cdots &\tilde{\beta}^k_N
\end{bmatrix} = [\ell_{ij}]_{i,j=1}^{N}.
\end{split}
\end{equation}

Direct effects (diagonal elements) show how changes in a firm's own \(k\)-th characteristic influence its own outcomes. Conversely, indirect spillin effects (row off-diagonal sums) represent the cumulative impact of changes in the \(k\)-th characteristic of peer firms on a firm's outcomes, while indirect spillout effects (column off-diagonal sums) demonstrate how changes in a firm's \(k\)-th characteristic affect peers' outcomes. We note that \(G=1\) produces homogeneous models, allowing for standard scalar summaries of the effects at the aggregate level -- direct effect (\(DE\)), indirect effect (\(IE\)), and total effect (\(TE = DE + IE\)) -- as described in LeSage and Pace (\citeproc{ref-lesage2009introduction}{2009}). When \(G>1\), the models become heterogeneous due to the heterogeneity in \(\tilde{\lambda}\) and \(\tilde{\beta}^{k}\). Consequently, the effects of interest are reported at the observation level for each firm in the sample. In this case, the \(N\)-dimensional vectors of firm-level spillin and spillout effects generally differ, even when a doubly-stochastic weight matrix is used, making it possible to distinguish between the two types of indirect effects. These quantities depend not only on firm type but also on the firm's position within the network. From a spatial perspective, a detailed discussion on the interpretation of such heterogeneous models can be found in LeSage and Chih (\citeproc{ref-lesage2016interpreting}{2016}), which pertains to our proposed framework.

\subsection{Estimation Results}\label{estimation-results}

\begin{table}[H]
\centering\centering
\caption{\label{tab:tab-RDNet-mix2-para}Parameter Estimates for the HSAR and SCHSAR Models}
\centering
\fontsize{7.5}{9.5}\selectfont
\begin{threeparttable}
\begin{tabular}[t]{lccccclc}
\toprule
\multicolumn{2}{c}{ } & \multicolumn{1}{c}{SAR} & \multicolumn{1}{c}{SCSAR} & \multicolumn{2}{c}{HSAR (G=2)} & \multicolumn{2}{c}{SCHSAR (G=2)} \\
\cmidrule(l{3pt}r{3pt}){3-3} \cmidrule(l{3pt}r{3pt}){4-4} \cmidrule(l{3pt}r{3pt}){5-6} \cmidrule(l{3pt}r{3pt}){7-8}
 &  & G=1 & G=1 & $1^{\text{st}}$ Type & $2^{\text{nd}}$ Type & $1^{\text{st}}$ Type & $2^{\text{nd}}$ Type\\
\midrule
\addlinespace[0.3em]
\multicolumn{8}{l}{\textbf{Network Interaction}}\\
\hspace{1em}Assignment Probability & $\pi$ & \makecell[c]{1.000\\$[$1.00, 1.00$]$} & \makecell[c]{1.000\\$[$1.00, 1.00$]$} & \makecell[c]{0.588\\$[$0.50, 0.67$]$} & \makecell[c]{0.412\\$[$0.33, 0.50$]$} & \makecell[c]{0.660\\$[$0.58, 0.74$]$} & \makecell[c]{0.340\\$[$0.26, 0.42$]$}\\
\hspace{1em}Network Effect & $\lambda$ & \makecell[c]{0.190\\$[$0.16, 0.22$]$} & \makecell[c]{0.157\\$[$0.12, 0.19$]$} & \makecell[c]{0.150\\$[$0.10, 0.20$]$} & \makecell[c]{0.226\\$[$0.16, 0.29$]$} & \makecell[c]{0.127\\$[$0.08, 0.17$]$} & \makecell[c]{0.215\\$[$0.14, 0.29$]$}\\
\hspace{1em}Intercept & $\beta_1$ & \makecell[c]{1.236\\$[$0.86, 1.62$]$} & \makecell[c]{2.664\\$[$2.23, 3.10$]$} & \makecell[c]{1.315\\$[$0.77, 1.86$]$} & \makecell[c]{-3.171\\$[$-4.08, -2.23$]$} & \makecell[c]{3.153\\$[$2.54, 3.79$]$} & \makecell[c]{-3.077\\$[$-4.23, -1.87$]$}\\
\hspace{1em}logTaxPrice & $\beta_2$ & \makecell[c]{-8.332\\$[$-9.53, -7.15$]$} & \makecell[c]{-8.063\\$[$-9.21, -6.92$]$} & \makecell[c]{-10.637\\$[$-12.66, -8.65$]$} & \makecell[c]{-2.428\\$[$-4.08, -0.77$]$} & \makecell[c]{-9.525\\$[$-11.28, -7.76$]$} & \makecell[c]{-2.223\\$[$-4.23, -0.29$]$}\\
\hspace{1em}logCapitalExpense & $\beta_3$ & \makecell[c]{0.616\\$[$0.54, 0.69$]$} & \makecell[c]{0.540\\$[$0.47, 0.61$]$} & \makecell[c]{0.685\\$[$0.57, 0.80$]$} & \makecell[c]{0.292\\$[$0.16, 0.43$]$} & \makecell[c]{0.589\\$[$0.48, 0.69$]$} & \makecell[c]{0.263\\$[$0.12, 0.42$]$}\\
\hspace{1em}EBIT & $\beta_4$ & \makecell[c]{0.071\\$[$0.04, 0.11$]$} & \makecell[c]{0.042\\$[$0.01, 0.08$]$} & \makecell[c]{0.059\\$[$0.01, 0.11$]$} & \makecell[c]{0.052\\$[$-0.00, 0.11$]$} & \makecell[c]{0.045\\$[$0.00, 0.09$]$} & \makecell[c]{0.058\\$[$-0.01, 0.13$]$}\\
\hspace{1em}logEmployment & $\beta_5$ & \makecell[c]{-0.025\\$[$-0.12, 0.07$]$} & \makecell[c]{-0.073\\$[$-0.16, 0.02$]$} & \makecell[c]{-0.102\\$[$-0.27, 0.05$]$} & \makecell[c]{-0.617\\$[$-0.79, -0.45$]$} & \makecell[c]{-0.086\\$[$-0.22, 0.05$]$} & \makecell[c]{-0.639\\$[$-0.82, -0.45$]$}\\
\hspace{1em}logRevenue & $\beta_6$ & \makecell[c]{-0.983\\$[$-1.06, -0.91$]$} & \makecell[c]{-0.962\\$[$-1.04, -0.89$]$} & \makecell[c]{-1.099\\$[$-1.20, -1.00$]$} & \makecell[c]{0.109\\$[$-0.08, 0.29$]$} & \makecell[c]{-1.080\\$[$-1.17, -0.99$]$} & \makecell[c]{0.135\\$[$-0.07, 0.33$]$}\\
\hspace{1em}Correlation & $\kappa$ & -- & \makecell[c]{0.802\\$[$0.67, 0.93$]$} & -- & -- & \makecell[c]{0.933\\$[$0.72, 1.15$]$} & \makecell[c]{0.083\\$[$-0.17, 0.34$]$}\\
\hspace{1em}Unobserved Heterogeneity & $\sigma^2_a$ & -- & \makecell[c]{3.092\\$[$2.84, 3.36$]$} & -- & -- &  & \hspace{-6em}\makecell[c]{3.093\\$[$2.85, 3.37$]$}\\
\addlinespace[0.3em]
\multicolumn{8}{l}{\textbf{Network Formation}}\\
\hspace{1em}SIC homophily & $\gamma_1$ & -- & \makecell[c]{0.722\\$[$0.69, 0.76$]$} & -- & -- &  & \hspace{-6em}\makecell[c]{0.724\\$[$0.69, 0.76$]$}\\
\hspace{1em}Tech homophily & $\gamma_2$ & -- & \makecell[c]{0.691\\$[$0.67, 0.71$]$} & -- & -- &  & \hspace{-6em}\makecell[c]{0.692\\$[$0.67, 0.71$]$}\\
\addlinespace[0.3em]
\multicolumn{8}{l}{\textbf{Criteria}}\\
\hspace{1em}Log likelihood &  & \makecell[c]{-1484.93\\} & \makecell[c]{-45471.58\\} &  & \hspace{-6em}\makecell[c]{-1958.68\\} &  & \hspace{-6em}\makecell[c]{-44949.63\\}\\
\hspace{1em}AICM &  & \makecell[c]{2978.05\\} & \makecell[c]{122146268.90\\} &  & \hspace{-6em}\makecell[c]{4794.66\\} &  & \hspace{-6em}\makecell[c]{189418920.84\\}\\
\hspace{1em}Observations &  & 1150 & 1150 &  & \hspace{-6em}1150 &  & \hspace{-6em}1150\\
\bottomrule
\end{tabular}
\begin{tablenotes}[para]
\item \textit{Notes:} This table presents the estimation results for the HSAR and SCHSAR models with $G=1$ and $G=2$. MCMC sampling was run for $50,491$ iterations, discarding the first 500 as burn-in and retaining every 10th draw thereafter, yielding $5,000$ effective draws. Posterior means and $95\%$ equal-tailed credible intervals are computed using these MCMC draws.
\end{tablenotes}
\end{threeparttable}
\end{table}

We estimate the model parameters using the MCMC procedure described in Section \ref{SCHSARsection3}. The algorithm is run for \(50,\!491\) iterations, with the first \(500\) draws discarded as burn-in and every \(10\)th of the remaining draws retained to conduct the posterior analysis. Table \ref{tab:tab-RDNet-mix2-para} presents estimation results for the HSAR and SCHSAR models under two specifications of the number of mixture components in the outcome equation (\(G=1\) and \(G=2\)).\footnote{Under a three-component model (i.e., \(G=3\)), the parameter estimates for the third type, including the assignment probability and network effect parameter, are not significantly distinguishable from zero.}

When \(G=1\), a homogeneous network effect is imposed. The first two columns of Table \ref{tab:tab-RDNet-mix2-para} present results for the homogeneous SAR and SCSAR models. Parameter estimates for the standard SAR specification suggests that a majority of firm-specific characteristics are statistically significant determinants of R\&D intensity. Furthermore, the estimated network effect (\(\lambda\)) is positive and significant, confirming the theoretical conjecture that corporate R\&D efforts are positively influenced by their collaborators. Concerning the endogenous adjacent matrix inherent in this benchmark model, the SCSAR model takes into account the network formation process among the firms. Results show strong evidence of homophily in link formation: firms are significantly more likely to collaborate with others in the same industry (SIC homophily) or technology class (tech homophily). Moreover, unobserved degree heterogeneity underpins a selection-on-gains pattern: firms with a higher latent propensity to form collaborations tend to invest more in R\&D. Controlling for firms' unobserved heterogeneity in both link formation and R\&D intensity reduces the estimated network effect from \(0.190\) to \(0.156\), though the effect remains positive and statistically significant. This highlights the upward bias in \(\lambda\) when ignoring selection due to endogenous network formation.

The models with \(G=2\) allow for heterogeneous network effects across two latent firm types. The SCHSAR and HSAR results are displayed in the last four columns of Table \ref{tab:tab-RDNet-mix2-para}. For the SCHSAR model, estimates of the network formation parameters (\(\gamma\)) and the random effect variance (\(\sigma_a^2\)) are virtually unchanged relative to the homogeneous SCSAR. This aligns with our modeling approach when the first-stage network formation is specified at the population level rather than at the type level. In terms of the second-stage outcome equation, the correlation coefficient \(\kappa\) remains positive in both types and is statistically significant for the first type. Selection correction again reduces the network effects overestimated by the HSAR model. The two groups are ordered based on their estimated population shares, with the smaller group recording a markedly larger network effect (\(\hat{\lambda}_2 \approx 1.7 \hat{\lambda}_1\)) but a substantially lower elasticity with respect to the R\&D tax price (\(\hat\beta_{22} \approx 4.28 \hat\beta_{21}\)). This interesting distinction indicates that some firms are highly responsive to their peer R\&D activities but less susceptible to direct R\&D tax price intervention from the government, and vice versa. Perhaps the latter type includes more ``self-reliant'' innovators (and network peer effects are less important to these firms), while firms of the former type are more influenced by what others are doing (and hence less directly responsive to their own tax changes). For simplicity, we refer to these groups as high-\(\lambda\) (more peer-driven) and low-\(\lambda\) (more self-driven) types. Although the latent group labels themselves have no intrinsic interpretation\footnote{See Geweke (\citeproc{ref-geweke2007interpretation}{2007}) and Frühwirth-Schnatter (\citeproc{ref-fruhwirth2006finite}{2006}) for full discussion.}, some attributes of each group provide further insight into the underlying mechanisms driving the heterogeneity.

In Table \ref{tab:rtab-probreg}, we present regression results of the probability of belonging to the high-\(\lambda\) type on multiple firm characteristics. For the linear specification, we use a continuous dependent variable obtained from the posterior means of \(z_i\) in \eqref{postz}. For the logit specification, we use the binary type indicators derived from posterior median of \(z_i\). The findings consistently suggest that larger firms (by total assets) are more likely to be in the high-\(\lambda\) type. Technology class also plays a role in determining the latent types.

\begin{table}[h]
\centering
\footnotesize
\setlength{\tabcolsep}{5pt}
\renewcommand{\arraystretch}{1.1}
\begin{threeparttable}
\caption{Regressions of Posterior Probability of Inclusion on Firm Characteristics}
\label{tab:rtab-probreg}
\begin{tabular}{lllll}
\toprule
& \multicolumn{2}{c}{Probability (Linear)} & \multicolumn{2}{c}{Binary (Logistic)} \\
\cmidrule(lr){2-3}\cmidrule(lr){4-5}
& \multicolumn{1}{c}{(1)} & \multicolumn{1}{c}{(2)} & \multicolumn{1}{c}{(3)} & \multicolumn{1}{c}{(4)} \\
\midrule
logTotalAsset
  & 0.016 (0.003)\textsuperscript{***} & 0.013 (0.003)\textsuperscript{***} & 0.150 (0.033)\textsuperscript{***} & 0.129 (0.040)\textsuperscript{***} \\
Computers \& Communications
  & 0.210 (0.021)\textsuperscript{***} & 0.205 (0.021)\textsuperscript{***} & 1.109 (0.273)\textsuperscript{***} & 1.077 (0.275)\textsuperscript{***} \\
Drugs \& Medical
  & 0.065 (0.022)\textsuperscript{***} & 0.059 (0.022)\textsuperscript{***} & 0.682 (0.292)\textsuperscript{**} & 0.643 (0.295)\textsuperscript{**} \\
Electrical \& Electronic
  & 0.191 (0.025)\textsuperscript{***} & 0.189 (0.025)\textsuperscript{***} & 1.177 (0.303)\textsuperscript{***} & 1.156 (0.304)\textsuperscript{***} \\
Mechanical
  & 0.135 (0.029)\textsuperscript{***} & 0.140 (0.029)\textsuperscript{***} & 0.862 (0.340)\textsuperscript{**} & 0.887 (0.341)\textsuperscript{***} \\
Others
  & 0.035 (0.030) & 0.039 (0.030) & 0.506 (0.372) & 0.527 (0.373) \\
Degree Centrality
  &      & 0.981 (0.501)\textsuperscript{*} &      & 4.751 (5.061) \\
Intercept
  & 0.115 (0.027)\textsuperscript{***} & 0.132 (0.028)\textsuperscript{***} & -2.955 (0.350)\textsuperscript{***} & -2.843 (0.367)\textsuperscript{***} \\
\midrule
Observations & \multicolumn{1}{c}{1150} & \multicolumn{1}{c}{1150} & \multicolumn{1}{c}{1150} & \multicolumn{1}{c}{1150} \\
Adjusted $R^{2}$ & \multicolumn{1}{c}{0.138} & \multicolumn{1}{c}{0.140} & & \\
AIC & & & \multicolumn{1}{c}{1228.42} & \multicolumn{1}{c}{1229.54} \\
\bottomrule
\end{tabular}
\begin{tablenotes}[para,flushleft]
\scriptsize
\item \textit{Notes:} Standard errors are reported in parentheses; coefficients marked with \textsuperscript{*}, \textsuperscript{**}, and \textsuperscript{***} are significant at the 10\%, 5\%, and 1\% levels, respectively.
\end{tablenotes}
\end{threeparttable}
\end{table}

\subsection{Policy Analysis}\label{policy-analysis}

We now examine how corporate R\&D efforts in the firm-level collaboration network respond to government tax incentives for R\&D. As discussed earlier, in the case of heterogeneous model, our focus of inference are the direct and indirect (spillin and spillout) effects at individual firm level. For each \(i=1,\ldots,N\), the direct effect shows the elasticity of \(i\)-th firm's R\&D intensity with respect to its own R\&D tax price. The indirect spillin effect measures the cumulative impact of changes in peer firms' R\&D tax prices on \(i\)-th firm's R\&D intensity, whereas the indirect spillout effect captures the cumulative impact of changes in R\&D tax prices of \(i\)-th firm on the R\&D intensity of its peers. These quantities reveal fine-grained heterogeneity across firms in policy responsiveness, thus enabling a more effective intervention design.

To illustrate, we simulate a policy in which the government seeks to encourage corporate R\&D investments by lowering the costs incurred by firms. We assess the effects of a \(1\%\) reduction in firm-specific R\&D tax prices using the SCHSAR estimates. We summarize the observational-level effects of interest succinctly using histograms and colored network graphs: histograms depict the empirical distribution of posterior mean estimates across \(1,\!150\) firms, while network graphs color firm nodes by effect magnitude, grouped into five quintile bins. Overall, all firms display values significantly different from zero, as indicated by \(95\%\) credible intervals derived from the MCMC draws.

\begin{figure}[H]

{\centering \includegraphics[width=1\linewidth]{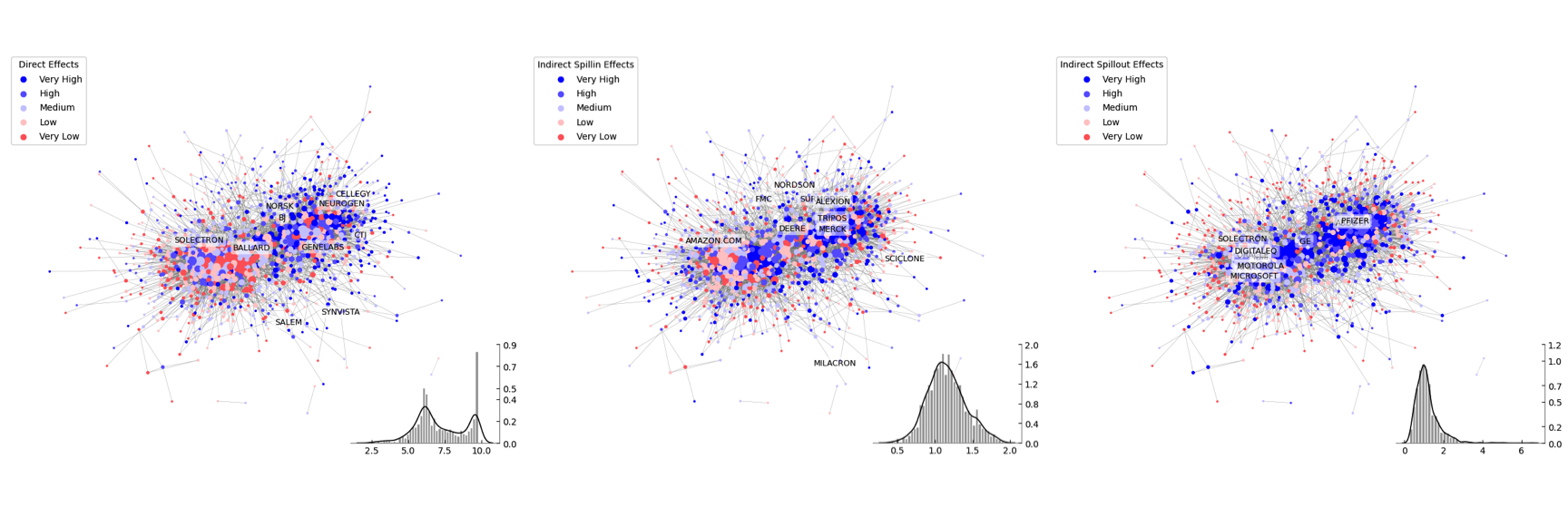} 

}

\caption{(a) Direct effects of a $1\%$ reduction in a firm's own R$\&$D tax price. (b) Indirect spillin effects to each firm from a $1\%$ reduction in peers' R$\&$D tax price. (c) Indirect spillout effects from each firm to its peers due to a $1\%$ reduction in the firm's own R$\&$D tax price. The histogram shows the distribution of the effects of interest across firms. In the network graph, firms are represented as nodes colored by the effect magnitude.}\label{fig:fig-hete-3effects}
\end{figure}

Direct effect estimates in Figure \ref{fig:fig-hete-3effects}(a) exhibit a clear bimodal distribution, consistent with the predictions of the two-component mixture SCHSAR model. Most values are significantly positive, ranging from about \(2\) to \(10\), implying substantial variation in elasticity to own tax changes. Firms with the largest direct effects are typically low-\(\lambda\) type, self-reliant innovators that respond strongly to their own R\&D incentives (large \(\beta_2\)) and rely less on network peers. Firms in the Drugs \& Medical and Chemical classes are heavily represented in this group. Further analysis shows that direct effects are negatively correlated with firm size and network centrality, suggesting that smaller, less central firms tend to be more directly responsive to tax incentives.

Indirect spillin effects across firms due to a \(1\%\) reduction in peers' R\&D tax prices are presented in Figure \ref{fig:fig-hete-3effects}(b). The distribution centers around \(1.0\)--\(1.3\), smaller in magnitude than the direct effects but still positive, indicating that firms gain when collaborators face lower R\&D costs. A firm's network position and network effect elasticity are pivotal in determining its responsiveness to peer-based incentives. Firms with high spillin effects predominantly belong to high-\(\lambda\) type. From a theoretical standpoint, we refer to these firms as ``responsive absorbers'', who rely more on external innovation like partnerships, and consequently exhibit greater peer dependency and weaker direct responses. The top-ranked firms span a wide range of technology classes and are generally larger in size.

Indirect spillout effects from each firm to its peers resulting from a \(1\%\) reduction in the firm's own R\&D tax price are presented in Figure \ref{fig:fig-hete-3effects}(c). While top spillin firms are typically high-\(\lambda\) type, spillout leaders are mostly low-\(\lambda\) type with high own-price elasticity. These firms tend to occupy central positions in the network, acting as ``influential transmitters'' that amplify strategic complementarities (\citeproc{ref-bulow1985multimarket}{Bulow et al. 1985}; \citeproc{ref-cooper1988coordinating}{Cooper and John 1988}) and propagate shocks through collaboration links. Top spillout performers are large (most rank in the \(90^{\text{th}}\)--\(100^{\text{th}}\) percentile of total assets) and are markedly concentrated in the Computers \& Communications class, which accounts for 13 of the top 20 firms. This pattern suggests technology-intensive sectors drive network-wide R\&D diffusion.

Taken together, the empirical results demonstrate that the SCHSAR framework flexibily captures the intended network mechanisms. While direct incentives have the largest magnitudes, indirect effects are economically meaningful and policy-relevant. Ignoring these indirect channels risks underestimating the total policy impact. The analysis also reveals that top spillin and spillout performers (responsive absorbers versus influential transmitters) differ systematically, highlighting their distinct roles in the network's innovation ecosystem.

\begin{figure}[H]

{\centering \includegraphics[width=0.67\linewidth]{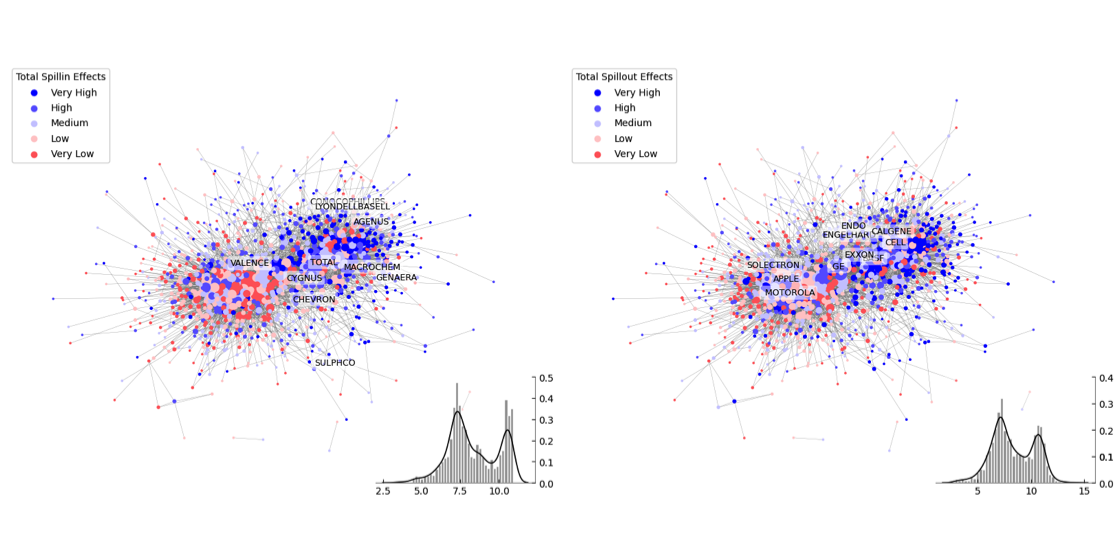} 

}

\caption{(a) Total spillin effects on each firm due to a $1\%$ reduction in the R$\&$D tax price for all firms. (b) Total spillout effects from each firm due to a $1\%$ reduction in the firm's own R$\&$D tax price. The histogram shows the distribution of the effects of interest across firms. In the network graph, firms are represented as nodes colored by the effect magnitude.}\label{fig:fig-heteTSITSO}
\end{figure}

Measuring firm-level total effects, which combine direct and indirect channels, offers the most comprehensive basis for designing policies that leverage both direct responsiveness and network effects. The \textit{total spillin effect} in Figure \ref{fig:fig-heteTSITSO}(a) measures the cumulative impact on a given firm's R\&D intensity of a 1\% tax cut applied to every firm in the network. This measure is particularly relevant under a uniform tax policy in which all firms receive the same marginal incentive from the government. In our context, direct effects outweigh indirect spillin effects, so top beneficiaries are typically self-reliant innovators---firms with strong own-price responsiveness---predominantly found in Drugs \& Medical and Chemical sectors. The \textit{total spillout effect} in Figure \ref{fig:fig-heteTSITSO}(b), on the contrary, captures the cumulative increase in network-wide R\&D intensity resulting from a \(1\%\) reduction in a given firm's tax price. This measure is the most valuable for designing targeted policies under budget constraints, as high-spillout firms act as policy multipliers, diffusing innovation widely at relatively low cost. These top firms tend to be innovation initiators and influencers, combining strong direct effects with extensive network reach. They are typically highly central in the network and concentrated in the Computers \& Communications, Chemical, or Drugs \& Medical sectors. While many are large industry leaders (e.g., GE, Apple, P\&G), some well-connected but much smaller firms also rank highly; conditional on sector and centrality, smaller firms tend to generate higher total spillout effects. Policymakers can maximize systemic diffusion from limited resources by prioritizing firms identified through this metric.

\section{Conclusion}\label{SCHSARsection6}

This paper introduces the Selection-corrected Heterogeneous Spatial Autoregressive (SCHSAR) framework to address two critical challenges in social and economic interaction models: heterogeneous peer effects and endogenous network formation. The proposed approach extends the standard SAR model by incorporating a finite mixture structure to capture rich heterogeneity in peer effects, while simultaneously modeling the network formation process to correct for potential selection bias. Three types of latent variables are integrated into the framework: individual-type indicators associated with mixture components in the outcome equation, dyadic utilities governing link formation in the network equation, and individual-specific random effects accounting for endogeneity. This hierarchical structure, while posing a challenge to the maximum likelihood estimators, is efficiently handled by our Bayesian algorithm based on data augmentation techniques.

The simulation study validates our fully Bayesian approach. It is computationally tractable and able to deliver reliable inference for the SCHSAR model across various data-generating processes. The proposed estimators exhibit near-unbiasedness and nominal coverage, especially when the signal level is favorable and endogenous network formation is mainly driven by unobserved degree heterogeneity. In contrast, reliance on either exogeneity or homogeneity results in significant bias and unreliable inference for peer effect coefficients when these assumptions are violated.

 Our empirical application to a technological collaboration network among U.S. firms provides several key insights. First, technological homophily significantly shapes the network structure, complemented by firm-specific latent advantages that affect both linkage ability and R\&D intensity. Second, positive network (peer) effects on corporate R\&D investments persist after correcting for selection bias due to endogenous network formation. Our heterogeneous model reveals two distinct types of firms, with varying levels of peer effects and tax price responsiveness. Third, we find substantial firm-level heterogeneity in both the direct and indirect (spillin and spillout) effects of R\&D tax incentives. Different firm behaviors are revealed: ``self-reliant innovators'' with strong direct effects, ``responsive absorbers'' with high spillin effects, and ``influential transmitters'' with significant spillout effects. This highlights the role of network position and firm attributes, such as firm size, in driving innovation dynamics. Finally, evaluating firm-level total effects facilitates the design of targeted policy interventions. For example, prioritizing firms with high total spillout effects could be a cost-effective strategy to maximize systemic innovation diffusion across the network.

The SCHSAR framework provides a flexible and robust approach to drawing credible conclusions about peer effects by effectively accommodating both heterogeneity and endogeneity. Future research directions include extending the framework to dynamic panel settings, automatically selecting the number of mixture components, and improving unobserved homophily handling. Although we focus on the analysis of firm network data, a wide range of potential applications can be explored using this method.

\setstretch{1}

\section*{References}\label{references}
\addcontentsline{toc}{section}{References}

\phantomsection\label{refs}

\newpage

\setstretch{1.15}

\appendix

\section*{APPENDIX}\label{appendix}
\addcontentsline{toc}{section}{APPENDIX}

\section{Details of Computational Algorithms}\label{details-of-computational-algorithms}

\subsection{Adaptation of Random-Walk Metropolis}\label{SCHSAR-APDX_AdaptiveMCMC}

Adaptation of random-walk Metropolis algorithm (\citeproc{ref-andrieu2008tutorial}{Andrieu and Thoms 2008}; \citeproc{ref-atchade2005adaptive}{Atchadé and Rosenthal 2005}; \citeproc{ref-roberts2009examples}{Roberts and Rosenthal 2009}; \citeproc{ref-vihola2022}{Vihola 2022}) is designed to improve the efficiency of standard random-walk Metropolis algorithm, by automatically tuning the proposal distribution during the simulation based on historical MCMC samples. The goal is to achieve better convergence and mixing without requiring manual calibration.

Formally, let \(\theta \in  \Theta \subset \mathbbm{R}^d\) be the parameter of interest with target probability density \(\pi(\theta)\). Our goal is to generate a dependent random sequence \(\{\theta^{[0]},\ldots,\theta^{[T]}\}\), called a chain, whose stationary density matches \(\pi\). At iteration \(s\), the standard algorithm uses a normal random walk proposal \(\mathcal{N}\left(\theta^{[t]},\tau^{[t]}\Sigma_{\theta}^{[t]}\right)\), where \(\theta^{[t]}\) is the current state, \(\tau^{[t]}\) is a scalar scaling factor (i.e., how far the proposal jumps in the parameter space), and \(\Sigma_{\theta}^{[t]}\) is an empirical estimate of the covariance matrix of the target \(\pi\). The acceptance probability of a candidate draw \(\theta^{\star}\) is \(\alpha = \min\bigl\{1,\pi\bigl(\theta^{\star}\bigr)/\pi\bigl(\theta^{[t]}\bigr)\bigr\}\) . The shape of the proposal distribution has a substantial effect on the algorithm's mixing behavior\footnote{Intuitively, if \(\tau^{[t]}\Sigma_{\theta}^{[t]}\) is either too large in some directions or too small in all directions the algorithm has either a very small or a very large acceptance probability, which results in a very poor exploration of the target distribution as the algorithm mix poorly.}, while choosing \(\tau\) and \(\Sigma_{\theta}\) in each iteration by trial and error is both time-consuming and problem-specific. To overcome this tuning challenge, adaptation allows the sampler learn \(\tau\) and \(\Sigma_{\theta}\) on-the-fly using past draws. The core idea is to adaptively update these parameters to target a prespecified ``optimal'' acceptance rate (\(\bar{\alpha}_{opt}\)).

Among several variants, we present in Algorithm \ref{alg:ASMgeneral} the Adaptive Metropolis algorithm with global adaptive scaling (\citeproc{ref-atchade2005adaptive}{Atchadé and Rosenthal 2005}). We note that, for the adaptation step, equation \eqref{updatescaling} aims to keep the asymptotic acceptance rate of the algorithm close to an optimal value \(\bar{\alpha}_{opt}\). Commonly, \(\bar{\alpha}_{opt} = 0.234\) for a multivariate target (\(d>1\)) and \(\bar{\alpha}_{opt} = 0.44\) when \(d=1\). The (recommended) adaptation stepsize is a decreasing sequence \(\nu^{[t+1]} = (t+1)^{-2/3}\). This approach is backed by Robbins-Monro recursive and more generally, the stochastic approximation framework (\citeproc{ref-benveniste2012adaptive}{Benveniste et al. 2012}), and aims to optimize the efficiency of the MCMC sampler by monitoring the acceptance rate toward a theoretically optimal region. The updates are designed to vanish over time (i.e., \(\nu^{[t]} \rightarrow 0\)) to ensure the ergodicity and convergence of the chain.

For our sampling problem for network interaction effects \(\lambda_g\) (\(g=1,\ldots,G\)) in Section \ref{SCHSARsection3}, we apply a univariate version of Algorithm \ref{alg:ASMgeneral}. Specifically, we employ Algorithm \ref{alg:ASMunivariate} which corresponds to \(d=1\) and the optimal acceptance rate \(\alpha_{opt} = 0.44\). In practice, whenever \(\alpha\left(\theta^{\star},\theta^{[t]}\right)\) falls below \(\bar{\alpha}_{opt}\) for most transition attempts, the log-scale update helps increase the scaling parameter \(\tau^{[t]}\), and vice versa. This automatic feedback allows the sampler to recover quickly from poor initialization, enabling an efficient exploration of the posteriors of interest. We embed this Adaptive Scaling Metropolis sampling step for each \(\lambda_g\) into our main MCMC samplers for the SCHSAR model (Algorithms \ref{alg:SCHSAR1} and \ref{alg:SCHSAR2}).

\begin{center}
\scalebox{0.95}{
\small
\begin{minipage}{1\linewidth}
\begin{algorithm}[H]
\caption{Adaptive Scaling Metropolis (ASM) algorithm - \textit{General case}}
\label{alg:ASMgeneral}

\vspace{3mm}

\textbf{Procedure}

\refstepcounter{algoline}\nlset{\thealgoline}\quad \textbf{Step 0:} Initialize $t=0, \theta^{[0]}, \mu_{\theta}^{[0]} = \theta^{[0]}, \Sigma_{\theta}^{[0]}, \tau^{[0]}$.

\refstepcounter{algoline}\nlset{\thealgoline}\quad \textbf{while} $t<T$ \textbf{do} 

\refstepcounter{algoline}\nlset{\thealgoline}\qquad \textbf{Step 1} (Proposal step): 

\refstepcounter{algoline}\nlset{\thealgoline}\qquad\qquad Sample a new candidate $\theta^{\star} \sim \mathcal{N}\left(\theta^{[t]},\tau^{[t]}\Sigma_{\theta}^{[t]}\right)$ 

\refstepcounter{algoline}\nlset{\thealgoline}\qquad\qquad Accept $\theta^{\star}$ with probability $\alpha\left(\theta^{\star},\theta^{[t]}\right)$. If accepted, $\theta_{t+1}=\theta^{\star}$; otherwise, $\theta^{[t+1]}=\theta^{[t]}$.

\refstepcounter{algoline}\nlset{\thealgoline}\qquad \textbf{Step 2} (Adaptation step): 

\refstepcounter{algoline}\nlset{\thealgoline}\qquad\qquad Update the scaling 
\begin{equation}
\label{updatescaling}
\text{log}\left(\tau^{[t+1]}\right) = \text{log}\left(\tau^{[t]}\right) + \nu^{[t+1]}\left[\alpha\left(\theta^{\star},\theta^{[t]}\right) -\bar{\alpha}_{opt}\right],
\end{equation}

\refstepcounter{algoline}\nlset{\thealgoline}\qquad\qquad Update the empirical covariance
\begin{equation}
\begin{split}
\mu_{\theta}^{[t+1]} &= \mu_{\theta}^{[t]} + \nu^{[t+1]} \left[\theta^{[t+1]}-\mu_{\theta}^{[t]}\right],\\
\Sigma_{\theta}^{[t+1]} &= \Sigma_{\theta}^{[t]} +  \nu^{[t+1]} \left[\left(\theta^{[t+1]}-\mu_{\theta}^{[t]}\right)\left(\theta^{[t+1]}-\mu_{\theta}^{[t]}\right)^\top - \Sigma_{\theta}^{[t]}\right].
\end{split}
\end{equation}

\refstepcounter{algoline}\nlset{\thealgoline}\qquad \textbf{return} $\theta^{[t+1]}, \mu_{\theta}^{[t+1]}, \Sigma_{\theta}^{[t+1]}, \tau^{[t+1]}$

\refstepcounter{algoline}\nlset{\thealgoline}\qquad $t \leftarrow t+1$

\refstepcounter{algoline}\nlset{\thealgoline}\quad \textbf{end while} 

\textbf{end procedure}

\end{algorithm}
\end{minipage}%
}
\end{center}

\begin{center}
\scalebox{0.95}{
\small
\begin{minipage}{1\linewidth}
\begin{algorithm}[H]
\caption{Adaptive Scaling Metropolis (ASM) algorithm - \textit{Univariate case}}
\label{alg:ASMunivariate}

\vspace{3mm}

\textbf{Procedure}

\refstepcounter{algoline}\nlset{\thealgoline}\quad \textbf{Step 0:} Initialize $t=0, \theta^{[0]}, \tau^{[0]}$.

\refstepcounter{algoline}\nlset{\thealgoline}\quad \textbf{while} $t<T$ \textbf{do} 

\refstepcounter{algoline}\nlset{\thealgoline}\qquad \textbf{Step 1} (Proposal step): 

\refstepcounter{algoline}\nlset{\thealgoline}\qquad\qquad Sample a new candidate $\theta^{\star} \sim \mathcal{N}\left(\theta^{[t]},\tau^{[t]}\right)$ 

\refstepcounter{algoline}\nlset{\thealgoline}\qquad\qquad Accept $\theta^{\star}$ with probability $\alpha\left(\theta^{\star},\theta^{[t]}\right)$. If accepted, $\theta_{t+1}=\theta^{\star}$; otherwise, $\theta^{[t+1]}=\theta^{[t]}$.

\refstepcounter{algoline}\nlset{\thealgoline}\qquad \textbf{Step 2} (Adaptation step): Update the scaling 
\begin{equation*}
\text{log}\left(\tau^{[t+1]}\right) = \text{log}\left(\tau^{[t]}\right) + \nu^{[t+1]}\left[\alpha\left(\theta^{\star},\theta^{[t]}\right) -\bar{\alpha}_{opt}\right].
\end{equation*}

\refstepcounter{algoline}\nlset{\thealgoline}\qquad \textbf{return} $\theta^{[t+1]}, \tau^{[t+1]}$

\refstepcounter{algoline}\nlset{\thealgoline}\qquad $t \leftarrow t+1$

\refstepcounter{algoline}\nlset{\thealgoline}\quad \textbf{end while} 

\textbf{end procedure}

\end{algorithm}
\end{minipage}%
}
\end{center}

\subsection{Community Detection Algorithms}\label{SCHSAR-APDX_CommunityDetection}

When unobserved homophily is present (see Subsection \ref{SCHSARhomophily}), obtaining a reliable initialization for the latent individual heterogeneity \(\{a_i\}_{i=1}^N\) can significantly improve the MCMC convergence. In the ordered-type specification, we assume that \(a_i\) takes values in \(\{1,\ldots,S\}\), where categories are ordered and homophily enters network formation through \(f(a_i,a_j)=-\vert a_i-a_j \vert\). A practical strategy is to initialize \(\mathbf{a}\) using available community detection algorithms to the observed network \(\mathbf W=[w_{ij}]\). Below we adopt two spectral-clustering variants that exploit the eigenstructure of graph matrices to reveal communities. In both algorithms \ref{alg:GCSpectralClust1} and \ref{alg:GCSpectralClust2}, we extract the top \(T\) eigenvectors of a chosen matrix (e.g., the Laplacian or modularity matrix), embed the \(N\) nodes into a \(T\)‑dimensional spectral subspace, and then apply a standard \(K\)-means method to partition them into \(S\) communities (with labels \(\tilde a_i\in\{1,\ldots,S\}\)). Because spectral clustering produces unordered cluster labels, we subsequently assign ordered categories by ranking the clusters according to the average outcome level within each cluster. Specifically, letting
\[
\bar{Y}_{(s)} \coloneqq \frac{1}{\sum_{i=1}^N \mathbbm{1}\{\tilde{a}_i=s\}} \sum_{i:\,\tilde{a}_i=s} Y_i , \quad s=1,\ldots,S,
\]
we relabel the clusters so that
\[
\bar{Y}_{(1)} \leq \bar{Y}_{(2)} \leq \ldots \leq \bar{Y}_{(S)},
\]
thereby obtaining ordered initial labels. This ordering step resolves the reflection indeterminacy of the latent types.

\begin{center}
\scalebox{0.95}{
\begin{minipage}{1\linewidth}
\begin{algorithm}[H]
\caption{Modularity‐Matrix Spectral Clustering algorithm}
\label{alg:GCSpectralClust1}

\vspace{3mm}

\textbf{Data:} Adjacency matrix $\mathbf{W} \in \mathbbm{R}^{N\times N}$, \# of clusters $S$, \# of top eigenvectors $T$ (typically $T=S$).\\

\textbf{Result:} Initial ordered labels $\{\tilde{a}_i\}_{i=1}^N$. \\

\refstepcounter{algoline}\nlset{\thealgoline}Compute degree vector $\mathbf{d} \in \mathbbm{R}^{N}$ where $d_{i} \coloneqq \sum_j w_{ij}$ and total edges $m = \frac{1}{2}\sum_{i=1}^{N}d_i$. \\

\refstepcounter{algoline}\nlset{\thealgoline}Form the modularity matrix
$$
\mathbf{M} \coloneqq \mathbf{W} - \frac{\mathbf{d}\mathbf{d}^\top}{2m}.
$$

\refstepcounter{algoline}\nlset{\thealgoline}Compute $T$ eigenvectors $\{\,\mathbf{v}_1,\dots,\mathbf{v}_T\}$ of $\mathbf{M}$ with largest positive eigenvalues.\\

\refstepcounter{algoline}\nlset{\thealgoline}Stack these eigenvectors into $\mathbf{V} \coloneqq [\,\mathbf{v}_1,\dots,\mathbf{v}_T\,]\in\mathbbm{R}^{N\times T}$.\\

\refstepcounter{algoline}\nlset{\thealgoline}Apply $K$-means clustering to the rows of $\mathbf{V}$, yielding labels $\{\tilde{a}_i\}_{i=1}^N$.\\

\refstepcounter{algoline}\nlset{\thealgoline}Order the cluster labels according to the cluster-wise mean outcome values $\bar{Y}_{(s)}$.\\

\end{algorithm}
\end{minipage}%
}
\end{center}

\begin{center}
\scalebox{0.95}{
\begin{minipage}{1\linewidth}
\begin{algorithm}[H]
\caption{Normalized‐Laplacian Spectral Clustering algorithm}
\label{alg:GCSpectralClust2}

\vspace{3mm}

\textbf{Data:} Adjacency matrix $\mathbf{W} \in \mathbbm{R}^{N\times N}$, \# of clusters $K$, \# of top eigenvectors $T$ (set $T=K=2$).\\

\textbf{Result:} Community assignments $\{\tilde{a}_i\}_{i=1}^N$. \\

\refstepcounter{algoline}\nlset{\thealgoline}Compute degree matrix $\mathbf{D} \coloneqq \mathrm{diag}(d_1,\dots,d_N)$ with $d_{i} \coloneqq \sum_j w_{ij}$. \\

\refstepcounter{algoline}\nlset{\thealgoline}Form the normalized Laplacian matrix
$$
\mathbf{L}_{\mathrm{norm}} \coloneqq \mathbf{I} - \mathbf{D}^{-\frac{1}{2}}\mathbf{W}\mathbf{D}^{-\frac{1}{2}}.
$$

\refstepcounter{algoline}\nlset{\thealgoline}Compute $T$ eigenvectors $\{\,\mathbf{v}_1,\dots,\mathbf{v}_T\}$ of $\mathbf{L}_{\mathrm{norm}}$ with smallest nonzero eigenvalues.\\

\refstepcounter{algoline}\nlset{\thealgoline}Stack these eigenvectors into $\mathbf{V} \coloneqq [\,\mathbf{v}_1,\dots,\mathbf{v}_T\,]\in\mathbbm{R}^{N\times T}$.\\

\refstepcounter{algoline}\nlset{\thealgoline}Row‐normalize $\mathbf{V}$
$$
\widetilde V_{i,:} \coloneqq \frac{V_{i,:}}{\|V_{i,:}\|_2}, 
\quad i=1,\dots,N.
$$

\refstepcounter{algoline}\nlset{\thealgoline}Apply $K$-means clustering to the rows of $\widetilde{\mathbf{V}}$, yielding labels $\{\tilde{a}_i\}_{i=1}^N$.\\

\refstepcounter{algoline}\nlset{\thealgoline}Order the cluster labels according to the cluster-wise mean outcome values $\bar{Y}_{(s)}$.\\

\end{algorithm}
\end{minipage}%
}
\end{center}

\medskip

By projecting nodes into a low-dimensional spectral subspace (\(T\ll N\)) instead of clustering directly in the original node space, these approaches effectively capture global connectivity and reveal nodes that are ``structurally similar'' in the same community. Alternative algorithms are also available in the \textsf{MATLAB} toolbox for community detection by Kehagias (\citeproc{ref-kehagias2018community}{2018}). In Monte Carlo experiments, we found that perfect clustering is unnecessary for our proposed MCMC sampler in Algorithm \ref{alg:SCHSAR2}. Moderate misclassification in the starting values is corrected throughout the MCMC procedure, which refines \(a_i\) using both network and outcome information.

\newpage

\section{On Simulation Study}\label{on-simulation-study}

\subsection{Additional Simulations for SCHSAR Model with Contextual Effects}\label{SCHSAR-APDX_extra_simulations}

This appendix presents additional Monte Carlo simulations for the full SCHSAR model that incorporates contextual effects through the inclusion of the spatially lagged covariate \(\mathbf{W}X\). The objective is to assess the robustness of the proposed Bayesian MCMC estimator when the outcome equation includes both endogenous peer effects and exogenous contextual interactions. The generation of the network data \(\{\mathbf{D}_i\}_{i=1}^N\) follows the same procedure as described in the main text. Specifically, the link function is probit and the network formation equation incorporates unobserved heterogeneity through the function \(f(a_i,a_j)\). The outcome data are generated according to the following equation:
\begin{equation}
\mathbf{Y} = \left[\mathbf{I}_N -\mathcal{L}\mathbf{W}\right]^{-1}\left(\tilde{\beta} \circ X + \tilde{\delta} \circ \mathbf{W}X +\tilde{\kappa} \circ \mathbf{a} + \mathbf{u}\right),
\end{equation}
where the covariate \(X_i \sim \mathcal{N}(0, 4)\), the error term \(u_i \sim \mathcal{N}(0, \sigma^2_{ug})\) for \(i\in \mathcal{I}_g\). The parameters \(\tilde{\beta}\), \(\tilde{\delta}\), and \(\tilde{\kappa}\) are generated as products of the matrix of latent type indicators \(\mathbf{z}\) and the corresponding coefficient vectors \(\beta\), \(\delta\), and \(\kappa\), respectively. Specifically,
\[
z_{i} \sim \mathcal{M}ult(1, \boldsymbol{\pi}), \quad \boldsymbol{\pi} = (0.4, 0.35, 0.25),
\]
\[
\beta = (-1.6, -0.5, 1.4), \quad \delta = (-1.1,-0.4,0.9), \quad \kappa = (1.15, 0.65, 0.15),
\]
\[
\tilde{\beta} = \mathbf{z}\beta, \quad \tilde{\delta} = \mathbf{z}\delta, \quad \tilde{\kappa} = \mathbf{z}\kappa,
\]
\[
\sigma_u^2 = 0.1^2\times[1,0.075,0.05].
\]
We examine two alternative specifications of the function \(f(a_i,a_j)\) to capture different forms of unobserved heterogeneity in the network formation process:

\begin{itemize}
\item[(i)] Unobserved degree heterogeneity: $f(a_i,a_j) = a_i+a_j, \text{ where } a_i \sim \mathcal{N}(0,\sigma_a^2), \sigma_a^2 = 2$.
\item[(ii)] Unobserved homophily: $f(a_i,a_j) = -\vert a_i-a_j\vert, \text{ where } a_i \sim \mathcal{C}at\left(\frac{1}{2},\frac{1}{2}\right)$.
\end{itemize}

Tables \ref{tab:tab-MCresults-DGPI-UDH} and \ref{tab:tab-MCresults-DGPII-UH} report the simulation results corresponding to these two cases, respectively, based on a sample size of \(N=1000\) and \(R=100\) Monte Carlo replicates. The results for the SCHSAR model are presented in the left panel, while the results for the HSAR model are shown in the right panel for comparison. Compared to the simulation design in the main text, the presence of contextual effects substantially increases the difficulty of estimation for the SCHSAR model, primarily because it strengthens collinearity among regressors and makes it more challenging to separately identify endogenous peer effects, contextual spillovers, and latent-type heterogeneity. This increased complexity is reflected in higher RMSE and weaker coverage rates for some parameters, particularly \(\lambda_g\), \(\beta_g\), and \(\delta_g\). However, the deterioration in performance remains considerably smaller than under the HSAR specification, indicating that the selection-corrected mixture structure continues to provide substantial efficiency gains even in the richer model. These simulations therefore serve as a robustness check for the general applicability of the SCHSAR estimation framework.

\begin{table}[H]
\centering
\caption{\label{tab:tab-MCresults-DGPI-UDH}DGP I: N= 1000, unobserved degree heterogeneity}
\centering
\resizebox{\ifdim\width>\linewidth\linewidth\else\width\fi}{!}{
\fontsize{10}{12}\selectfont
\begin{threeparttable}
\begin{tabular}[t]{ccccccccccccc}
\toprule
\multicolumn{3}{c}{\textbf{ }} & \multicolumn{5}{c}{\textbf{SCHSAR}} & \multicolumn{5}{c}{\textbf{HSAR}} \\
\cmidrule(l{3pt}r{3pt}){4-8} \cmidrule(l{3pt}r{3pt}){9-13}
SNR & Parameter & True Value & Mean & Std & Bias & RMSE & Coverage & Mean & Std & Bias & RMSE & Coverage\\
\midrule
 & $\lambda_1$ & -0.45 & -0.390 & 0.106 & 0.060 & 0.122 & 0.81 & -0.914 & 0.104 & -0.464 & 0.475 & 0.10\\

 & $\lambda_2$ & 0.15 & 0.098 & 0.095 & -0.052 & 0.109 & 0.81 & -0.662 & 0.273 & -0.812 & 0.856 & 0.09\\

 & $\lambda_3$ & 0.35 & 0.348 & 0.010 & -0.002 & 0.011 & 0.98 & 0.108 & 0.134 & -0.242 & 0.277 & 0.03\\

 & $\omega_1$ & 0.40 & 0.401 & 0.014 & 0.001 & 0.014 & 0.96 & 0.423 & 0.024 & 0.023 & 0.034 & 0.88\\

 & $\omega_2$ & 0.35 & 0.346 & 0.013 & -0.004 & 0.013 & 0.97 & 0.328 & 0.023 & -0.022 & 0.032 & 0.87\\

 & $\omega_3$ & 0.25 & 0.252 & 0.014 & 0.002 & 0.014 & 0.94 & 0.249 & 0.013 & -0.001 & 0.013 & 0.99\\

 & $\beta_{1}$ & -1.60 & -1.486 & 0.172 & 0.114 & 0.206 & 0.97 & -1.387 & 0.226 & 0.213 & 0.311 & 0.85\\

 & $\beta_{2}$ & -0.50 & -0.609 & 0.174 & -0.109 & 0.205 & 1.00 & -0.556 & 0.381 & -0.056 & 0.385 & 0.98\\

 & $\beta_{3}$ & 1.40 & 1.396 & 0.014 & -0.004 & 0.015 & 0.96 & 1.280 & 0.245 & -0.120 & 0.272 & 0.98\\

 & $\delta_{1}$ & -1.10 & -1.012 & 0.284 & 0.088 & 0.297 & 0.97 & -1.179 & 3.502 & -0.079 & 3.503 & 0.59\\

 & $\delta_{2}$ & -0.40 & -0.500 & 0.268 & -0.100 & 0.286 & 0.97 & -0.980 & 3.387 & -0.580 & 3.437 & 0.60\\

 & $\delta_{3}$ & 0.90 & 0.892 & 0.065 & -0.007 & 0.065 & 0.96 & 0.652 & 1.296 & -0.248 & 1.319 & 0.35\\

 & $\kappa_1$ & 1.15 & 1.103 & 0.076 & -0.047 & 0.090 & 0.96 & -- & -- & -- & -- & --\\

 & $\kappa_2$ & 0.65 & 0.693 & 0.079 & 0.043 & 0.090 & 0.98 & -- & -- & -- & -- & --\\

 & $\kappa_3$ & 0.15 & 0.151 & 0.005 & 0.001 & 0.005 & 0.96 & -- & -- & -- & -- & --\\

 & $\gamma$ & 1.50 & 1.504 & 0.009 & 0.004 & 0.010 & 0.79 & -- & -- & -- & -- & --\\

\multirow{-17}{*}{\centering\arraybackslash High} & $\sigma^2_a$ & 2.00 & 1.998 & 0.084 & -0.002 & 0.084 & 0.95 & -- & -- & -- & -- & --\\
\bottomrule
\end{tabular}
\begin{tablenotes}[para]
\item \textit{Notes:} This table displays results based on $R=100$ replicates. The values include the average and standard deviation of the point estimates; the average bias (Bias), the Root Mean Squared Error (RMSE), and the coverage rate (Coverage).
% across replicates; where $\text{Bias}=R^{-1}\sum_{r=1}^R (\hat{\alpha}_r-\alpha),\text{ RMSE}=\sqrt{R^{-1}\sum_{r=1}^R (\hat{\alpha}_r-\alpha)^2},$ and $\text{ Coverage}=R^{-1}\sum_{r=1}^R \mathbbm{1}\{\alpha\in \widehat{CI}_{0.95,r}\}$.
\end{tablenotes}
\end{threeparttable}}
\end{table}

\begin{table}[H]
\centering
\caption{\label{tab:tab-MCresults-DGPII-UH}DGP II: N= 1000, unobserved homophily}
\centering
\resizebox{\ifdim\width>\linewidth\linewidth\else\width\fi}{!}{
\fontsize{10}{12}\selectfont
\begin{threeparttable}
\begin{tabular}[t]{ccccccccccccc}
\toprule
\multicolumn{3}{c}{\textbf{ }} & \multicolumn{5}{c}{\textbf{SCHSAR}} & \multicolumn{5}{c}{\textbf{HSAR}} \\
\cmidrule(l{3pt}r{3pt}){4-8} \cmidrule(l{3pt}r{3pt}){9-13}
SNR & Parameter & True Value & Mean & Std & Bias & RMSE & Coverage & Mean & Std & Bias & RMSE & Coverage\\
\midrule
 & $\lambda_1$ & -0.45 & -0.353 & 0.111 & 0.097 & 0.147 & 0.57 & 0.993 & 0.017 & 1.443 & 1.444 & 0.00\\

 & $\lambda_2$ & 0.15 & 0.055 & 0.115 & -0.095 & 0.149 & 0.60 & 0.993 & 0.013 & 0.843 & 0.843 & 0.00\\

 & $\lambda_3$ & 0.35 & 0.346 & 0.021 & -0.004 & 0.022 & 0.96 & 0.571 & 0.021 & 0.221 & 0.222 & 0.00\\

 & $\omega_1$ & 0.40 & 0.400 & 0.014 & 0.000 & 0.014 & 0.94 & 0.401 & 0.015 & 0.001 & 0.015 & 0.98\\

 & $\omega_2$ & 0.35 & 0.347 & 0.013 & -0.003 & 0.013 & 0.95 & 0.347 & 0.013 & -0.003 & 0.014 & 0.96\\

 & $\omega_3$ & 0.25 & 0.252 & 0.014 & 0.002 & 0.015 & 0.95 & 0.252 & 0.015 & 0.002 & 0.015 & 0.94\\

 & $\beta_{1}$ & -1.60 & -1.476 & 0.187 & 0.124 & 0.224 & 0.98 & -1.452 & 0.193 & 0.148 & 0.244 & 0.96\\

 & $\beta_{2}$ & -0.50 & -0.618 & 0.189 & -0.118 & 0.223 & 1.00 & -0.641 & 0.198 & -0.141 & 0.243 & 0.93\\

 & $\beta_{3}$ & 1.40 & 1.395 & 0.017 & -0.005 & 0.017 & 0.94 & 1.395 & 0.021 & -0.005 & 0.021 & 0.92\\

 & $\delta_{1}$ & -1.10 & -1.015 & 0.131 & 0.085 & 0.156 & 0.98 & -0.385 & 1.533 & 0.715 & 1.692 & 0.34\\

 & $\delta_{2}$ & -0.40 & -0.502 & 0.155 & -0.102 & 0.185 & 0.98 & -0.057 & 0.985 & 0.343 & 1.043 & 0.40\\

 & $\delta_{3}$ & 0.90 & 0.900 & 0.076 & 0.000 & 0.076 & 0.90 & 0.998 & 0.195 & 0.098 & 0.218 & 0.56\\

 & $\kappa_1$ & 1.15 & 1.060 & 0.089 & -0.090 & 0.126 & 0.56 & -- & -- & -- & -- & --\\

 & $\kappa_2$ & 0.65 & 0.737 & 0.093 & 0.086 & 0.127 & 0.57 & -- & -- & -- & -- & --\\

 & $\kappa_3$ & 0.15 & 0.154 & 0.016 & 0.004 & 0.017 & 0.96 & -- & -- & -- & -- & --\\

\multirow{-16}{*}{\centering\arraybackslash High} & $\gamma$ & 1.50 & 1.500 & 0.005 & 0.000 & 0.005 & 0.85 & -- & -- & -- & -- & --\\
\bottomrule
\end{tabular}
\begin{tablenotes}[para]
\item \textit{Notes:} This table displays results based on $R=100$ replicates. The values include the average and standard deviation of the point estimates; the average bias (Bias), the Root Mean Squared Error (RMSE), and the coverage rate (Coverage).
% across replicates; where $\text{Bias}=R^{-1}\sum_{r=1}^R (\hat{\alpha}_r-\alpha),\text{ RMSE}=\sqrt{R^{-1}\sum_{r=1}^R (\hat{\alpha}_r-\alpha)^2},$ and $\text{ Coverage}=R^{-1}\sum_{r=1}^R \mathbbm{1}\{\alpha\in \widehat{CI}_{0.95,r}\}$.
\end{tablenotes}
\end{threeparttable}}
\end{table}

\end{document}